\documentclass[10pt,a4paper,twocolumn]{article}
\usepackage[T1]{fontenc}
\usepackage[utf8]{inputenc}
\usepackage{newtxtext,newtxmath}
\usepackage[margin=1.8cm,columnsep=0.7cm]{geometry}
\usepackage{graphicx,booktabs,array,amsmath,microtype,longtable,placeins,etoolbox}
\usepackage[super,sort&compress]{natbib}
\usepackage{xcolor}
\usepackage{xurl}
\usepackage[colorlinks=true,allcolors=teal]{hyperref}

\usepackage{authblk}

\newcommand{\GAPW}{GAPW}
\newcommand{\GTH}{GTH}

\newcommand{\uvec}{\boldsymbol{u}}
\newcommand{\gvec}{\boldsymbol{g}}
\title{Native implementation of the machine-learned Skala exchange--correlation functional in CP2K: Unified one-centre reconstruction for molecular and condensed-phase calculations}
\author[1,2]{Johann Pototschnig}
\author[1,2]{Franz P\"oschel}
\author[3]{J\"urg Hutter}
\author[1,2,4]{Thomas D. K\"uhne\textsuperscript{\dag}}
\affil[1]{Center for Advanced Systems Understanding (CASUS), G\"orlitz, Germany}
\affil[2]{Helmholtz-Zentrum Dresden-Rossendorf, Dresden, Germany}
\affil[3]{Department of Chemistry, University of Zurich, Zurich, Switzerland}
\affil[4]{Institute of Artificial Intelligence, Technische Universit\"at Dresden, Dresden, Germany}
\date{}
\graphicspath{{figures/}{../figures/}}

\hypersetup{
  pdftitle={Native implementation of the machine-learned Skala exchange--correlation functional in CP2K: Unified one-centre reconstruction for molecular and condensed-phase calculations},
  pdfauthor={Johann Pototschnig, Franz P\"oschel, J\"urg Hutter, Thomas D. K\"uhne},
  pdfsubject={Manuscript with supplementary information}
}
\begin{document}
\addtocontents{toc}{\protect\setcounter{tocdepth}{-1}}

\raggedbottom
\twocolumn[\maketitle
\begin{center}\small
\textsuperscript{\dag}E-mail: \href{mailto:tkuehne@cp2k.org}{tkuehne@cp2k.org}
\end{center}
\begin{abstract}
Machine-learned exchange--correlation (XC) functionals combine high molecular accuracy with the prospect of efficient condensed-phase simulation. We implement the Skala XC functional natively in CP2K using its Gaussian and plane-wave (GPW) and Gaussian and augmented-plane-wave (GAPW) methods. The central development is a joint one-centre reconstruction of density, density gradients, and kinetic-energy density before functional evaluation. This preserves mixed gradient terms and nonlocal couplings across the smooth and atom-local contributions for all-electron and pseudopotential calculations. The XC-specific GAPW representation resolves rapidly varying local contributions on atom-centred grids, reducing the required plane-wave resolution for pseudopotential calculations as well. The implementation incorporates Brillouin-zone sampling and point-group symmetry reduction.
With sufficiently flexible orbital bases, the native implementation retains the molecular benchmark accuracy of our earlier GauXC formulation. Crystalline CO$_2$, NH$_3$, and urea probe complementary interaction regimes, ranging from dispersion and quadrupolar electrostatics to hydrogen bonding and cooperative hydrogen-bond networks. All-electron Skala--D3(BJ) calculations give a mean absolute error of 1.54~kJ~mol$^{-1}$ against diffusion Monte Carlo for this three-crystal set. Across all thirteen DMC-ICE13 phases, the corresponding errors are 1.16~kJ~mol$^{-1}$ for absolute lattice energies and 0.82~kJ~mol$^{-1}$ for the twelve relative energies to ice Ih. At fixed experimental geometries, the all-electron/mixed-core calculations yield a band-gap MAE of 0.43~eV on the 15-material comparison set, substantially lower than those of commonly used semilocal functionals and comparable to those of widely used hybrid functionals. The LC10 benchmark nevertheless reveals systematically underestimated equilibrium lattice constants, indicating structural overbinding.  The resulting framework connects molecular Skala to condensed-phase electronic structure and enables future development using periodic many-body reference data.
\end{abstract}
\vspace{0.5cm}]

\section{Introduction}

The accuracy of Kohn--Sham density-functional theory (DFT) is governed to a large extent by the approximation to the exchange--correlation (XC) functional.\cite{Kohn1999Nobel,Jones2015DFT} Learning this contribution from accurate electronic-structure data offers an alternative to conventional functional forms, as exemplified by the DM21 functional.\cite{Kirkpatrick2021DM21} Skala combines local density information with a learned nonlocal representation and has demonstrated strong performance across molecular energetics.\cite{Skala2025} The Skala-1.1 model used here was trained on molecular and atomic data, with no periodic electron densities in its reported training set.\cite{Skala2025,Ehlert2026ACC} Applying the unchanged model to condensed phases therefore raises two distinct questions. Can the functional be evaluated consistently within a periodic electronic-structure method? Does its molecular accuracy transfer to the resulting electronic environments?

Condensed phases introduce coordination, polarization, screening, and collective response beyond the environments sampled by isolated molecules. Although the exact functional is universal, a learned approximation trained on a finite molecular dataset may not fully describe these effects.

CP2K's broader framework connects electronic structure to dynamics, transport, and spectroscopic response,\cite{CP2KDynamics2026} whereas its \textsc{Quickstep} module provides the underlying density-functional electronic-structure methods.\cite{CP2K2020} Our molecular implementation of Skala in CP2K through GauXC provides an independent reference for this development.\cite{Poeschel2026MolecularSkala,Williams2020} There, a Gaussian-basis density matrix is evaluated on an atom-centred quadrature. Here, the native implementation instead uses density fields constructed by CP2K's own Gaussian and plane-wave (GPW) and Gaussian and augmented-plane-wave (GAPW) machinery.\cite{Lippert1997,Lippert1999,CP2KMadeSimple2026} This gives direct access to the existing treatment of boundary conditions, reciprocal-space electrostatics, Brillouin-zone sampling, and symmetry. The important methodological issue is not merely replacing one quadrature library with another. A learned nonlocal functional must receive a jointly reconstructed density, including the one-centre contributions of the augmented representation.

We develop this reconstruction for all-electron (AE), pseudopotential, and mixed AE/pseudopotential calculations and describe its variational coupling to the electronic degrees of freedom. A molecular comparison establishes the connection to the existing Skala implementations.\cite{Skala2025,Poeschel2026MolecularSkala} We then consider three compact molecular crystals with complementary bonding motifs and the thirteen crystalline ice phases of DMC-ICE13, using diffusion Monte Carlo lattice energies as high-level references.\cite{DellaPia2022Ice,DellaPia2024X23} The LC10 benchmark assesses equilibrium lattice constants and bulk moduli. Finally, we examine electronic band gaps at fixed experimental geometries and compare them with experimental references and published density-functional calculations. Detailed numerical tests are collected in the Electronic Supplementary Information (ESI). The aim is to establish a practical native implementation while making the distinction between implementation consistency and condensed-phase model accuracy explicit.

\section{Theory and Implementation}
\label{sec:theory}

\subsection{From the density matrix to primitive fields}

In a Gaussian basis, the spin density at a point is obtained from the occupied Bloch orbitals or, equivalently, the density matrix as
\begin{equation}
 \rho_\sigma(\mathbf r)=\sum_{\mathbf k}w_{\mathbf k}
 \sum_{\mu\nu}P^\sigma_{\mu\nu}(\mathbf k)
 \chi_{\mu\mathbf k}(\mathbf r)\chi^*_{\nu\mathbf k}(\mathbf r),
 \label{eq:rho}
\end{equation}
where the index $\sigma\in\{\alpha,\beta\}$ labels the two spin channels. Here, $\mathbf r$ is the spatial coordinate and $\rho_\sigma(\mathbf r)$ is the spin-resolved electron density. The wavevectors $\mathbf k$ sample the Brillouin zone with weights $w_{\mathbf k}$ normalized as $\sum_{\mathbf k}w_{\mathbf k}=1$. The indices $\mu$ and $\nu$ label atom-centred Gaussian basis functions, whose Bloch sums are denoted by $\chi_{\mu\mathbf k}(\mathbf r)$, with the asterisk indicating complex conjugation. Although the elements $P^\sigma_{\mu\nu}(\mathbf k)$ of the Hermitian spin-density matrix $P^\sigma(\mathbf k)$ may be complex, their contraction with the Bloch basis functions yields a real density. The primitive input to the native interface is
\begin{equation}
 \begin{aligned}
 \uvec_\sigma(\mathbf r)&=\big(\rho_\sigma,\nabla\rho_\sigma,\tau_\sigma\big),\\
 \tau_\sigma&=\tfrac12\sum_{n\mathbf k}w_{\mathbf k}f_{n\mathbf k\sigma}
 |\nabla\psi_{n\mathbf k\sigma}|^2.
 \end{aligned}
 \label{eq:primitive}
\end{equation}
The field tuple $\uvec_\sigma(\mathbf r)$ combines the spin density and its spatial gradient $\nabla\rho_\sigma(\mathbf r)$ with the positive-definite kinetic-energy density $\tau_\sigma(\mathbf r)$. In atomic units, the latter is one half of the sum of the squared norms $|\nabla\psi_{n\mathbf k\sigma}|^2$ of the gradients of the Kohn--Sham Bloch orbitals $\psi_{n\mathbf k\sigma}(\mathbf r)$, weighted by their spin-channel occupations $f_{n\mathbf k\sigma}$ and the Brillouin-zone weights $w_{\mathbf k}$. Here, $n$ labels the bands and $\nabla$ differentiates with respect to $\mathbf r$.
Thus, the native Skala interface receives densities $\rho_\sigma$, gradients $\nabla\rho_\sigma$, and kinetic-energy densities $\tau_\sigma$, not the density matrix itself. Our molecular CP2K--GauXC interface instead passes the Gaussian-basis density matrix to GauXC, which constructs these fields on its atom-centred quadrature before model evaluation.\cite{Poeschel2026MolecularSkala,Williams2020} Both implementations evaluate the model on fields, but the native implementation keeps their construction and the corresponding potential-matrix assembly completely within CP2K.

On quadrature points $\mathbf r_i$ with weights $W_i$, the model energy has the form\cite{Skala2025}
\begin{equation}
 E_{\rm xc}^{\theta}=-C_{\rm x}\sum_i W_i
 \left(\rho_{\alpha i}^{4/3}+\rho_{\beta i}^{4/3}\right)
 f_{\theta}[\mathbf x]_i.
 \label{eq:skala-energy}
\end{equation}
Here, $C_{\rm x}=\tfrac34(6/\pi)^{1/3}$ is the spin-resolved exchange prefactor, and $\mathbf x$ contains spin densities, gradient invariants, and kinetic-energy densities. The parameters $\theta$ specify the trained model, and $\rho_{\sigma i}=\rho_\sigma(\mathbf r_i)$ denotes the density at quadrature point $i$. The enhancement factor $f_\theta$, evaluated at that point as $f_\theta[\mathbf x]_i$, depends on features at multiple points within an atom-associated block, rather than only at point $i$. Consequently, both the reconstructed fields and their spatial organization enter the discrete functional. Our contribution is the consistent construction and differentiation of this functional within GPW/GAPW, not a modification or retraining of the Skala model.

\subsection{GPW and the augmented density decomposition}

In GPW, Gaussian orbital products are collocated on a hierarchy of regular real-space grids and represented by plane waves for operations such as solving the Poisson equation.\cite{Lippert1997} A finer plane-wave grid resolves more rapidly varying products, but increases the cost throughout the cell. This is particularly restrictive for the sharply varying core density of an AE calculation. GAPW separates the global smooth density from atom-local contributions so that these two length scales need not be resolved on the same grid.\cite{Lippert1999}

For the primitive fields this separation is written as
\begin{equation}
 \uvec_\sigma=\widetilde{\uvec}_\sigma+
 \sum_{A,\mathbf L}\left(\uvec^h_{A\mathbf L,\sigma}
                  -\uvec^s_{A\mathbf L,\sigma}\right),
 \label{eq:reconstruct}
\end{equation}
where $A$ labels atoms and $\mathbf L$ the periodic images required by the boundary conditions. The hard field $\uvec^h_{A\mathbf L,\sigma}$ restores the local representation, while the soft field $\uvec^s_{A\mathbf L,\sigma}$ subtracts the corresponding contribution already included in the global smooth field $\widetilde{\uvec}_\sigma$. The subtraction prevents double counting. A one-centre correction is not an independent atomic energy added to the system.

The same construction is needed for all components of $\uvec_\sigma$. For example, in a real atom-local basis $\{\phi^q_{Aa}\}$, with $q\in\{h,s\}$ denoting the hard and soft representations, respectively, the one-centre density and kinetic-energy density have the structure
\begin{align}
 \rho^q_{A\sigma}(\mathbf r)&=\sum_{ab}D^{q,\sigma}_{A,ab}
                    \phi^q_{Aa}(\mathbf r)\phi^q_{Ab}(\mathbf r),\nonumber\\
 \tau^q_{A\sigma}(\mathbf r)&=\tfrac12\sum_{ab}D^{q,\sigma}_{A,ab}
                \nabla\phi^q_{Aa}(\mathbf r)\!\cdot\!\nabla\phi^q_{Ab}(\mathbf r).
 \label{eq:one-centre-fields}
\end{align}
The indices $a$ and $b$ label basis functions on atom $A$. The local coefficients $D^{q,\sigma}_{A,ab}$ form the corresponding spin-resolved density matrix and are obtained from the electronic density matrix through the GAPW projection. Density gradients follow by differentiating the orbital products. Reconstructing $\rho$ alone is therefore insufficient because $\tau$ is an orbital-dependent field and cannot be inferred by differentiating the reconstructed density. Finite one-centre expansions and finite grids introduce representation and integration errors into Eq.~\eqref{eq:reconstruct}. Their convergence must be checked together with the energy and its derivatives.

\subsection{AE and pseudopotential GAPW}

For AE GAPW (GAPW-AE), this reconstruction contains the explicitly treated core and valence electrons. With a Goedecker--Teter--Hutter (GTH) pseudopotential (PP), it reconstructs the explicit valence representation without reintroducing core electrons removed by the pseudopotential Hamiltonian.\cite{Goedecker1996} The field reconstruction is not specific to the GTH form and also applies to other effective core potentials (ECPs). This distinction between AE and pseudopotential fields is essential when comparing their results. Their difference is not, in general, an error of numerical integration.

In GAPW-AE, the hard-minus-soft contribution $\sum_{A,\mathbf L}(\uvec^h_{A\mathbf L,\sigma}-\uvec^s_{A\mathbf L,\sigma})$ in Eq.~\eqref{eq:reconstruct} is intrinsic to the chosen GAPW representation. Passing only its smooth part to Skala would omit part of the AE fields. An AE model evaluation without an \emph{explicit} one-centre correction is nevertheless possible through direct evaluation of the complete Gaussian-basis density matrix on an adequately resolved atom grid, as in the molecular GauXC implementation. There the full fields are already present. Whether reconstruction is needed depends on how the fields are represented, not on whether the integration library is GauXC or native CP2K.

For GTH atoms within ordinary GAPW, we compare direct-valence and one-centre reconstructed-valence representations, labelled \textit{direct} and \textit{one-centre}, respectively. The direct representation retains the unsmoothed valence basis and supplies GPW-like regular-grid fields, whereas the one-centre representation combines smooth fields with the atom-local hard-minus-soft contributions in Eq.~\eqref{eq:reconstruct}. The local reconstruction is analogous to the projector augmented-wave (PAW) method in its algebraic form.\cite{Bloechl1994PAW,Bloechl2026CPPAW} Here, however, it is neither a change to a PAW Hamiltonian nor an AE reconstruction of the removed core. The same PP and explicit electron count are retained. At finite numerical resolution, the two representations may supply different fields to the nonlinear model. Their agreement must therefore be assessed for observables rather than inferred solely from a common pseudopotential.

In mixed AE/GTH calculations, the direct/one-centre choice applies only to the GTH atoms. The AE one-centre reconstruction is retained in both variants. A mixed direct calculation therefore still contains the AE hard-minus-soft contributions. Here, mixed AE/GTH denotes the element-wise combination of AE and GTH core treatments.

\subsection{The XC-specific GAPW representation}

The XC-specific GAPW representation (GAPW-XC) uses the reconstruction in Eq.~\eqref{eq:reconstruct} with an XC-specific smooth field $\widetilde{\uvec}_{{\rm xc},\sigma}$ and matching local hard and soft contributions, allowing the XC and electrostatic representations to be chosen separately. CP2K derives the separate, softer XC density $\widetilde\rho_{\rm xc}$ and its corresponding gradient and kinetic-energy density from the same orbital density matrix. The complementary atom-local fields restore the structure shifted out of this regular-grid contribution.

The associated hard/soft projections and smooth collocation must be used consistently. The smooth term $\widetilde{\uvec}_{{\rm xc},\sigma}$ cannot simply be exchanged while leaving its reverse mapping unchanged. This XC-specific decomposition does not change the selected pseudopotential or add core electrons.

The practical advantage is that rapidly varying contributions to the XC input fields can be resolved on atom-centred grids without requiring an equally fine plane-wave grid throughout the cell. GAPW-XC is therefore our recommended default starting representation for native Skala calculations with pseudopotentials. The conformational-energy cutoff comparison between GPW and GAPW-XC in the molecular convergence tests (ESI Sec.~\ref{esi-sec:cutoff}) supports its faster convergence relative to GPW. The one-centre reconstruction makes this efficiency possible while retaining the model's response to the complete augmented fields. Standard GAPW and GAPW-XC distribute the regular-grid and atom-local work differently, so their finite-resolution errors are assessed through complete energy differences and derivatives. This numerical advantage is separate from the physical approximation introduced by replacing core electrons with a pseudopotential.

Table~\ref{tab:representations} separates the electronic Hamiltonian, the regular-grid XC field, and the local reconstruction. GPW uses direct regular-grid fields, whereas GAPW-AE and GAPW-XC include their intrinsic one-centre contributions. These three representations are not split into direct/one-centre variants in this work. The distinction applies only to GAPW-GTH and to the GTH atoms in mixed GAPW-AE/GTH. Merely deleting the matching local terms from Eq.~\eqref{eq:reconstruct} would leave the softer XC field incomplete. Selecting direct-valence GTH fields instead requires the corresponding regular-grid representation and does not by itself replace the remaining GAPW electrostatic machinery.

\begin{table*}[t]
\centering\small
\caption{Density representations in the native implementation. The direct/one-centre distinction concerns only GTH atoms within ordinary GAPW. AE reconstruction is retained in both mixed variants. Where present, one-centre fields are added before descriptor formation and Skala evaluation. No removed core electrons are restored by valence reconstruction.}
\label{tab:representations}
\vspace{3pt}
\begin{tabular}{@{}>{\raggedright\arraybackslash}p{0.215\textwidth}>{\raggedright\arraybackslash}p{0.13\textwidth}>{\raggedright\arraybackslash}p{0.27\textwidth}>{\raggedright\arraybackslash}p{0.295\textwidth}@{}}
\toprule
Representation & Hamiltonian & Regular-grid XC field & Atom-local contribution\\
\midrule
GPW & PP & Direct valence & None\\
GAPW-AE & AE & Smooth GAPW & Hard minus soft, including explicit core and valence\\
GAPW-GTH, direct & PP & Direct valence & None for PP kinds\\
GAPW-GTH, one-centre & PP & Smooth GAPW & Hard-minus-soft valence fields\\
Mixed AE/GTH, direct & AE + PP & Smooth AE and direct GTH fields & Hard minus soft for AE atoms only\\
Mixed AE/GTH, one-centre & AE + PP & Smooth AE and GTH fields & Hard minus soft for AE and GTH atoms\\
GAPW-XC/GTH & PP & Softer XC-specific field & Matching hard-minus-soft valence fields\\
\bottomrule
\end{tabular}
\end{table*}

\subsection{One reconstruction, one joint functional}
\label{sec:joint-functional}

Conventional GAPW organizes local or semilocal XC integration as a smooth-grid contribution plus atom-centred hard-minus-soft energy integrals. Its validity relies on the construction and matching of the local fields, not on linearity of the XC functional. For a nonlinear nonlocal model, the expression
\begin{equation}
 E_{\rm xc}[\widetilde{\uvec}]+\sum_A
 \big(E_{\rm xc}[\uvec^h_A]-E_{\rm xc}[\uvec^s_A]\big)
 \label{eq:wrong}
\end{equation}
is not interchangeable with evaluating the functional on Eq.~\eqref{eq:reconstruct}. Spin and image indices are suppressed in this schematic expression. The Skala model is nonlinear and highly nonlocal. Separate evaluations omit both mixed local descriptors and couplings between different contributions to the reconstructed environment.

Even a quadratic gradient invariant illustrates the issue. For one spin channel, let $\gvec=\nabla\rho_\sigma$, with the tilde and superscripts $h,s$ denoting its smooth, hard, and soft contributions. Writing $\gvec=\widetilde{\gvec}+\gvec^h-\gvec^s$ gives
\begin{align}
 |\gvec|^2={}&|\widetilde{\gvec}|^2+|\gvec^h|^2+|\gvec^s|^2
 +2\widetilde{\gvec}\!\cdot\!\gvec^h \nonumber\\
 &-2\widetilde{\gvec}\!\cdot\!\gvec^s-2\gvec^h\!\cdot\!\gvec^s.
 \label{eq:cross}
\end{align}
The cross terms exist wherever the corresponding gradient fields overlap, particularly in augmentation regions. The reconstructed invariant also contains the positive $|\gvec^s|^2$ term. Subtracting a separately evaluated soft energy is not equivalent to subtracting a field before forming its nonlinear descriptors. The spin-coupled case follows the same principle. The six smooth, hard, and soft spin-gradient vectors admit fifteen unordered off-diagonal products, twelve of which couple different components. The full expansion is given in ESI Sec.~\ref{esi-sec:spin-gradients}. A larger common grid alone does not recover these terms if the three densities are still passed to separate functional evaluations.

The nonlocal difficulty goes beyond this local algebra. A hard contribution near one centre can alter the learned response at another point in the same descriptor block, where the smooth density and neighbouring augmentation fields also contribute. Separate hard and soft model evaluations do not include this combined environment. Even exact quadrature of the separated expressions would not restore these missing dependencies.

\begin{figure}[t]
 \centering
 \includegraphics[width=\columnwidth]{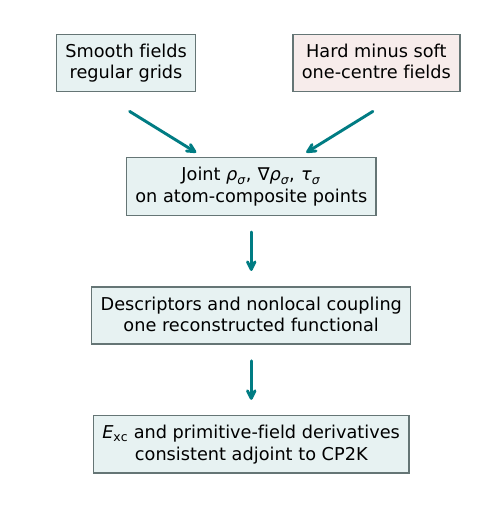}
 \caption{The native implementation reconstructs the primitive fields before descriptor formation and Skala evaluation. The reverse operation follows the adjoint of the same reconstruction, preserving the hard-minus-soft signs. The atom-centred quadrature resolves local structure without requiring a uniformly fine regular grid.}
 \label{fig:reconstruction}
\end{figure}

We therefore interpolate the smooth fields and add the hard-minus-soft fields \emph{pointwise}, before forming model descriptors. Skala then evaluates the jointly reconstructed fields (Fig.~\ref{fig:reconstruction}). Here, ``one evaluation'' means one consistent functional of the combined fields, not a restriction to one physical inference call. Atom-block batching and parallel execution remain possible. The distinction is between computational partitioning of the same functional and replacing it by a sum of different functionals.

\subsection{Accurate one-centre integration and joint reconstruction}

The accurate XC integration scheme in conventional GAPW improves cancellation between smooth-grid and soft atom-centred integrals evaluated on different quadratures.\cite{CP2K2020} The scheme uses the atom-local switch
\begin{equation}
 s_A(\mathbf r)=\exp[-\alpha_A|\mathbf r-\mathbf R_A|^2],
 \label{eq:xc-switch}
\end{equation}
where $\mathbf R_A$ is the nuclear position and $\alpha_A>0$ controls the inverse squared decay length of the dimensionless switch $s_A$. With periodic images implicit, the local or semilocal XC energy can be written schematically as
\begin{align}
 E_{\rm xc}^{\rm acc}\simeq{}&\int\!\left(1-\sum_A s_A\right)
                         e_{\rm xc}[\widetilde{\uvec}]\,d\mathbf r\nonumber\\
 &+\sum_A\int\!\left\{e_{\rm xc}[\uvec_A^h]
                  -(1-s_A)e_{\rm xc}[\uvec_A^s]\right\}d\mathbf r.
 \label{eq:accurate-integration}
\end{align}
Here, $e_{\rm xc}$ is a pointwise energy density, not an independent nonlocal Skala calculation, and $E_{\rm xc}^{\rm acc}$ denotes the energy in the accurate-integration scheme. Near a nucleus, the hard atom-centred term $e_{\rm xc}[\uvec_A^h]$ carries the rapidly varying contribution, while the nearly cancelling smooth and soft terms involving $\widetilde{\uvec}$ and $\uvec_A^s$ are suppressed. This rearrangement relies on GAPW's local matching assumptions. The switching weights $s_A$, quadrature, and their geometric derivatives must remain consistent.

Equation~\eqref{eq:accurate-integration} does not supply the mixed descriptors missing from Eq.~\eqref{eq:wrong}. The native Skala term instead uses its own jointly reconstructed fields and atom-composite quadrature. We retain accurate integration for the applicable conventional GAPW terms, but do not multiply the Skala reconstruction by the corresponding switching weights or add separately evaluated hard/soft Skala energies to it. This distinction separates an improvement of GAPW integration from the extension required for a learned nonlocal functional.

\subsection{Periodic atom-composite quadrature}

For a radial--angular point $\mathbf r_{Ai}$ centred on atom $A$, the discrete reconstruction is
\begin{equation}
 \begin{aligned}
 \uvec_{Ai,\sigma}={}&\sum_g I_{Ai,g}\widetilde{\uvec}_{g,\sigma}\\
 &+\sum_{B,\mathbf L}\left[\uvec^h_{B\mathbf L,\sigma}(\mathbf r_{Ai})
                        -\uvec^s_{B\mathbf L,\sigma}(\mathbf r_{Ai})\right].
 \end{aligned}
 \label{eq:composite}
\end{equation}
The interpolation $I$ maps the regular-grid fields to the target atom grid, with $I_{Ai,g}$ connecting regular-grid point $g$ to point $i$ of atom block $A$. The second sum includes neighbouring centres $B$ and periodic images $\mathbf L$, not only the target atom. Thus, ``one-centre'' describes the source expansion. It does not mean that its contribution is confined to a separate model call or omitted from neighbouring descriptor blocks. The sum is absent for direct-valence PP kinds, while intrinsic AE contributions are retained.

Let $w_{Ai}$ denote the unpartitioned radial--angular quadrature weight. Smooth atom-image partition functions define the energy weights through
\begin{equation}
 p_{A\mathbf L}(\mathbf r)=
 \frac{q_{A\mathbf L}(\mathbf r)}{\sum_{B,\mathbf L'}q_{B\mathbf L'}(\mathbf r)},
 \qquad \sum_{A,\mathbf L}p_{A\mathbf L}(\mathbf r)=1,
 \label{eq:partition}
\end{equation}
so that a target grid carries $W_{Ai}=w_{Ai}p_{A\mathbf0}(\mathbf r_{Ai})$, where $\mathbf0$ selects the central image of atom $A$. The differentiable shape functions $q$ provide a Becke-like partition.\cite{Becke1988} Periodic images distribute the integration over equivalent cells without double counting.

For the nonlocal descriptors, the partition is rebuilt using only the images $\mathcal S_A$ of the target atom $A$, giving
\begin{equation}
 \begin{aligned}
 d_A(\mathbf r)&=\frac{q^{\mathcal S_A}_{A\mathbf0}(\mathbf r)}
 {\sum_{(A,\mathbf L)\in\mathcal S_A}q^{\mathcal S_A}_{A\mathbf L}(\mathbf r)},\\
 D_{Ai}&=w_{Ai}\,T\!\left[d_A(\mathbf r_{Ai})\right].
 \end{aligned}
 \label{eq:descriptor-weights}
\end{equation}
The superscript identifies the centres used to construct the shape functions. The descriptor partition $d_A$ is not obtained by merely renormalizing the all-atom weights $p$. The smooth window $T$ suppresses negligible self-image weights and is unity away from that tail. Thus, $W_{Ai}$ weights the energy integral, whereas $D_{Ai}$ weights the model's internal nonlocal processing. An atom block sees the jointly reconstructed fields, including neighbouring atoms and their required images, throughout its windowed radial--angular grid. Neighbouring atoms affect these fields but do not partition the descriptor domain into separate atomic regions. The image sums are evaluated on finite target-block layouts. Normalization within a finite layout does not establish image-shell convergence. ESI Sec.~\ref{esi-sec:periodic-weights} defines the shape functions, window, image layouts, and interpolation. The geometric derivatives include the dependence on both sets of weights.

The role of $D_{Ai}$ can be made explicit by writing one nonlocal layer schematically as\cite{Skala2025}
\begin{equation}
 \begin{aligned}
 \mathbf m_A&=\sum_i D_{Ai}\,
     \mathsf K_\theta(\mathbf r_{Ai}-\mathbf R_A)\,\mathbf h_{Ai},\\
 \mathbf h'_{Ai}&=\mathcal U_\theta\!\left(
     \mathbf h_{Ai},\mathbf r_{Ai}-\mathbf R_A,\mathbf m_A\right).
 \end{aligned}
 \label{eq:nonlocal-layer}
\end{equation}
Here, $\mathbf h_{Ai}$ denotes transformed field features, or hidden features in later layers. The vector $\mathbf m_A$ collects their coarse atom-centred representation, and $\mathbf h'_{Ai}$ denotes the updated fine-grid features, with the prime indicating a layer update rather than a derivative. The kernel $\mathsf K_\theta$ comprises radial functions, spherical harmonics, and channel projections onto the atom-centred coarse point. The update $\mathcal U_\theta$ represents equivariant mixing and nonlinear contraction of the coarse features, their projection back onto the same atom's fine grid, and a local update. Repeating these layers yields the enhancement factor in Eq.~\eqref{eq:skala-energy}. The final energy uses $W_{Ai}$, not $D_{Ai}$. This schematic does not introduce message passing between different atomic coarse points. Neighbour information instead enters through the jointly reconstructed fields sampled by each block. The window therefore defines a nonlocal integration domain, not an isolated-atom density approximation.

The same construction reduces to a molecular atom-composite quadrature when no periodic images are required. A regular common grid can also support joint reconstruction, but resolving hard local fields everywhere is expensive. On the atom-composite grid, radial and angular resolution control the local structure independently of the plane-wave cutoff for the smooth contribution. This is an efficiency distinction, not a reason to evaluate different functionals on the two layouts.

Particle number provides an additional consistency check. The number of explicitly treated electrons is obtained from the density and overlap matrices as
\begin{equation}
 N=\sum_{\mathbf k\sigma}w_{\mathbf k}
           \mathrm{Tr}[P^\sigma(\mathbf k)S(\mathbf k)].
 \label{eq:particle-number}
\end{equation}
Here, $S(\mathbf k)$ is the Bloch-basis overlap matrix and $\mathrm{Tr}$ denotes the matrix trace. Integrating the reconstructed density gives $\widetilde N+\sum_A(N_A^h-N_A^s)$, where $\widetilde N$ and $N_A^{h,s}$ are the spin-summed integrals of the smooth and local hard/soft densities. The finite model quadrature instead gives $N_Q=\sum_{Ai\sigma}W_{Ai}\rho_{Ai,\sigma}$. A discrepancy between $N_Q$ and Eq.~\eqref{eq:particle-number} tests the field representation and quadrature. It does not, by itself, imply a change in the self-consistent-field (SCF) occupations. No density-dependent renormalization is applied, since it would alter both the functional and its variational derivative. Cutoff, radial/angular, and one-centre convergence are consequently distinct from changing the core treatment.

Finally, a nonlinear core correction (NLCC) must not be confused with hard-minus-soft reconstruction. An NLCC potential supplies a separate frozen-core density that can be added to the valence density and gradient before descriptor formation.\cite{Louie1982NLCC,Willand2013NLCC} It does not by itself provide a corresponding core kinetic-energy density. This is not equivalent to the explicit core orbitals of GAPW-AE, and it is not the origin of the one-centre corrections compared here.

\subsection{Variational coupling, forces, and stress}
\label{sec:variational}

Let $\mathcal R$ denote the discrete reconstruction from the density matrix $P$ to the model fields. The XC contribution to the generalized Kohn--Sham matrix is obtained by the chain rule as
\begin{equation}
 V_{\rm xc}=\left(\frac{\partial\mathcal R}{\partial P}\right)^{\!*}
             \frac{\partial E_{\rm xc}}{\partial\uvec}.
 \label{eq:adjoint}
\end{equation}
The asterisk denotes the adjoint under the chosen discrete contractions. The symbols $P$, $\uvec$, and $V_{\rm xc}$ collect the density matrices, model fields, and XC potential matrices, with spin indices implicit and k-point indices suppressed on the matrices. Overbars denote energy derivatives with respect to the indicated fields or local density-matrix coefficients. With $\overline{\uvec}_{Ai}=\partial E_{\rm xc}/\partial\uvec_{Ai}$, the component-wise adjoints from Eq.~\eqref{eq:composite} are
\begin{align}
 \overline{\widetilde{\uvec}}_g&=\sum_{Ai}I_{Ai,g}\overline{\uvec}_{Ai},\nonumber\\
 \overline D^h_A&=\left(\frac{\partial\uvec^h_A}{\partial D^h_A}\right)^{\!*}
                       \overline{\uvec},\qquad
 \overline D^s_A=-\left(\frac{\partial\uvec^s_A}{\partial D^s_A}\right)^{\!*}
                       \overline{\uvec}.
 \label{eq:component-adjoints}
\end{align}
The contractions include every target point to which a source contributes. The quadrature weights are already included in $\overline{\uvec}$, so applying them a second time would be incorrect. The local coefficients are then projected back to the orbital density matrix. Thus, the reverse path is the transpose of the actual discrete forward map, rather than an independently interpolated potential. This consistency is particularly important for the cancellation of hard and soft contributions and for SCF convergence near the numerical noise floor.
Kernel-level scalar-product tests verify the adjoint identity for the atom-local field interpolation, while equivalent back-projection variants agree at floating-point precision (ESI Table~\ref{esi-tab:adjoint-validation}). These checks complement the self-consistent force and stress tests below.

More explicitly, write the field adjoints as
\begin{equation}
 \begin{aligned}
 a_{Ai\sigma}&=\frac{\partial E_{\rm xc}}{\partial\rho_{Ai\sigma}},\qquad
 \mathbf b_{Ai\sigma}=\frac{\partial E_{\rm xc}}{\partial\nabla\rho_{Ai\sigma}},\\
 c_{Ai\sigma}&=\frac{\partial E_{\rm xc}}{\partial\tau_{Ai\sigma}}.
 \end{aligned}
 \label{eq:field-adjoints}
\end{equation}
Skala supplies the field derivatives through LibTorch automatic differentiation, while CP2K performs the adjoint reconstruction and assembles the XC matrix. These derivatives are obtained by back-propagating through all nonlocal layers and spin-gradient invariants of the full discrete energy, rather than differentiating an isolated local energy density. At fixed geometry, define the discrete field-response kernels $B^\rho$, $\mathbf B^g$, and $B^\tau$ for the density, its gradient, and the kinetic-energy density, respectively, by
\begin{equation}
 \delta\uvec_{Ai\sigma}=\sum_{\mathbf k\mu\nu}w_{\mathbf k}
 \begin{pmatrix}B^\rho_{Ai,\mu\nu}(\mathbf k)\\
 \mathbf B^g_{Ai,\mu\nu}(\mathbf k)\\B^\tau_{Ai,\mu\nu}(\mathbf k)\end{pmatrix}
 \delta P^\sigma_{\nu\mu}(\mathbf k).
 \label{eq:field-response}
\end{equation}
The symbol $\delta$ denotes a first-order variation at fixed geometry. With the convention $\delta E_{\rm xc}=\sum_{\mathbf k\sigma}w_{\mathbf k}
\mathrm{Tr}[V^\sigma_{\rm xc}(\mathbf k)\delta P^\sigma(\mathbf k)]$, the matrix is
\begin{equation}
 \begin{aligned}
 \bigl[V^\sigma_{\rm xc}(\mathbf k)\bigr]_{\mu\nu}
 &=\sum_{Ai}a_{Ai\sigma}B^\rho_{Ai,\mu\nu}(\mathbf k)\\
 &\quad+\sum_{Ai}\mathbf b_{Ai\sigma}\!\cdot\!\mathbf B^g_{Ai,\mu\nu}(\mathbf k)\\
 &\quad+\sum_{Ai}c_{Ai\sigma}B^\tau_{Ai,\mu\nu}(\mathbf k).
 \end{aligned}
 \label{eq:explicit-xc-matrix}
\end{equation}
For direct evaluation of orbital fields, these kernels would be
\begin{equation}
 \begin{aligned}
 B^\rho_{\mu\nu}&=\chi_{\mu\mathbf k}^*\chi_{\nu\mathbf k},\\
 \mathbf B^g_{\mu\nu}&=\nabla(\chi_{\mu\mathbf k}^*\chi_{\nu\mathbf k}),\\
 B^\tau_{\mu\nu}&=\tfrac12\nabla\chi_{\mu\mathbf k}^*\!\cdot\!\nabla\chi_{\nu\mathbf k},
 \end{aligned}
 \label{eq:orbital-field-kernels}
\end{equation}
evaluated at the target point $\mathbf r_{Ai}$. In the native GAPW implementation, however, the kernels in Eq.~\eqref{eq:field-response} include the smooth-grid interpolation, hard-minus-soft fields, and one-centre density-matrix projections of the actual reconstruction. Replacing them by direct orbital products would generally change the finite discretization. The real one-centre products and their derivatives are given in Eq.~\eqref{eq:one-centre-fields}. Equation~\eqref{eq:explicit-xc-matrix} adds no second quadrature factor or k-point weight. Its kinetic-energy-density term is a generalized Kohn--Sham matrix contribution rather than a multiplicative density potential.

At fixed orbital density matrix, the explicit XC derivative with respect to a nuclear coordinate or cell-strain parameter $\lambda$ includes geometry-dependent mappings and takes the form
\begin{equation}
 \left.\frac{\partial E_{\rm xc}}{\partial\lambda}\right|_P=
 \sum_{Ai}\overline{\uvec}_{Ai}\!\cdot\!
                 \left.\frac{\partial\uvec_{Ai}}{\partial\lambda}\right|_P
 +\left.\frac{\partial E_{\rm xc}}{\partial\lambda}\right|_{\uvec}.
 \label{eq:geometric-derivative}
\end{equation}
The restrictions $|_P$ and $|_{\uvec}$ hold the density matrix and model fields fixed, respectively, and the dot product contracts the field components and spin channels. The first term includes interpolation and hard/soft-field derivatives, while the second includes explicit model-coordinate and quadrature-weight dependence, including both periodic partitions. Stationarity of the complete electronic energy and the basis/overlap Pulay terms are then used to obtain the total force and stress. Because the XC contribution alone is not stationary with respect to the density matrix, model differentiation does not provide the full derivative of the self-consistent total energy. The nuclear force is $-dE/d\mathbf R_A$, and the strain derivative per cell volume gives the stress in the adopted sign convention. Self-consistent finite-difference (FD) checks of a force and diagonal stress component in a periodic water test support the derivative consistency of all five representations. Across the five tested representations and both FD step sizes, the maximum absolute discrepancies are approximately $2.9\times10^{-6}$~$E_h$/bohr and $4.7\times10^{-4}$~GPa. These results concern one Cartesian force component and one diagonal stress component of the periodic $\Gamma$-point water system. The FD protocol, numerical errors, and tested scope are documented in ESI Sec.~\ref{esi-sec:fd} and Table~\ref{esi-tab:derivatives}.

\subsection{K-points, symmetry reduction and parallelism}
\label{sec:kpoints}

Brillouin-zone integration on regular k-point meshes\cite{Monkhorst1976} enters through the density matrix and Eq.~\eqref{eq:rho}. Symmetry reduction must transform orbital components and Bloch phases consistently, together with the integration weights. Multiplying a scalar result by a symmetry count is insufficient. Once the real primitive fields are assembled, Skala need not know whether they originated from a full k-point mesh or a symmetry-reduced one. Their adjoints are returned to the corresponding complex Hermitian Kohn--Sham matrices. The native implementation uses CP2K's k-point and symmetry infrastructure,\cite{CP2KDynamics2026,Alizadeh2026PeriodicGFN2} with full point-group reduction through spglib in the molecular-crystal calculations.\cite{Togo2024Spglib} Numerical full/reduced-mesh comparisons are given in ESI Sec.~\ref{esi-sec:symmetry-meshes} and Table~\ref{esi-tab:symmetry}.

CP2K supplies fields, atom-associated coordinates, energy weights, and descriptor weights directly to the same Skala TorchScript model used by the molecular implementation. LibTorch returns the energy and input derivatives without a GauXC integration step. Complete atom blocks can be distributed over Message Passing Interface (MPI) ranks or batched on a device because the nonlocal processing retains the atom-block structure. Threaded field construction and back-projection complement inference on central processing units (CPUs) or graphics processing units (GPUs). Batching preserves the functional only if each block's descriptor domain and all contributing reconstructed fields are retained. This provides parallel execution without reverting to the separated smooth/hard/soft functional in Eq.~\eqref{eq:wrong}.

\section{Computational Strategy and Numerical Validation}
\label{sec:strategy}

We use the Skala-1.1 model distributed through Hugging Face\cite{SkalaModelCard} with its accompanying D3(BJ) dispersion contribution.\cite{Skala2025,Grimme2011D3BJ} We report the mean absolute error (MAE) and complementary error statistics, retaining each source's recorded dispersion contribution. The native Gaussian basis sets and GTH pseudopotentials use the 2026.2 parameter definitions of the University of Zurich (UZH) protocol.\cite{Mirhosseini2026UZH} The UZH AE bases are def2-derived bases optimized for GAPW using the molecularly optimized (MOLOPT) strategy, rather than unchanged def2 sets.\cite{Weigend2005,VandeVondele2007,Mirhosseini2026UZH} The native AE calculations use the triple- and quadruple-zeta valence plus double-polarization bases TZVPP-ae and QZVPP-ae, respectively, rather than the def2-TZVP/ma-def2-TZVP protocol of the earlier molecular GauXC study.\cite{Poeschel2026MolecularSkala} Bases with the same nominal zeta level can differ in accuracy between these families.

The molecular AE series for the dietGMTKN55 subset and the AE calculations for CO$_2$, NH$_3$, and urea treat every atom all-electron, as do the ICE13 crystals and their isolated-water references. The molecular and molecular-crystal comparisons also include calculations using GTH pseudopotentials for all atoms. Mixed AE/GTH calculations serve as additional core-treatment controls in LC10 and extend the band-gap study to compounds not fully covered by the available AE bases.

The numerical tests below guide the choice of settings for the molecular and condensed-phase applications presented in Sec.~\ref{sec:results}. For isolated molecules, each cell dimension is the corresponding molecular extent plus 25~\AA{}. We use nonperiodic boundary conditions and an analytic isolated-system Poisson solver, which evaluates an analytical reciprocal-space Green's function for zero-dimensional electrostatics using a truncated Coulomb kernel instead of the fully periodic one. Thus, molecular isolation does not rely on vacuum separation alone. The cell-padding study in ESI Sec.~\ref{esi-sec:padding} compares added extents of 22, 24, and 25~\AA{} under the same isolated-system electrostatics. These checks support the selected 25~\AA{} padding for our high-accuracy molecular calculations.

The regular-grid cutoff tests in ESI Sec.~\ref{esi-sec:cutoff} start with GAPW-AE for H$_2$O using the QZVPP-ae basis and a fixed 50/50 atom grid (Table~\ref{esi-tab:ae-cutoff}). Relative to the 1200~Ry reference, the total-energy deviations are substantially smaller at 600--1000~Ry than at 400~Ry. On this basis, we choose 640~Ry for the molecular AE calculations, which is slightly above the 600~Ry test point.

For GTH pseudopotentials, we compare the GPW and GAPW-XC alkane ACONF conformational energies, calculated with the TZV2P-GTH basis, against their respective 800~Ry results. The weaker cutoff dependence of GAPW-XC supports the choice of 400~Ry for the molecular GAPW-XC/GTH calculations (ESI Table~\ref{esi-tab:aconf-cutoff}). GTH pseudopotentials remove the explicit core electrons and smooth the valence orbitals, reducing the regular-grid resolution requirements. GAPW-XC further lowers these requirements by treating rapidly varying local contributions on atom-centred grids. At 800~Ry, the GPW and GAPW-XC conformational energies agree to within 0.009~kJ~mol$^{-1}$, supporting the accuracy of the combined GAPW density-decomposition and one-centre reconstruction approximations within GAPW-XC.

The plane-wave cutoff sets the resolution of the finest density grid, whereas the relative cutoff controls the assignment of Gaussian products to the multigrid levels. Increasing the latter assigns more products to finer grids. The higher AE cutoff is a conservative choice. GAPW treats the rapidly varying near-core density on atom-centred grids, but tighter Gaussian exponents retained in its smooth density representation can lead to slower plane-wave cutoff convergence.\cite{CP2KMadeSimple2026}

ESI Sec.~\ref{esi-sec:quadrature} independently tests radial and angular quadrature for GPW/GTH water at 800/60~Ry (Table~\ref{esi-tab:water-quadrature}). These tests show that increasing angular sampling alone is insufficient. Both radial and angular sampling must be refined to reduce the Skala quadrature error. Although the self-consistent density matrix has the prescribed electron number, fixed by its occupations and overlap-matrix trace, the finite atom-centred quadrature can still produce a different integral. We therefore monitor the integrated electron count separately to assess quadrature accuracy, without renormalizing the density.

Orbital-basis flexibility and integration accuracy are separate requirements. An AE basis must represent compact core states as well as the valence orbitals, whereas GTH pseudopotentials remove the explicit core states and smooth the valence orbitals. The molecular and lattice-energy comparisons motivate QZVPP-ae as the preferred AE basis for these energy studies, with the TZVPP-ae basis retained to quantify basis sensitivity (ESI Tables~\ref{esi-tab:urea-basis}, \ref{esi-tab:common62-summary}, and \ref{esi-tab:ice-basis}--\ref{esi-tab:ice-relative}). GTH calculations instead use the smaller valence-only TZV2P-GTH basis.

For lattice energies, numerical integration errors in the crystal and gas-phase molecular energies need not cancel. We therefore use the same refined integration settings for both phases. Crystal and molecule also use the same orbital basis, with isolated-system electrostatics for the gas-phase reference.

 In the molecular-crystal tests, the finer AE atom quadrature resolves the more rapidly varying core-density structure, while the smoother GTH valence density permits the less demanding quadrature used here. This local integration requirement is distinct from the choice of orbital basis and plane-wave cutoff.  The AE basis and quadrature comparisons are reported in ESI Sec.~\ref{esi-sec:crystal-numerical-controls}, Table~\ref{esi-tab:urea-basis}.

For the molecular crystals, the tested atom-centred quadrature refinement changes the lattice energy more than the tested k-point and symmetry-tolerance adjustments. CPU/CUDA lattice-energy differences are below 0.001~kJ~mol$^{-1}$ for CO$_2$ with GAPW-XC/GTH and urea with GAPW-AE. These controls and the crystal-only cutoff test are listed in ESI Table~\ref{esi-tab:x23_controls_si}. These fixed-input comparisons test device consistency. Full/reduced-mesh checks assess symmetry reduction at fixed k-point sampling (ESI Sec.~\ref{esi-sec:symmetry-meshes}), while self-consistent FD tests assess forces and stress (ESI Sec.~\ref{esi-sec:fd}).

\section{Results and Discussion}
\label{sec:results}

\subsection{Molecular energetics and the GauXC connection}
\label{sec:molecular-results}

We use a selected set of 62 dietGMTKN55 reactions\cite{Gould2018,Goerigk2017} to test whether the periodic native-grid Skala implementation retains the molecular benchmark accuracy of our earlier molecular GauXC implementation in the isolated-molecule limit.\cite{Poeschel2026MolecularSkala} Each molecule is placed in a large supercell with nonperiodic boundary conditions and isolated-system electrostatics. On the same reactions, we compare GauXC GPW/GTH, GauXC GAPW-AE, native GAPW-XC/GTH, native GAPW-AE with the TZVPP-ae basis, and native GAPW-AE with the QZVPP-ae basis. PySCF provides an additional independent all-electron Skala comparison using the same def2-TZVP/ma-def2-TZVP basis protocol as GauXC GAPW-AE. 

 The production settings follow the numerical tests in Sec.~\ref{sec:strategy} and are specified in ESI Sec.~\ref{esi-sec:molecular-data}.

The 62-reaction set was selected to ensure fully all-electron treatment of every constituent species in all four AE protocols. The two pure-GTH protocols are evaluated on this identical reaction set. 

Table~\ref{tab:molecular-errors} summarizes the reference-error statistics for this common reaction set. Native QZVPP-ae, GauXC-AE, and PySCF give closely comparable mean absolute errors against the benchmark references. Native GAPW-XC/GTH and GauXC GPW/GTH show similarly good agreement in this measure. The mean absolute differences (MADs) between the corresponding native and GauXC reaction energies are reported separately in ESI Table~\ref{esi-tab:common62-pairs}. These MADs measure agreement between predictions, whereas the MAEs measure accuracy against the benchmark references. Complete statistics and reaction energies appear in ESI Tables~\ref{esi-tab:common62-summary}--\ref{esi-tab:common62-external}.

\begin{table}[tb]
\centering\small
\setlength{\tabcolsep}{3pt}
\caption{Unweighted mean absolute errors (MAEs) and root-mean-square errors (RMSEs) against the references for the same selected 62 dietGMTKN55 reactions, in kJ~mol$^{-1}$. The def2 entries use def2-TZVP/ma-def2-TZVP. The GTH entries use pseudopotentials.}
\label{tab:molecular-errors}
\begin{tabular}{@{}llrr@{}}
\toprule
Protocol & Basis & MAE & RMSE \\
\midrule
GauXC GAPW-AE & def2 & 4.05 & 9.94 \\
PySCF AE & def2 & 4.02 & 9.92 \\
Native GAPW-AE & TZVPP-ae & 7.14 & 14.72 \\
Native GAPW-AE & QZVPP-ae & 4.29 & 11.19 \\
\midrule
GauXC GPW/GTH & TZVP-GTH & 4.98 & 9.62 \\
Native GAPW-XC/GTH & TZV2P-GTH & 5.07 & 10.12 \\
\bottomrule
\end{tabular}
\end{table}

The slightly larger errors of the GTH protocols relative to the QZVPP-ae and def2 AE protocols are consistent with the trend reported in our molecular GauXC study.\cite{Poeschel2026MolecularSkala} In both GTH approaches, Skala receives the valence pseudodensity, without the explicit core-electron density and with a smoothed near-nuclear valence profile. This change from the AE training densities for light elements provides a plausible explanation for the trend. One-centre reconstruction of GTH valence fields does not recover the removed core electrons.

The effect of the native AE basis change on each reaction, including both improvements and deteriorations, is shown in ESI Fig.~\ref{esi-fig:common62-basis-improvements}.

The nominal triple-zeta def2-TZVP/ma-def2-TZVP protocol gives lower reference errors than TZVPP-ae on this selected set. However, with the QZVPP-ae basis, the native mean absolute error approaches those of GauXC and PySCF.

\subsection{Three molecular crystals with complementary bonding}
\label{sec:crystal-results}

CO$_2$, NH$_3$, and urea provide a compact test of transfer from isolated molecules to crystals. In nonpolar CO$_2$, cohesion combines London dispersion with electrostatic interactions between molecular quadrupoles, without hydrogen bonds. Polar NH$_3$ molecules form a network of directional N--H$\cdots$N hydrogen bonds. In urea, the carbonyl oxygen accepts hydrogen bonds from neighbouring N--H groups, connecting the polar molecules through an extended N--H$\cdots$O network.\cite{Morrison2003HydrogenBonds} The selection therefore contrasts dispersion-dominated binding with hydrogen-bonded environments that differ in donor--acceptor chemistry and network connectivity.

Crystal and independently supplied gas-phase molecular geometries are taken from the versioned supporting dataset of Della Pia \emph{et al.}\cite{DellaPia2024X23} We compare the electronic lattice energies with the corresponding diffusion Monte Carlo (DMC) references. For a unit cell containing $Z$ molecules, the lattice energy per molecule is
\begin{equation}
 E_{\rm latt}=E_{\rm crystal}/Z-E_{\rm molecule},
 \label{eq:lattice}
\end{equation}
where $E_{\rm crystal}$ is the total energy of the unit cell and $E_{\rm molecule}$ that of the isolated molecule, with negative values denoting binding. The comparison concerns electronic energies without thermal or zero-point contributions.

 The AE and GTH calculations use the QZVPP-ae and TZV2P-GTH bases, respectively. The full numerical protocol is specified in ESI Sec.~\ref{esi-sec:crystal-protocol}.

\begin{table*}[t]
\centering
\caption{Electronic lattice energies in kJ~mol$^{-1}$ per molecule at the supplied geometries, including D3(BJ) for all Skala results. All AE entries use the QZVPP-ae basis, while GTH entries retain their molecularly optimized triple-zeta bases. The AE triple-zeta calculations using the same geometries and integration grids are reported in the ESI. DMC reference values are from Ref.~\citenum{DellaPia2024X23}. Parentheses give its statistical uncertainties, not a complete systematic uncertainty. Negative energies denote binding.}
\label{tab:crystals}
\begin{tabular}{lrrrrr}
\toprule
Crystal & GAPW-AE & GAPW-GTH one-centre & GAPW-GTH direct & GAPW-XC/GTH & DMC\\
\midrule
CO$_2$ & -32.067 & -37.392 & -37.376 & -37.408 & -29.4(2)\\
NH$_3$ & -38.185 & -40.076 & -40.094 & -40.074 & -38.2(1)\\
Urea & -110.427 & -113.719 & -113.805 & -113.725 & -108.5(3)\\
\bottomrule
\end{tabular}
\end{table*}

The comparison with DMC benchmarks the accuracy of each complete Skala--D3(BJ) protocol. Differences between AE and GTH reflect the combined effects of core treatment, orbital basis, and atom quadrature. In contrast, direct and one-centre GAPW-GTH retain the same pseudopotentials, basis, and integration settings, providing a targeted test of valence-field reconstruction.

The three GTH representations give stronger binding than DMC and agree closely with one another (Table~\ref{tab:crystals}). The direct and one-centre GTH lattice energies differ by less than 0.09~kJ~mol$^{-1}$, making the numerical effect of the one-centre approximation negligible on the scale of the deviations from DMC. In contrast, the larger-basis AE results approach the DMC references, with essentially zero residual for ammonia.

The QZVPP-ae calculations reduce the overbinding relative to GAPW-XC/GTH for all three crystals. Ammonia is the most accurate case, while CO$_2$ retains the largest absolute and fractional error in every representation. Enlarging both the crystal and molecular AE bases at fixed fine quadrature reduces the three-crystal MAE from 7.95 to 1.54~kJ~mol$^{-1}$.  For AE urea, the $+14.42$~kJ~mol$^{-1}$ shift from TZVPP-ae to QZVPP-ae at fixed 200/974 quadrature is much larger than the $-0.21$~kJ~mol$^{-1}$ shift from refining the quadrature from 150/770 to 200/974 with the TZVPP-ae basis. This identifies basis flexibility as the dominant tested contribution and motivates the larger AE basis independently of the integration-grid choice. Thus, the native implementation delivers highly accurate molecular-crystal energies when sufficient AE basis flexibility is supplied, without changing the functional, geometries, or one-centre reconstruction. The basis-set and integration-grid convergence tests and the DMC errors at both basis levels are documented in ESI Sec.~\ref{esi-sec:crystal-numerical-controls}, Tables~\ref{esi-tab:urea-basis} and \ref{esi-tab:crystal-basis-errors}.

The material-dependent errors may reflect different balances of intermolecular attraction, repulsion, and molecular deformation. Dispersion-dominated CO$_2$ tests a delicate balance of attractive correlation, quadrupolar electrostatics, and short-range repulsion. Earlier correlated calculations also find a larger gas-to-crystal intramolecular deformation contribution for urea.\cite{Sharkas2014Crystals}

For CO$_2$, excessive attraction or insufficient short-range repulsion could each produce the remaining overbinding. Its smaller reference lattice energy also amplifies the fractional error.  This contrasts with the consistently small NH$_3$ residual, which may involve favourable cancellation between crystal and molecular errors.

The tested integration-grid and CPU/GPU effects are too small to explain the remaining CO$_2$ discrepancy. The close agreement among the GTH representations also shows that the choice of density reconstruction has little influence on this residual.

We compare Skala--D3(BJ) electronic lattice energies with DMC references, avoiding the zero-point and thermal corrections required when using experimental sublimation enthalpies.\cite{DellaPia2024X23,Dolgonos2019X23b}

Published dispersion-inclusive DFT results provide a broader comparison.\cite{DellaPia2025MLIP} On this three-crystal set, native GAPW-AE has a lower MAE than the listed dispersion-inclusive DFT methods, while GAPW-XC/GTH is competitive for ammonia and urea. Periodic MP2/CBS, SCS(MI)-MP2/CBS, and the multi-level LNO-CCSD(T)/RPA+ph approach with a periodic Hartree--Fock correction provide complementary wavefunction comparisons.\cite{Liang2023CrystalMP2,Syty2025CrystalCC} MP2/CBS and the multi-level approach give smaller MAEs than native AE, but the differences are on the scale of the estimated overall DMC uncertainty.\cite{DellaPia2024X23} ESI Sec.~\ref{esi-sec:crystal-literature} and Table~\ref{esi-tab:x23-dft-context} collect the individual energies and statistics and document the differing geometries and numerical protocols.

\subsection{Ice polymorphs and condensed-phase transferability}
\label{sec:transfer-results}

The DMC-ICE13 calculations compare the TZVPP-ae and QZVPP-ae bases at fixed crystal geometries using the same integration settings at both basis levels. The full numerical protocol and gas-phase reference geometry are specified in ESI Sec.~\ref{esi-sec:ice-protocol}.

\begin{figure*}[t]
\centering
\includegraphics[width=\textwidth]{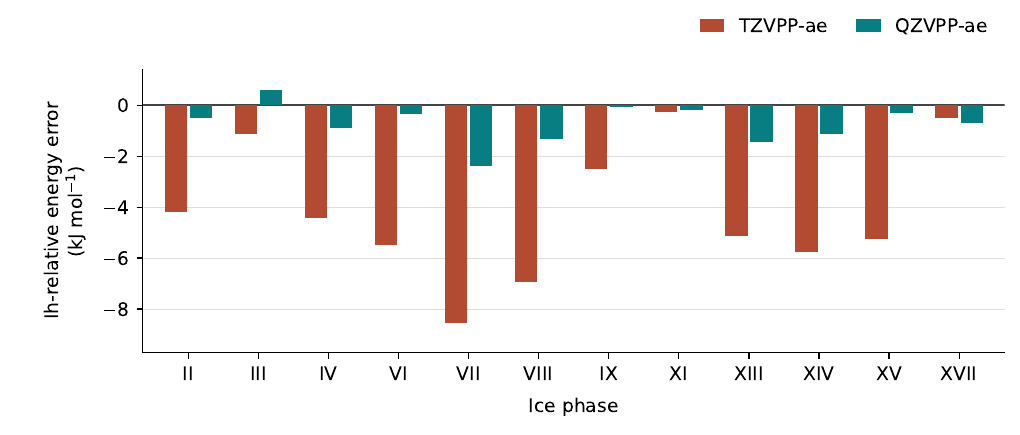}
\caption{Deviations of Ih-relative native Skala--D3(BJ) GAPW-AE energies from DMC-ICE13 references\cite{DellaPia2022Ice} for the TZVPP-ae and QZVPP-ae bases. The bars show $\Delta E_p-\Delta E_p^{\rm DMC}$ for the twelve phases other than Ih, where $\Delta E_p=E_{{\rm latt},p}-E_{{\rm latt},{\rm Ih}}$ is the lattice-energy difference per water molecule between phase $p$ and Ih. Negative deviations indicate underestimated energy differences. Full energies and DMC statistical uncertainties are given in ESI Tables~\ref{esi-tab:ice-basis} and \ref{esi-tab:ice-relative}.}
\label{fig:ice13-errors}
\end{figure*}

The complete DMC-ICE13 comparison provides a broader test of hydrogen-bonded condensed phases.\cite{DellaPia2022Ice} Across the thirteen phases calculated at both basis levels, enlarging the AE basis reduces the absolute lattice-energy MAE from 21.35 to 1.16~kJ~mol$^{-1}$.  The systematic TZVPP-ae overbinding is not retained with the QZVPP-ae basis. The MAE of the twelve energy differences relative to Ih also decreases, from 4.17 to 0.82~kJ~mol$^{-1}$. Figure~\ref{fig:ice13-errors} shows their deviations from DMC by phase. This comparison cancels the gas-phase reference exactly, showing that the improvement is not merely a common shift in binding. Nevertheless, QZVPP-ae underestimates eleven of these twelve relative energies, with the largest residual for VII.  Differences in the ordering of nearly degenerate phases also remain. ESI Sec.~\ref{esi-sec:ice-basis} and Tables~\ref{esi-tab:ice-basis} and \ref{esi-tab:ice-relative} give the phase-resolved comparison.

Taken together, the larger-basis molecular-crystal and ice results demonstrate accurate lattice energies across these condensed-phase environments.

\subsection{Equilibrium crystal structures}
\label{sec:structure-results}

The ten-solid equation-of-state comparison (LC10) probes equilibrium structures rather than energies at fixed reference geometries. All ten lattice constants lie below the static-lattice reference values in each of the four native representations, with MAEs of 0.073--0.084~\AA{} (Table~\ref{tab:lc10-primary} and ESI Table~\ref{esi-tab:lc10-reconstruction}). This systematic lattice contraction indicates structural overbinding, distinct from the accurate fixed-geometry molecular-crystal and ice lattice energies. The EOS results and numerical protocol are given in ESI Sec.~\ref{esi-sec:solids}. Enlarging the AE basis to QZVPP-ae reduces the diamond lattice error but increases the Si contraction, with a smaller change for MgO (ESI Sec.~\ref{esi-sec:solid-basis}).

\begin{table*}[t]
\centering\small
\caption{Equilibrium lattice constants $a_0$ and bulk moduli $B_0$ for the ten LC10 solids using GAPW-XC/GTH and GAPW-AE. Experimental static-lattice references follow Ref.~\citenum{Goldzak2022Solids}. Each MAE is calculated over all ten solids in the unit of the corresponding column.}
\label{tab:lc10-primary}
\begin{tabular}{lrrrrrr}
\toprule
& \multicolumn{3}{c}{$a_0$ (\AA{})} & \multicolumn{3}{c}{$B_0$ (GPa)} \\
\cmidrule(lr){2-4}\cmidrule(lr){5-7}
Solid & Experiment & \shortstack{GAPW-XC/\\GTH} & GAPW-AE & Experiment & \shortstack{GAPW-XC/\\GTH} & GAPW-AE \\
\midrule
AlN & 4.3680 & 4.2811 & 4.3025 & 206.00 & 239.07 & 225.89 \\
AlP & 5.4480 & 5.3622 & 5.3781 & 87.00 & 94.34 & 101.35 \\
BN & 3.5930 & 3.5777 & 3.5675 & 388.50 & 423.39 & 420.87 \\
BP & 4.5250 & 4.4439 & 4.4696 & 176.50 & 187.93 & 183.64 \\
C & 3.5530 & 3.5345 & 3.5292 & 453.30 & 457.33 & 491.35 \\
LiCl & 5.0720 & 4.9081 & 4.9155 & 37.30 & 47.98 & 46.37 \\
MgO & 4.1890 & 4.1059 & 4.0935 & 173.00 & 188.28 & 198.71 \\
MgS & 5.1880 & 5.0689 & 5.0536 & 81.00 & 87.90 & 90.27 \\
Si & 5.4210 & 5.3386 & 5.3740 & 100.30 & 92.76 & 100.56 \\
SiC & 4.3470 & 4.2517 & 4.2920 & 228.90 & 263.89 & 241.68 \\
\midrule
MAE & -- & 0.0831 & 0.0728 & -- & 16.61 & 16.89 \\
\bottomrule
\end{tabular}
\end{table*}

The pure-AE series has the smallest lattice-constant MAE, but every core-treatment protocol retains the contraction. At fixed element-wise core treatment, the direct and one-centre GAPW-AE/GTH variants give nearly identical equilibrium lattice constants. Across the five genuinely mixed and three all-GTH systems, their differences remain below $6\times10^{-4}$~\AA{}. The bulk moduli are more sensitive to the representation, particularly for MgO (ESI Table~\ref{esi-tab:lc10-reconstruction}). Thus, the GTH valence-reconstruction choice has little influence on the equilibrium lattice contraction, although the curvature of the energy--volume curve is more affected.

This structural bias should be considered in light of the training domain. Skala-1.1 was developed from molecular and atomic reference data rather than periodic-solid densities and energies.\cite{Skala2025,Ehlert2026ACC} Limited coverage of extended coordination, screening, and cooperative polarization is therefore a possible contributor to the LC10 contraction, alongside basis, core-treatment, and dispersion effects. High-accuracy periodic reference data are needed to assess these contributions and to guide a targeted extension or reparameterization of the XC functional.

High-accuracy reference data can be obtained with periodic wavefunction-based methods,\cite{Azadi2020Hydrogen} including periodic coupled-cluster theory\cite{Hummel2017,Gruber2018} and full configuration-interaction quantum Monte Carlo (FCIQMC).\cite{Booth2013} Complementary real-space quantum Monte Carlo (QMC) approaches use flexible trial wavefunctions based on shadow\cite{Calcavecchia2014Shadow,Calcavecchia2015Shadow,Calcavecchia2018} and resonating-valence-bond constructions,\cite{Azadi2015Ozone} or neural networks.\cite{Kessler2021,Hermann2020PauliNet,Pfau2020FermiNet}

\subsection{Electronic band gaps}
\label{sec:bandgap-results}

We assess Skala's electronic band gaps using triple-zeta GAPW-AE, GAPW-XC/GTH, and mixed AE/GTH one-centre calculations for the experimentally referenced materials considered by Lee \emph{et al.}\cite{Lee2021Gaps,Lee2022Hybrids}

To separate electronic-structure errors from geometry-induced shifts, the band-gap calculations use fixed experimental geometries rather than Skala-optimized structures, with a uniform triple-zeta protocol and fixed insulating occupations.  ESI Sec.~\ref{esi-sec:bandgap-protocol} records the phase-specific experimental geometries, basis sets, core choices, Brillouin-zone sampling, and convergence thresholds.

We report 50 band-gap results for 28 materials, comprising 20 AE, 23 GAPW-XC/GTH, and seven mixed one-centre results.

The combined AE-or-mixed one-centre series contains fifteen pure-AE and seven mixed results on 22 materials also available for GAPW-XC/GTH. The respective MAEs are 0.543 and 0.816~eV. Both native series have smaller MAEs than PBE, SCAN, and B97M-rV on exactly these 22 materials. ESI Tables~\ref{esi-tab:bandgap-outcomes} and \ref{esi-tab:bandgap-paired-combined-Lee2021} report the individual gaps, experimental references, and all ten local or semilocal comparators.

Fifteen of these materials also occur in the published hybrid-functional comparison, with twelve AE and three mixed native results. On this common population, the AE-or-mixed and GAPW-XC/GTH MAEs are 0.432 and 0.778~eV. Table~\ref{tab:bandgap-method-errors} compares the native results with all ten local or semilocal and twelve hybrid XC functionals on exactly these fifteen materials. The AE-or-mixed MAE is comparable to those of HSE, PBE0, and B3LYP, although its RMSE is larger. On the same fifteen-material set, its MAE is substantially lower than that of M06-2X and those of all four range-separated hybrid functionals included in the table. The ordering of the closely spaced Skala and HSE MAEs reverses when the BAs reference is changed from 1.46~eV to the independently reported 1.82~eV absorption estimate (ESI Table~\ref{esi-tab:bandgap-BAs-sensitivity}).\cite{Yue2020BAs}

\begin{table}[tb]
\centering\small
\setlength{\tabcolsep}{2.5pt}
\caption{Band-gap errors in eV for the fifteen materials common to the native calculations and both literature benchmarks.\cite{Lee2021Gaps,Lee2022Hybrids} Skala AE/mixed comprises twelve AE and three mixed AE/GTH one-centre results. Skala GTH denotes GAPW-XC/GTH. All ten local or semilocal and twelve hybrid XC comparators use the same set of BAs, BN, BP, C, CdS, CdSe, GaAs, GaP, Ge, InP, LiCl, Si, ZnS, ZnSe, $\beta$-GaN. ME is calculation minus experiment averaged over the set. mGGA denotes meta-GGA. RS denotes range separation. Individual gaps are given in ESI Table~\ref{esi-tab:bandgap-outcomes}.}
\label{tab:bandgap-method-errors}
\begin{tabular}{@{}llrrr@{}}
\toprule
Method & XC class & ME & MAE & RMSE \\
\midrule
Skala AE/mixed & Native & -0.365 & 0.432 & 0.602 \\
Skala GTH & Native & -0.776 & 0.778 & 0.926 \\
\midrule
LDA & LDA & -1.411 & 1.411 & 1.574 \\
PBE & GGA & -1.181 & 1.181 & 1.339 \\
PBEsol & GGA & -1.327 & 1.327 & 1.483 \\
revPBE & GGA & -1.097 & 1.097 & 1.249 \\
BLYP & GGA & -1.066 & 1.085 & 1.248 \\
B97-D & GGA & -1.026 & 1.043 & 1.180 \\
SCAN & mGGA & -0.947 & 0.961 & 1.086 \\
M06-L & mGGA & -0.793 & 0.845 & 1.010 \\
MN15-L & mGGA & -1.083 & 1.095 & 1.248 \\
B97M-rV & mGGA & -0.845 & 0.872 & 1.016 \\
\midrule
B3LYP & Hybrid GGA & +0.097 & 0.405 & 0.510 \\
PBE0 & Hybrid GGA & +0.285 & 0.402 & 0.502 \\
revPBE0 & Hybrid GGA & +0.355 & 0.451 & 0.521 \\
B97-3 & Hybrid GGA & +0.640 & 0.692 & 0.739 \\
M06-2X & Hybrid mGGA & +2.057 & 2.057 & 2.123 \\
MN15 & Hybrid mGGA & +0.894 & 0.931 & 1.080 \\
SCAN0 & Hybrid mGGA & +0.399 & 0.453 & 0.545 \\
HSE & Screened hybrid & -0.387 & 0.435 & 0.578 \\
CAM-B3LYP & RS hybrid & +2.411 & 2.411 & 2.453 \\
$\omega$B97X-rV & RS hybrid & +3.988 & 3.988 & 4.011 \\
$\omega$B97M-rV & RS hybrid & +3.492 & 3.492 & 3.528 \\
CAM-QTP01 & RS hybrid & +4.203 & 4.203 & 4.231 \\
\bottomrule
\end{tabular}
\end{table}

The mixed one-centre protocol reduces the gap underestimation relative to GAPW-XC/GTH on all seven compounds shared by these two series, lowering the MAE from 0.766 to 0.650~eV (ESI Table~\ref{esi-tab:bandgap-core-subsets}). This comparison includes changes in core treatment and the corresponding element-specific orbital bases. For Ge and MgS, the sampled gaps are indirect with GAPW-AE but direct with GAPW-XC/GTH, illustrating that the AE/GTH protocol affects not only the gap magnitude but also the relative positions of the band extrema (ESI Table~\ref{esi-tab:bandgap-edges}).

The separate direct/one-centre controls retain the AE reconstruction and test the GTH valence representation in six genuinely mixed compounds and one all-GTH control. The one-centre reconstruction of GTH valence fields changes the calculated gaps by less than 0.009~eV across the seven tested systems (ESI Table~\ref{esi-tab:bandgap-controls}).

\section{Conclusions}

The joint one-centre reconstruction brings learned nonlocal XC into CP2K's native GPW/GAPW framework without GauXC. Reconstructing density, gradients, and kinetic-energy density before evaluation preserves the mixed terms and nonlocal couplings required by GAPW-XC, GAPW-AE, and augmented GAPW-GTH. GAPW-XC is the recommended default starting representation for pseudopotential calculations because rapidly varying contributions to the XC input fields are resolved on atom-centred grids independently of the global plane-wave cutoff. The same variational construction connects the functional to CP2K's forces, stress, k-point sampling, and point-group reduction, and to its boundary-condition framework for systems of different dimensionality. The present validation covers molecules and three-dimensional crystals.

With the QZVPP-ae basis, native GAPW-AE reaches a molecular MAE close to that of our earlier def2-based GauXC implementation on the same selected set of 62 reactions from dietGMTKN55. At the nominal triple-zeta level, however, the def2 protocol has the more favourable error balance. The improvement with the larger AE basis identifies basis flexibility as an important source of the initial deviations. Together with the numerical consistency tests, these results show that the native implementation preserves Skala's molecular predictive accuracy while extending its application to condensed-phase electronic structure.

Skala with D3(BJ) achieves excellent agreement with DMC for the three molecular crystals when the AE basis is enlarged uniformly to QZVPP-ae. The resulting MAE is 1.54~kJ~mol$^{-1}$. The complete thirteen-phase ice comparison independently reaches MAEs of 1.16~kJ~mol$^{-1}$ for absolute lattice energies and 0.82~kJ~mol$^{-1}$ for the twelve relative energies to Ih. Residual phase-dependent errors and a bias towards underestimated relative energies remain. These results highlight the importance of basis flexibility when assessing the functional's condensed-phase accuracy. The triple-zeta GTH calculations give errors comparable to those of established dispersion-inclusive approximations for ammonia and urea. Together with the numerical controls, the molecular-crystal and ice results support both the native implementation and accurate transfer beyond the gas phase, while motivating broader condensed-phase validation with many-body references.

The band-gap calculations assess transferability to electronic structure at fixed experimental geometries. On the matched 22-material population, the preselected AE-or-mixed series has smaller mean absolute gap errors than PBE, SCAN, and B97M-rV. On the distinct fifteen-material set shared with the hybrid-functional study, its 0.432~eV MAE is comparable to those of HSE, PBE0, and B3LYP. These results complement the accurate molecular-crystal and ice energies at fixed geometries. In contrast, the systematically underestimated LC10 equilibrium lattice constants reveal a remaining structural bias.

Periodic coupled-cluster and QMC calculations, including FCIQMC where feasible, can supply the required many-body reference data. Reference energies and densities spanning periodic environments and volumes would allow us to test directly whether the molecular training data adequately cover condensed-phase environments. They would also provide a basis for targeted retraining or reparameterization of Skala aimed at reducing the remaining structural bias while preserving its molecular accuracy. The native GPW/GAPW implementation allows this development within the same electronic-structure framework used for subsequent applications.

\section*{Conflicts of interest}
There are no conflicts to declare.

\section*{Data availability}
\label{sec:data-availability}
The supporting reaction energies, lattice-energy comparisons, equation-of-state data, band-gap tables, and numerical controls are included in the ESI. Absolute crystal and molecular energies, geometries, computational inputs, analysis scripts, and source-version metadata for the molecular, molecular-crystal, ice, and solid benchmarks are collected in the native-grid Skala benchmark repository at \url{https://github.com/DCM-Uni-Paderborn/native-grid-skala-benchmarks}. The separate band-gap dataset accompanies the manuscript source and is not included in that public archive. A versioned public release of the complete article dataset will accompany submission.

\section*{Acknowledgements}
The authors thank the AI for Science team at Microsoft Research for valuable discussions and the CP2K and GauXC development communities. Part of the research was funded by the German Research Foundation (DFG), project numbers 417590517/Collaborative Research Centre 1415 and 519869949. OpenAI tools assisted with language editing, literature organization, and document preparation. Responsibility for the scientific content remains with the authors.

\bibliographystyle{rsc}
\bibliography{references}

@article{Kohn1999Nobel,
  author = {Kohn, W.},
  title = {Nobel Lecture: Electronic structure of matter---wave functions and density functionals},
  journal = {Rev. Mod. Phys.},
  volume = {71},
  number = {5},
  pages = {1253--1266},
  year = {1999},
  doi = {10.1103/RevModPhys.71.1253}
}

@article{Jones2015DFT,
  author = {Jones, R. O.},
  title = {Density functional theory: Its origins, rise to prominence, and future},
  journal = {Rev. Mod. Phys.},
  volume = {87},
  number = {3},
  pages = {897--923},
  year = {2015},
  doi = {10.1103/RevModPhys.87.897}
}

@article{Kirkpatrick2021DM21,
  author = {Kirkpatrick, James and McMorrow, Brendan and Turban, David H. P. and Gaunt, Alexander L. and Spencer, James S. and Matthews, Alexander G. D. G. and Obika, Annette and Thiry, Louis and Fortunato, Meire and Pfau, David and Rom{\'a}n Castellanos, Lara and Petersen, Stig and Nelson, Alexander W. R. and Kohli, Pushmeet and Mori-S{\'a}nchez, Paula and Hassabis, Demis and Cohen, Aron J.},
  title = {Pushing the frontiers of density functionals by solving the fractional electron problem},
  journal = {Science},
  volume = {374},
  number = {6573},
  pages = {1385--1389},
  year = {2021},
  doi = {10.1126/science.abj6511}
}

@article{Lippert1997,
  author = {Lippert, G. and Hutter, J. and Parrinello, M.},
  title = {A hybrid Gaussian and plane wave density functional scheme},
  journal = {Mol. Phys.},
  volume = {92},
  pages = {477--488},
  year = {1997},
  doi = {10.1080/002689797170220}
}

@article{Lippert1999,
  author = {Lippert, G. and Hutter, J. and Parrinello, M.},
  title = {The Gaussian and augmented-plane-wave density functional method for ab initio molecular dynamics simulations},
  journal = {Theor. Chem. Acc.},
  volume = {103},
  pages = {124--140},
  year = {1999},
  doi = {10.1007/s002140050523}
}

@article{Bloechl1994PAW,
  author = {Bl{\"o}chl, Peter E.},
  title = {Projector Augmented-Wave Method},
  journal = {Phys. Rev. B},
  volume = {50},
  pages = {17953--17979},
  year = {1994},
  doi = {10.1103/PhysRevB.50.17953}
}

@article{Bloechl2026CPPAW,
  author = {Bl{\"o}chl, Peter E. and Schade, Robert and Allen-Rump, Lukas and Rajpurohit, Sangeeta and Rathnakaran, Amrith and Tamoev, Konstantin and Lokamani, Mani and K{\"u}hne, Thomas D.},
  title = {The {CP-PAW} Code Package for First-Principles Calculations from a User's Perspective},
  journal = {J. Phys. Chem. A},
  volume = {130},
  number = {27},
  pages = {5261--5281},
  year = {2026},
  doi = {10.1021/acs.jpca.6c00342}
}

@article{CP2K2020,
  author = {K{\"u}hne, T. D. and others},
  title = {{CP2K}: An electronic structure and molecular dynamics software package - Quickstep: Efficient and accurate electronic structure calculations},
  journal = {J. Chem. Phys.},
  volume = {152},
  pages = {194103},
  year = {2020},
  doi = {10.1063/5.0007045}
}

@article{CP2KMadeSimple2026,
  author = {Iannuzzi, Marcella and Wilhelm, Jan and Stein, Frederick and Bussy, Augustin and Elgabarty, Hossam and Golze, Dorothea and Hehn, Anna and Graml, Maximilian and Marek, Stepan and Sertcan G{\"o}kmen, Beliz and Schran, Christoph and Forbert, Harald and Khaliullin, Rustam Z. and Kozhevnikov, Anton and Taillefumier, Mathieu and Meli, Rocco and Rybkin, Vladimir V. and Brehm, Martin and Schade, Robert and Sch{\"u}tt, Ole and Pototschnig, Johann V. and Mirhosseini, Hossein and Kn{\"u}pfer, Andreas and Marx, Dominik and Krack, Matthias and Hutter, J{\"u}rg and K{\"u}hne, Thomas D.},
  title = {The {CP2K} Program Package Made Simple},
  journal = {J. Phys. Chem. B},
  volume = {130},
  pages = {1237--1310},
  year = {2026},
  doi = {10.1021/acs.jpcb.5c05851}
}

@misc{CP2KDynamics2026,
  author = {Wilhelm, Jan and Hehn, Anna-Sophia and Elgabarty, Hossam and others},
  title = {{CP2K}: An electronic structure and molecular dynamics software package -- Dynamics, Transport, and Spectroscopic Response},
  year = {2026},
  howpublished = {arXiv:2607.22916},
  eprint = {2607.22916},
  archivePrefix = {arXiv},
  primaryClass = {physics.chem-ph},
  doi = {10.48550/arXiv.2607.22916},
  url = {https://arxiv.org/abs/2607.22916}
}

@article{Louie1982NLCC,
  author = {Louie, Steven G. and Froyen, Sverre and Cohen, Marvin L.},
  title = {Nonlinear ionic pseudopotentials in spin-density-functional calculations},
  journal = {Phys. Rev. B},
  volume = {26},
  number = {4},
  pages = {1738--1742},
  year = {1982},
  doi = {10.1103/PhysRevB.26.1738}
}

@article{Willand2013NLCC,
  author = {Willand, Alex and Kvashnin, Yaroslav O. and Genovese, Luigi and V{\'a}zquez-Mayagoitia, {\'A}lvaro and Deb, Arpan Krishna and Sadeghi, Ali and Deutsch, Thierry and Goedecker, Stefan},
  title = {Norm-conserving pseudopotentials with chemical accuracy compared to all-electron calculations},
  journal = {J. Chem. Phys.},
  volume = {138},
  number = {10},
  pages = {104109},
  year = {2013},
  doi = {10.1063/1.4793260}
}

@article{Goedecker1996,
  author = {Goedecker, S. and Teter, M. and Hutter, J.},
  title = {Separable dual-space Gaussian pseudopotentials},
  journal = {Phys. Rev. B},
  volume = {54},
  pages = {1703--1710},
  year = {1996},
  doi = {10.1103/PhysRevB.54.1703}
}

@article{Weigend2005,
  author = {Weigend, Florian and Ahlrichs, Reinhart},
  title = {Balanced basis sets of split valence, triple zeta valence and quadruple zeta valence quality for H to Rn: Design and assessment of accuracy},
  journal = {Phys. Chem. Chem. Phys.},
  volume = {7},
  pages = {3297--3305},
  year = {2005},
  doi = {10.1039/B508541A}
}

@article{VandeVondele2007,
  author = {VandeVondele, Joost and Hutter, Juerg},
  title = {Gaussian basis sets for accurate calculations on molecular systems in gas and condensed phases},
  journal = {J. Chem. Phys.},
  volume = {127},
  pages = {114105},
  year = {2007},
  doi = {10.1063/1.2770708}
}

@article{Mirhosseini2026UZH,
  author = {Mirhosseini, Hossein and M{\"u}ller, Tiziano M. A. and Krack, Matthias and K{\"u}hne, Thomas D. and Hutter, J{\"u}rg},
  title = {The {UZH} protocol: Separating errors and constructing improved {CP2K} basis sets and pseudopotentials},
  journal = {J. Chem. Phys.},
  volume = {165},
  number = {10},
  pages = {104103},
  year = {2026},
  doi = {10.1063/5.0347392}
}

@misc{Skala2025,
  author = {Luise, Giulia and Huang, Chin-Wei and Vogels, Thijs and Kooi, Derk P. and Ehlert, Sebastian and Lanius, Stephanie and Giesbertz, Klaas J. H. and Karton, Amir and Gunceler, Deniz and Battaglia, Stefano and Simm, Gregor N. C. and Szab{\'o}, P. Bern{\'a}t and Stanley, Megan and Bruinsma, Wessel P. and Huang, Lin and Wei, Xinran and Garrido Torres, Jos{\'e} and Katbashev, Abylay and Chavez Zavaleta, Rodrigo and M{\'a}t{\'e}, B{\'a}lint and Kaba, S{\'e}kou-Oumar and Sordillo, Roberto and Chen, Yingrong and Williams-Young, David B. and Bishop, Christopher M. and Hermann, Jan and van den Berg, Rianne and Gori-Giorgi, Paola},
  title = {Accurate and scalable exchange-correlation with deep learning},
  howpublished = {arXiv:2506.14665},
  year = {2026},
  eprint = {2506.14665},
  archivePrefix = {arXiv},
  url = {https://arxiv.org/abs/2506.14665}
}

@article{Ehlert2026ACC,
  author = {Ehlert, Sebastian and Hermann, Jan and Vogels, Thijs and Garcia Satorras, Victor and Lanius, Stephanie and Segler, Marwin and Giesbertz, Klaas J. H. and Kooi, Derk P. and Takeda, Kenji and Huang, Chin-Wei and Luise, Giulia and van den Berg, Rianne and Gori-Giorgi, Paola and Karton, Amir},
  title = {Accurate Chemistry Collection: Coupled cluster atomization energies for broad chemical space},
  journal = {Sci. Data},
  volume = {13},
  pages = {951},
  year = {2026},
  doi = {10.1038/s41597-026-07200-8}
}

@misc{SkalaModelCard,
  author = {{Microsoft Research AI for Science}},
  title = {{Skala}-1.1 model card},
  year = {2026},
  url = {https://huggingface.co/microsoft/skala-1.1},
  note = {Accessed September 2026}
}

@article{Grimme2011D3BJ,
  author = {Grimme, Stefan and Ehrlich, Stephan and Goerigk, Lars},
  title = {Effect of the damping function in dispersion corrected density functional theory},
  journal = {J. Comput. Chem.},
  volume = {32},
  pages = {1456--1465},
  year = {2011},
  doi = {10.1002/jcc.21759}
}

@article{Gould2018,
  title = {‘{{Diet GMTKN55}}’ Offers Accelerated Benchmarking through a Representative Subset Approach},
  author = {Gould, Tim},
  year = {2018},
  journal = {Phys. Chem. Chem. Phys.},
  volume = {20},
  number = {44},
  pages = {27735--27739},
  doi = {10.1039/C8CP05554H}
}

@article{Goerigk2017,
  title={A look at the density functional theory zoo with the advanced {GMTKN55} database for general main group thermochemistry, kinetics and noncovalent interactions},
  author={Goerigk, Lars and Hansen, Andreas and Bauer, Christoph and Ehrlich, Stephan and Najibi, Asim and Grimme, Stefan},
  journal={Phys. Chem. Chem. Phys.},
  volume={19},
  number={48},
  pages={32184--32215},
  year={2017},
  publisher={The Royal Society of Chemistry}
}

@article{Williams2020,
  author={David B. Williams--Young and Wibe A. de Jong and Hubertus J.J. van Dam and
          Chao Yang},
  title={On the Efficient Evaluation of the Exchange Correlation Potential on
         Graphics Processing Unit Clusters},
  journal={Frontiers in Chemistry},
  volume={8},
  pages={581058},
  year={2020},
  doi={10.3389/fchem.2020.581058},
  url={https://www.frontiersin.org/articles/10.3389/fchem.2020.581058/abstract},
  preprint={https://arxiv.org/abs/2007.03143}
}

@unpublished{Poeschel2026MolecularSkala,
  author = {P{\"o}schel, Franz and Pototschnig, Johann and Stein, Frederick and Kn{\"u}pfer, Andreas and Vogels, Thijs and Battaglia, Stefano and Ehlert, Sebastian and Hutter, J{\"u}rg and K{\"u}hne, Thomas D.},
  title = {Molecular Implementation of the Machine-Learned {Skala} Exchange--Correlation Functional in {CP2K} through {GauXC}},
  note = {{\href{https://arxiv.org/abs/2608.19033}{arXiv:2608.19033}}},
  year = {2026},
  url = {https://arxiv.org/abs/2608.19033}
}

@article{Becke1988,
  author = {Becke, Axel D.},
  title = {A Multicenter Numerical Integration Scheme for Polyatomic Molecules},
  journal = {J. Chem. Phys.},
  volume = {88},
  pages = {2547--2553},
  year = {1988},
  doi = {10.1063/1.454033}
}

@article{Monkhorst1976,
  author = {Monkhorst, Hendrik J. and Pack, James D.},
  title = {Special Points for {Brillouin-Zone} Integrations},
  journal = {Phys. Rev. B},
  volume = {13},
  number = {12},
  pages = {5188--5192},
  year = {1976},
  doi = {10.1103/PhysRevB.13.5188}
}

@article{Togo2024Spglib,
  author = {Togo, Atsushi and Shinohara, Kohei and Tanaka, Isao},
  title = {{Spglib}: A Software Library for Crystal Symmetry Search},
  journal = {Sci. Technol. Adv. Mater. Methods},
  volume = {4},
  number = {1},
  pages = {2384822},
  year = {2024},
  doi = {10.1080/27660400.2024.2384822}
}

@article{Goldzak2022Solids,
  author = {Goldzak, Tamar and Wang, Xiao and Ye, Hong-Zhou and Berkelbach, Timothy C.},
  title = {Accurate Thermochemistry of Covalent and Ionic Solids from Spin-Component-Scaled {MP2}},
  journal = {J. Chem. Phys.},
  volume = {157},
  pages = {174112},
  year = {2022},
  doi = {10.1063/5.0119633}
}

@article{Zhang2018Solids,
  author = {Zhang, Guo-Xu and Reilly, Anthony M. and Tkatchenko, Alexandre and Scheffler, Matthias},
  title = {Performance of Various Density-Functional Approximations for Cohesive Properties of 64 Bulk Solids},
  journal = {New J. Phys.},
  volume = {20},
  pages = {063020},
  year = {2018},
  doi = {10.1088/1367-2630/aac7f0}
}

@article{Alizadeh2026PeriodicGFN2,
  author = {Alizadeh, Vahideh and Pototschnig, Johann and Seidler, Leopold M. and Ehlert, Sebastian and K{\"u}hne, Thomas D.},
  title = {Periodic {GFN2-xTB} in {CP2K}: Multipolar Ewald Electrostatics, k-Point Sampling, and Transferability Benchmarks for Solids},
  journal = {J. Chem. Theory Comput.},
  volume = {22},
  number = {18},
  pages = {9688--9699},
  year = {2026},
  doi = {10.1021/acs.jctc.6c01034}
}

@article{Booth2013,
 author={Booth, George H. and Gr{\"u}neis, Andreas and Kresse, Georg and Alavi, Ali},
 title={Towards an exact description of electronic wavefunctions in real solids},
 journal={Nature}, year={2013}, volume={493}, pages={365--370}, doi={10.1038/nature11770}
}

@article{Hummel2017,
 author={Hummel, Felix and Tsatsoulis, Theodoros and Gr{\"u}neis, Andreas},
 title={Low rank factorization of Coulomb integrals for periodic coupled cluster theory},
 journal={J. Chem. Phys.}, year={2017}, volume={146}, pages={124105}, doi={10.1063/1.4977994}
}

@article{Gruber2018,
 author={Gruber, Thomas and Liao, Ke and Tsatsoulis, Theodoros and Hummel, Felix and Gr{\"u}neis, Andreas},
 title={Applying the Coupled-Cluster Ansatz to Solids and Surfaces in the Thermodynamic Limit},
 journal={Phys. Rev. X}, year={2018}, volume={8}, pages={021043}, doi={10.1103/PhysRevX.8.021043}
}

@article{Azadi2020Hydrogen,
 author={Azadi, Sam and Booth, George H. and K{\"u}hne, Thomas D.},
 title={Equation of state of atomic solid hydrogen by stochastic many-body wave function methods},
 journal={J. Chem. Phys.}, year={2020}, volume={153}, pages={204107}, doi={10.1063/5.0026499}
}

@article{Kessler2021,
 author={Kessler, Jan and Calcavecchia, Francesco and K{\"u}hne, Thomas D.},
 title={Artificial Neural Networks as Trial Wave Functions for Quantum Monte Carlo},
 journal={Adv. Theory Simul.}, year={2021}, volume={4}, pages={2000269}, doi={10.1002/adts.202000269}
}

@article{Calcavecchia2018,
 author={Calcavecchia, Francesco and K{\"u}hne, Thomas D.},
 title={Metal--Insulator Transition of Solid Hydrogen by the Antisymmetric Shadow Wave Function},
 journal={Z. Naturforsch. A}, year={2018}, volume={73}, pages={845--858}, doi={10.1515/zna-2018-0180}
}

@article{Azadi2015Ozone,
 author={Azadi, Sam and Singh, Ranber and K{\"u}hne, Thomas D.},
 title={Resonating Valence Bond Quantum Monte Carlo: Application to the Ozone Molecule},
 journal={Int. J. Quantum Chem.}, year={2015}, volume={115}, pages={1673--1677}, doi={10.1002/qua.25005}
}

@article{Dolgonos2019X23b,
 author={Dolgonos, Grygoriy A. and Hoja, Johannes and Boese, A. Daniel},
 title={Revised values for the X23 benchmark set of molecular crystals},
 journal={Phys. Chem. Chem. Phys.}, year={2019}, volume={21}, pages={24333--24344}, doi={10.1039/C9CP04488D}
}

@article{DellaPia2025MLIP,
 author = {Della Pia, Flaviano and Shi, Benjamin X. and Kapil, Venkat and Zen, Andrea and Alf{\`e}, Dario and Michaelides, Angelos},
 title = {Accurate and efficient machine learning interatomic potentials for finite temperature modelling of molecular crystals},
 journal = {Chemical Science},
 year = {2025}, volume = {16}, pages = {11419--11433},
 doi = {10.1039/D5SC01325A}
}

@article{Sharkas2014Crystals,
 author = {Sharkas, Kamal and Toulouse, Julien and Maschio, Lorenzo and Civalleri, Bartolomeo},
 title = {Double-hybrid density-functional theory applied to molecular crystals},
 journal = {J. Chem. Phys.},
 year = {2014}, volume = {141}, pages = {044105},
 doi = {10.1063/1.4890439}
}

@article{Morrison2003HydrogenBonds,
 author = {Morrison, Carole A. and Siddick, Muhammad M.},
 title = {Determining the Strengths of Hydrogen Bonds in Solid-State Ammonia and Urea: Insight from Periodic DFT Calculations},
 journal = {Chem. Eur. J.},
 year = {2003}, volume = {9}, pages = {628--634},
 doi = {10.1002/chem.200390067}
}

@misc{Battaglia2026DietData,
  author = {Battaglia, Stefano},
  title = {Reaction-resolved molecular {Skala} validation data},
  year = {2026},
  url = {https://github.com/DCM-Uni-Paderborn/Molecular-Skala-in-CP2K},
  note = {DietGMTKN55 validation dataset, accessed 11 September 2026}
}

@article{Hermann2020PauliNet,
 author={Hermann, Jan and Sch{\"a}tzle, Zeno and No{\'e}, Frank},
 title={Deep-neural-network solution of the electronic Schr{\"o}dinger equation},
 journal={Nat. Chem.}, year={2020}, volume={12}, pages={891--897},
 doi={10.1038/s41557-020-0544-y}
}

@article{Pfau2020FermiNet,
 author={Pfau, David and Spencer, James S. and Matthews, Alexander G. D. G. and Foulkes, W. M. C.},
 title={Ab initio solution of the many-electron Schr{\"o}dinger equation with deep neural networks},
 journal={Phys. Rev. Research}, year={2020}, volume={2}, pages={033429},
 doi={10.1103/PhysRevResearch.2.033429}
}

@article{Calcavecchia2014Shadow,
 author={Calcavecchia, Francesco and Pederiva, Francesco and Kalos, Malvin H. and K{\"u}hne, Thomas D.},
 title={Sign problem of the fermionic shadow wave function},
 journal={Phys. Rev. E}, year={2014}, volume={90}, pages={053304},
 doi={10.1103/PhysRevE.90.053304}
}

@article{Calcavecchia2015Shadow,
 author={Calcavecchia, Francesco and K{\"u}hne, Thomas D.},
 title={On Fermionic Shadow Wave Functions for strongly-correlated multi-reference systems based on a single Slater determinant},
 journal={EPL}, year={2015}, volume={110}, pages={20011},
 doi={10.1209/0295-5075/110/20011}
}

@article{DellaPia2024X23,
  author = {Della Pia, Flaviano and Zen, Andrea and Alf\`e, Dario and Michaelides, Angelos},
  title = {How Accurate Are Simulations and Experiments for the Lattice Energies of Molecular Crystals?},
  journal = {Phys. Rev. Lett.},
  volume = {133},
  pages = {046401},
  year = {2024},
  doi = {10.1103/PhysRevLett.133.046401}
}

@article{DellaPia2022Ice,
  author = {Della Pia, Flaviano and Zen, Andrea and Alf{\`e}, Dario and Michaelides, Angelos},
  title = {{DMC-ICE13}: Ambient and high pressure polymorphs of ice from diffusion Monte Carlo and density functional theory},
  journal = {J. Chem. Phys.},
  year = {2022},
  volume = {157},
  pages = {134701},
  doi = {10.1063/5.0102645}
}

@article{Gillan2012WaterClusters,
  author = {Gillan, M. J. and Manby, F. R. and Towler, M. D. and Alf{\`e}, D.},
  title = {Assessing the accuracy of quantum Monte Carlo and density functional theory for energetics of small water clusters},
  journal = {J. Chem. Phys.},
  year = {2012},
  volume = {136},
  pages = {244105},
  doi = {10.1063/1.4730035}
}

@article{Liang2023CrystalMP2,
  author = {Liang, Yu Hsuan and Ye, Hong-Zhou and Berkelbach, Timothy C.},
  title = {Can Spin-Component Scaled {MP2} Achieve kJ/mol Accuracy for Cohesive Energies of Molecular Crystals?},
  journal = {J. Phys. Chem. Lett.},
  year = {2023},
  volume = {14},
  pages = {10435--10441},
  doi = {10.1021/acs.jpclett.3c02411}
}

@article{Syty2025CrystalCC,
  author = {Syty, Krystyna and Czeka{\l}o, Grzegorz and Pham, Khanh Ngoc and Modrzejewski, Marcin},
  title = {Multi-Level Coupled-Cluster Description of Crystal Lattice Energies},
  journal = {J. Chem. Theory Comput.},
  year = {2025},
  volume = {21},
  pages = {5533--5544},
  doi = {10.1021/acs.jctc.5c00428}
}

@article{Lee2021Gaps,
 author = {Lee, Joonho and Feng, Xintian and Cunha, Leonardo A. and Gonthier, J{\'e}r{\^o}me F. and Epifanovsky, Evgeny and Head-Gordon, Martin},
 title = {Approaching the basis set limit in Gaussian-orbital-based periodic calculations with transferability: Performance of pure density functionals for simple semiconductors},
 journal = {J. Chem. Phys.}, year = {2021}, volume = {155}, pages = {164102},
 doi = {10.1063/5.0069177}, note = {Comparison data from Table 1 of arXiv:2108.12972v2}
}

@article{Lee2022Hybrids,
 author = {Lee, Joonho and Rettig, Adam and Feng, Xintian and Epifanovsky, Evgeny and Head-Gordon, Martin},
 title = {Faster Exact Exchange for Solids via occ-RI-K: Application to Combinatorially Optimized Range-Separated Hybrid Functionals for Simple Solids with Pseudopotentials Near the Basis Set Limit},
 journal = {J. Chem. Theory Comput.}, year = {2022}, volume = {18}, pages = {7336--7349},
 doi = {10.1021/acs.jctc.2c00742}, note = {Comparison data from Table AII of arXiv:2207.09028}
}

@article{Heyd2005Gaps,
 author = {Heyd, Jochen and Peralta, Juan E. and Scuseria, Gustavo E. and Martin, Richard L.},
 title = {Energy band gaps and lattice parameters evaluated with the Heyd--Scuseria--Ernzerhof screened hybrid functional},
 journal = {J. Chem. Phys.}, year = {2005}, volume = {123}, pages = {174101}, doi = {10.1063/1.2085170}
}

@article{Nolan2009LiH,
 author = {Nolan, S. J. and Gillan, M. J. and Alf{\`e}, D. and Allan, N. L. and Manby, F. R.},
 title = {Calculation of properties of crystalline lithium hydride using correlated wave function theory},
 journal = {Phys. Rev. B}, year = {2009}, volume = {80}, pages = {165109}, doi = {10.1103/PhysRevB.80.165109}
}

@article{Matsushita2011Gaps,
 author = {Matsushita, Yu-ichiro and Nakamura, Kazuma and Oshiyama, Atsushi},
 title = {Comparative study of hybrid functionals applied to structural and electronic properties of semiconductors and insulators},
 journal = {Phys. Rev. B}, year = {2011}, volume = {84}, pages = {075205}, doi = {10.1103/PhysRevB.84.075205}
}

@article{Yue2020BAs,
 author = {Yue, Shuai and Gamage, Geethal Amila and Mohebinia, Mohammadjavad and Mayerich, David and Talari, Vishal and Deng, Yu and Tian, Fei and Dai, Shen-Yu and Sun, Haoran and Hadjiev, Viktor G. and Zhang, Wei and Feng, Guoying and Hu, Jonathan and Liu, Dong and Wang, Zhiming and Ren, Zhifeng and Bao, Jiming},
 title = {Photoluminescence mapping and time-domain thermo-photoluminescence for rapid imaging and measurement of thermal conductivity of boron arsenide},
 journal = {Materials Today Physics}, year = {2020}, volume = {13}, pages = {100194},
 doi = {10.1016/j.mtphys.2020.100194}
}

@article{Rafferty1998Gaps,
 author = {Rafferty, B. and Brown, L. M.},
 title = {Direct and indirect transitions in the region of the band gap using electron-energy-loss spectroscopy},
 journal = {Phys. Rev. B}, year = {1998}, volume = {58}, pages = {10326--10337},
 doi = {10.1103/PhysRevB.58.10326}
}

@article{Wolverson2001MgS,
 author = {Wolverson, D. and Bird, D. M. and Bradford, C. and Prior, K. A. and Cavenett, B. C.},
 title = {Lattice dynamics and elastic properties of zinc-blende {MgS}},
 journal = {Phys. Rev. B}, year = {2001}, volume = {64}, pages = {113203},
 doi = {10.1103/PhysRevB.64.113203}
}

@article{Zollweg1958Absorption,
 author = {Zollweg, R. J.},
 title = {Optical Absorption and Photoemission of Barium and Strontium Oxides, Sulfides, Selenides, and Tellurides},
 journal = {Phys. Rev.}, year = {1958}, volume = {111}, pages = {113--119},
 doi = {10.1103/PhysRev.111.113}
}

\clearpage
\newgeometry{margin=2.2cm}
\onecolumn
\setcounter{section}{0}
\setcounter{subsection}{0}
\setcounter{subsubsection}{0}
\setcounter{table}{0}
\setcounter{figure}{0}
\setcounter{equation}{0}
\setcounter{footnote}{0}
\pagenumbering{arabic}
\renewcommand{\thesection}{S\arabic{section}}
\renewcommand{\thetable}{S\arabic{table}}
\renewcommand{\thefigure}{S\arabic{figure}}
\renewcommand{\theequation}{S\arabic{equation}}
\renewcommand{\thepage}{S\arabic{page}}
\renewcommand{\theHsection}{S\arabic{section}}
\renewcommand{\theHtable}{S\arabic{table}}
\renewcommand{\theHfigure}{S\arabic{figure}}
\renewcommand{\theHequation}{S\arabic{equation}}
\setlength{\emergencystretch}{2em}
\makeatletter
\patchcmd{\l@section}{1.5em}{2.2em}{}{}
\makeatother
\addtocontents{toc}{\protect\setcounter{tocdepth}{2}}
\pdfbookmark[0]{Supplementary information}{supplementary-information}
\begin{center}
{\LARGE Electronic Supplementary Information\\[6pt]\large Native implementation of the machine-learned Skala exchange--correlation functional in CP2K: Unified one-centre reconstruction for molecular and condensed-phase calculations\par}
\vspace{1.5em}
{\large Johann Pototschnig, Franz P\"oschel,\\J\"urg Hutter, and Thomas D. K\"uhne\par}
\end{center}
\vspace{1em}
\tableofcontents
\clearpage

\section{Notation and statistical measures}
\label{esi-sec:scope}

Unless stated otherwise, reported total energies and energy differences use kJ~mol$^{-1}$, forces use $E_h$/bohr, and stresses use GPa. Band gaps and their errors use eV. The molar entity is specified where energies refer to simulation cells or atoms rather than molecules or reactions. Plane-wave and relative cutoffs are given in Ry. A grid designation such as 150/770 gives the number of radial points and Lebedev angular points, respectively.

TZVPP-ae and QZVPP-ae denote the all-electron bases of the UZH protocol, while TZVP-GTH and TZV2P-GTH denote the bases used with GTH pseudopotentials.

Reference errors are calculated minus reference values. ME denotes the mean signed error, MAE the mean absolute error, and RMSE the root-mean-square error. Direct differences between two calculated protocols are reported as the mean difference (MD), mean absolute difference (MAD), and root-mean-square difference (RMSD). These measures are distinct from differences between reference-error statistics.

\section{Reconstruction, derivative and symmetry consistency}
\label{esi-sec:reconstruction}

\subsection{Spin-gradient cross terms}
\label{esi-sec:spin-gradients}

Write the two reconstructed gradient fields as $\mathbf g_\alpha=\widetilde{\mathbf g}_\alpha+\mathbf g^h_\alpha-\mathbf g^s_\alpha$ and analogously for $\beta$, where $\mathbf g_\sigma=\nabla\rho_\sigma$ for spin $\sigma\in\{\alpha,\beta\}$ and the tilde and superscripts $h,s$ identify the smooth, hard, and soft contributions. Their three spin invariants are $\gamma_{\alpha\alpha}=\mathbf g_\alpha^2$, $\gamma_{\beta\beta}=\mathbf g_\beta^2$, and $\gamma_{\alpha\beta}=\mathbf g_\alpha\cdot\mathbf g_\beta$, respectively. Expanding the cross-spin invariant gives
\begin{align}
\gamma_{\alpha\beta}={}&\widetilde{\mathbf g}_\alpha\cdot\widetilde{\mathbf g}_\beta
+\mathbf g^h_\alpha\cdot\mathbf g^h_\beta+\mathbf g^s_\alpha\cdot\mathbf g^s_\beta\nonumber\\
&+\widetilde{\mathbf g}_\alpha\cdot\mathbf g^h_\beta+\mathbf g^h_\alpha\cdot\widetilde{\mathbf g}_\beta
-\widetilde{\mathbf g}_\alpha\cdot\mathbf g^s_\beta-\mathbf g^s_\alpha\cdot\widetilde{\mathbf g}_\beta\nonumber\\
&-\mathbf g^h_\alpha\cdot\mathbf g^s_\beta-\mathbf g^s_\alpha\cdot\mathbf g^h_\beta.
\end{align}
The six component vectors give $\binom{6}{2}=15$ unordered off-diagonal dot products. Three are opposite-spin products within the same smooth, hard, or soft component, while twelve couple different components.  Separate smooth/hard/soft evaluations also combine diagonal terms with the wrong nonlinear structure. Reconstructing the fields first avoids this ambiguity and retains the nonlocal coupling of the model, which cannot be recovered by adding local algebraic corrections afterwards.

\subsection{Periodic weights and interpolation}
\label{esi-sec:periodic-weights}

This section specifies the discrete construction behind main-text Eqs.~\eqref{eq:partition} and \eqref{eq:descriptor-weights}.

\paragraph{Two centre sets.}
For a finite set $\mathcal C$ of atom-image centres, let $\mathbf R_a$ be the position of centre $a$, $r_a=|\mathbf r-\mathbf R_a|$, and $R_{ab}=|\mathbf R_a-\mathbf R_b|$. The unnormalized Becke-like shapes are
\begin{equation}
 q_a^{\mathcal C}(\mathbf r)=\prod_{b\in\mathcal C\setminus\{a\}}
 \frac{1-B(\mu_{ab})}{2},\qquad
 \mu_{ab}=\frac{r_a-r_b}{R_{ab}},\qquad
 B=f\circ f\circ f,\quad f(x)=\tfrac12x(3-x^2).
 \label{esi-eq:partition-shapes}
\end{equation}
The pair coordinate $\mu_{ab}$ enters the switching function $B$, obtained by composing the cubic polynomial $f$ three times, as indicated by $\circ$.
The pair coordinate is clipped to $[-1,1]$ to account for roundoff. No atomic-radius adjustment is applied. Normalizing these products over $\mathcal C$ yields its partition. The energy layout $\mathcal I_A$ contains the retained images of every atom around target $A$, whereas $\mathcal S_A\subset\mathcal I_A$ contains only images of $A$. The products are recomputed for each set. Removing other atoms from the normalization alone would leave their pair factors in the products, yielding a different descriptor window. The energy weight uses the normalized $q^{\mathcal I_A}$. The descriptor weight uses the window of the normalized $q^{\mathcal S_A}$ defined below.

\paragraph{Finite image layouts.}
Let $H$ be the cell matrix, $p_d\in\{0,1\}$ the image periodicity in direction $d$, and $\mathrm{nint}$ nearest-integer rounding. For each source atom $B$, the layout shared by all points of target block $A$ uses
\begin{equation}
 b_{BA,d}=p_d\,\mathrm{nint}\!\left([H^{-1}(\mathbf R_A-\mathbf R_B)]_d\right),
 \qquad n_d=b_{BA,d}-p_d,\ldots,b_{BA,d}+p_d,
 \qquad \mathbf R_{B\mathbf n}=\mathbf R_B+H\mathbf n.
 \label{esi-eq:partition-images}
\end{equation}
The columns of $H$ are the cell vectors, $d=1,2,3$ labels lattice directions, and $p_d$ is one for a periodic direction and zero otherwise. The integer $b_{BA,d}$ centres the image range, while the integer vector $\mathbf n$ identifies the image at $\mathbf R_{B\mathbf n}$.
Thus, the three-dimensional periodic layout contains 27 images per atom, with a one-shell extension about the rounded fractional displacement. Within a target block, the layout remains fixed instead of being rebuilt by nearest-image selection at each quadrature point. The self-image layout uses the same rule with $B=A$. The normalized weights sum to unity within each retained layout. Enlarging that layout is a distinct numerical convergence question. Products are evaluated with screening at floating-point precision and a logarithmic fallback to avoid underflow.

\paragraph{Descriptor window.}
The internal quadrature weight uses a window of the self-image partition, $D_{Ai}=w_{Ai}T(d_A)$, with
\begin{equation}
 T(d)=\begin{cases}
 0,&d\leq\epsilon,\\
 x^3(10-15x+6x^2),&\epsilon<d<10\epsilon,\\
 1,&d\geq10\epsilon,
 \end{cases}
 \qquad x=\frac{d-\epsilon}{9\epsilon},\qquad \epsilon=10^{-12}.
 \label{esi-eq:descriptor-window}
\end{equation}
The argument $d=d_A(\mathbf r_{Ai})$ is the target atom's self-image partition at grid point $i$, $\epsilon$ is its screening threshold, and $x$ rescales the transition interval to $[0,1]$.
The quintic taper has vanishing first and second derivatives at its endpoints. It retains the unpartitioned radial--angular weight over the interior and suppresses only the small-weight tail of the target atom's self-image partition. Other atoms and their augmentation fields remain visible inside this domain. For isolated molecular atom-composite calculations, the auxiliary FFT cell is still wrapped in the interpolation and image partition. This numerical construction does not change the nonperiodic electrostatic boundary condition. Increasing molecular cell padding tests its residual effect.

\paragraph{Weight derivatives on moving atom grids.}
Let $\overline W_{Ai}=\partial E_{\rm xc}/\partial W_{Ai}$ and $\overline D_{Ai}=\partial E_{\rm xc}/\partial D_{Ai}$ denote the model's weight adjoints at fixed fields and coordinates. For a geometric parameter $\lambda$, their contribution is
\begin{equation}
 \left(\frac{dE_{\rm xc}}{d\lambda}\right)_{\!\rm weights}
 =\sum_{Ai}\left(\overline W_{Ai}\frac{dW_{Ai}}{d\lambda}
                   +\overline D_{Ai}\frac{dD_{Ai}}{d\lambda}\right).
 \label{esi-eq:weight-chain-rule}
\end{equation}
At fixed radial--angular base weights, $dW_{Ai}/d\lambda=w_{Ai}\,dp_{A\mathbf0}/d\lambda$ and $dD_{Ai}/d\lambda=w_{Ai}T'(d_A)\,dd_A/d\lambda$, where $T'$ denotes the derivative of the window with respect to its argument. These adjoints are distinct from the field adjoints in main-text Eq.~\eqref{eq:field-adjoints}.

Write the moving quadrature point as $\mathbf r_{Ai}=\mathbf R_A+\mathbf s_{Ai}$ with fixed local offset $\mathbf s_{Ai}$. For a fixed cell and retained image layout, translation invariance gives the energy-partition derivative
\begin{equation}
 \frac{dp_{A\mathbf0}(\mathbf r_{Ai})}{d\mathbf R_C}
 =\left.\frac{\partial p_{A\mathbf0}}{\partial\mathbf R_C}\right|_{\mathbf r}
 -\delta_{AC}\sum_B
       \left.\frac{\partial p_{A\mathbf0}}{\partial\mathbf R_B}\right|_{\mathbf r},
 \label{esi-eq:moving-partition}
\end{equation}
where each partial derivative moves all retained images of its base atom at fixed evaluation point $\mathbf r$, $C$ labels the displaced atom, and the Kronecker delta $\delta_{AC}$ is one when $A=C$ and zero otherwise. The second term accounts for motion of the target grid. In contrast, the self-image window satisfies
\begin{equation}
 d_A(\mathbf R_A+\mathbf s_{Ai};\{\mathbf R_A+H\mathbf L\})
 =d_A(\mathbf s_{Ai};\{H\mathbf L\}),\qquad
 \frac{dD_{Ai}}{d\mathbf R_C}=0.
 \label{esi-eq:self-window-translation}
\end{equation}
Thus, its centre and target-grid derivatives cancel under nuclear translation, while the energy-partition, field, and explicit model-coordinate derivatives remain.

Cell strain changes the self-image separations. For $H\mapsto(1+\varepsilon)H$, $\partial(H\mathbf L)_a/\partial\varepsilon_{bc}=\delta_{ab}(H\mathbf L)_c$, where $\varepsilon$ is the infinitesimal strain tensor, $1$ denotes the identity matrix, $\mathbf L$ is an integer lattice-image vector, and $a,b,c$ are Cartesian component indices. For either normalized partition $f=p_{A\mathbf0}$ or $f=d_A$, the explicit image-translation term is
\begin{equation}
 \left.\frac{\partial f}{\partial\varepsilon_{ab}}\right|_{\rm images}
 =\sum_{(B,\mathbf L)\in\mathcal C}
    \frac{\partial f}{\partial R_{B\mathbf L,a}}(H\mathbf L)_b,
 \label{esi-eq:partition-image-strain}
\end{equation}
with the appropriate energy or self-image set $\mathcal C$. Both weight adjoints therefore contribute to the strain derivative, including the factor $T'(d_A)$ for the descriptor window. This is only the explicit image term. The remaining coordinate, interpolation, and reconstructed-field contributions must also be included as in main-text Eq.~\eqref{eq:geometric-derivative}. The total force and stress additionally require electronic stationarity and the basis/overlap Pulay terms. These identities apply to a fixed image layout and do not differentiate discrete changes in image selection.

\paragraph{Regular-grid interpolation.}
For the smooth fields, $I_{Ai,g}$ in main-text Eq.~\eqref{eq:composite} is a tensor product of twelve-node Lagrange interpolants. In each grid direction, if $t$ is the fractional coordinate relative to the lower grid point, the offsets are $j=-5,\ldots,6$ and
\begin{equation}
 L_j(t)=\prod_{\substack{m=-5\\m\ne j}}^{6}\frac{t-m}{j-m},
 \qquad I_{Ai,g}=\prod_{d=1}^{3}L_{j_d}(t_d).
 \label{esi-eq:smooth-interpolation}
\end{equation}
The polynomial $L_j$ is the Lagrange weight of stencil node $j$, with $m$ running over the other nodes. For regular-grid point $g$, $j_d$ is its stencil offset in direction $d$ and $t_d$ is the target point's local grid coordinate in that direction.
Periodic or wrapped auxiliary-grid indices are reduced modulo the grid dimensions. Density, Cartesian density-gradient components, and kinetic-energy density use the same stencil. The reverse map accumulates their adjoints using the exact transpose of the interpolation. Spatial derivatives use derivatives of the same interpolation polynomials, transformed from grid to Cartesian coordinates. An independently interpolated potential would not be that transpose.

\paragraph{Neighbouring one-centre fields.}
The hard-minus-soft fields of the target atom's central image are available directly on its grid. Other centres and noncentral images are transferred using radial quintic Hermite interpolation and the retained real-spherical-harmonic expansion on the source Lebedev grid. Radial slopes use neighbouring-node secants and curvatures use three-node differences, with one-sided endpoint treatment. The angular interpolation kernel has the form
\begin{equation}
 K_j(\widehat{\mathbf r})=\omega_j\sum_{\ell m}^{\rm retained}
 Y_{\ell m}(\widehat{\mathbf r}_j)Y_{\ell m}(\widehat{\mathbf r}),
 \label{esi-eq:angular-interpolation}
\end{equation}
where $\omega_j$ are the source angular weights, $Y_{\ell m}$ are real spherical harmonics of degree $\ell$ and order $m$, and $\widehat{\mathbf r}_j$ and $\widehat{\mathbf r}$ are the source-node and target directions relative to the source centre. Derivatives of $K_j$ and the differentiated radial interpolant provide the Cartesian derivatives. The same coefficients are used in reverse projection.

This field-image sum is distinct from the one-shell partition layout. Around the source image selected relative to each target point, the search extends by $\lceil r_c\lVert(H^{-1})_{d,:}\rVert\rceil+1$ images in every periodic direction, where $r_c$ is the retained radial support radius of the source fields, $\lVert(H^{-1})_{d,:}\rVert$ is the Euclidean norm of row $d$ of the inverse cell matrix, and $\lceil\cdot\rceil$ rounds upward to an integer. Only points inside that support and the outermost source radial node contribute. The central self contribution is omitted from this sum because it has already been included. Consequently, the descriptor domain is set by the self-image window, whereas the reconstructed fields include all contributing centres found within the separate finite-support search.

\paragraph{Kernel-level adjoint checks.}
We also test the atom-local interpolation and its hard-minus-soft back-projection using routines extracted from the CP2K source and retaining its atom-grid and spherical-harmonic types. For input fields $x$ and target-field adjoints $y$, the scalar-product check compares $L=\langle Ax,y\rangle$ with $R=\langle x,A^*y\rangle$, retaining the negative soft-field contributions. Here, $A$ denotes the linear interpolation map rather than an atom index, $A^*$ its adjoint, and $\langle\cdot,\cdot\rangle$ the discrete scalar product. The test includes density, Cartesian density gradients, and kinetic-energy density for one and two spin channels. The implemented acceptance criterion is $|L-R|/\max(1,|L|)\leq2\times10^{-13}$. The test reports a pass rather than the measured maximum of this quantity.

Separate comparisons of equivalent interpolation/back-projection variants vary the three field-component masks and use 1, 2, 4, 8, and 16 OpenMP threads. The test cases include nonperiodic, orthogonal periodic, and triclinic periodic cells, translated image points, target-ownership partitions, and empty local ownership. The normalized difference is $\epsilon_{\rm variant}=\lVert v-v_{\rm ref}\rVert_\infty/\max(1,\lVert v_{\rm ref}\rVert_\infty)$, where $v$ collects the hard and soft potential arrays, $v_{\rm ref}$ is the corresponding reference-variant result, and $\lVert\cdot\rVert_\infty$ selects the largest absolute component. Table~\ref{esi-tab:adjoint-validation} distinguishes these measured differences from the scalar-product tolerance. These kernel checks use fixed input fields.

\begin{table}[htbp]

\centering\small
\caption{Discrete atom-local interpolation and back-projection checks. The scalar-product entry gives the tolerance satisfied by the test, not a measured error. The variant comparisons report the maximum normalized differences.}
\label{esi-tab:adjoint-validation}
\begin{tabular}{@{}p{0.33\textwidth}p{0.32\textwidth}p{0.27\textwidth}@{}}
\toprule
Check & Scope & Result \\
\midrule
Adjoint scalar-product identity & $\rho$, $\nabla\rho$, $\tau$ in one or two spin channels & Passed $2\times10^{-13}$ criterion \\
Back-projection variants & Bounds/exception-checked build & $6.1088\times10^{-16}$ \\
Back-projection variants & Optimized build & $5.7513\times10^{-16}$ \\
Forward-field equivalence & Full/values-only interpolation & Bitwise identical fields \\
\bottomrule
\end{tabular}
\end{table}

\FloatBarrier

\subsection{Finite-difference protocol}
\label{esi-sec:fd}

Self-consistent total-energy finite-difference (FD) tests complement the fixed-field adjoint checks.

For a displaced coordinate $R_{A,x}$, the central FD estimate is
\begin{equation}
 F^{\rm FD}_{A,x}(h)=-\frac{E(R_{A,x}+h)-E(R_{A,x}-h)}{2h}.
\end{equation}
Here, $F^{\rm FD}_{A,x}$ is the force on atom $A$ along $x$, $h$ is the displacement step, and $E$ is the self-consistent total energy with all other nuclear coordinates fixed.
For a homogeneous $xx$ strain, both cell and Cartesian coordinates are transformed, while fractional coordinates remain fixed. The positive-pressure stress convention gives
\begin{equation}
 \sigma_{xx}^{\rm FD}(\eta)=-\frac{E(+\eta)-E(-\eta)}{2\eta\Omega}.
\end{equation}
The dimensionless step $\eta$ is the applied $xx$ strain, $\Omega$ is the unstrained cell volume, and $E(\pm\eta)$ are the total energies of the strained structures used to estimate the diagonal stress $\sigma_{xx}^{\rm FD}$.
Two step sizes distinguish truncation error from SCF noise.

Self-consistent checks use one water molecule in a periodically repeated 8~\AA{} cube at $\Gamma$, without D3, to isolate the native electronic derivatives. The grid settings are 400/60~Ry, five multigrids, and 100 radial/434 angular points. The basis/core and reconstruction definitions match the corresponding production representations. The first hydrogen is displaced along $x$ with a common reference step $h=10^{-3}$~bohr. The common homogeneous $xx$ strain is $\eta=10^{-4}$, a dimensionless quantity. Repeating the checks with $h/2=5\times10^{-4}$~bohr and $\eta/2=5\times10^{-5}$ tests step-size sensitivity. The OT threshold is $10^{-9}$. The regular-grid dimensions remain unchanged under these perturbations.

\begin{table}[htbp]
\centering\small
\caption{Signed FD minus analytic derivatives at common reference steps $h=10^{-3}$~bohr and $\eta=10^{-4}$. The half-step columns report step-size sensitivity.}
\label{esi-tab:derivatives}
\begin{tabular}{lrrrr}
\toprule
& \multicolumn{2}{c}{Force error ($E_h$/bohr)} & \multicolumn{2}{c}{Stress error (GPa)}\\
\cmidrule(lr){2-3}\cmidrule(lr){4-5}
Representation & $\Delta F(h)$ & $\Delta F(h/2)$ & $\Delta\sigma(\eta)$ & $\Delta\sigma(\eta/2)$\\
\midrule
GPW/GTH & $+2.28\times10^{-7}$ & $+2.239\times10^{-6}$ & $+3.8\times10^{-5}$ & $-2.08\times10^{-4}$ \\
GAPW-XC/GTH & $-3.77\times10^{-7}$ & $+2.380\times10^{-6}$ & $+1.2\times10^{-5}$ & $-3.32\times10^{-4}$ \\
GAPW-AE & $+2.855\times10^{-6}$ & $-2.404\times10^{-6}$ & $+7\times10^{-6}$ & $-4.72\times10^{-4}$ \\
GAPW-GTH direct & $+2.12\times10^{-7}$ & $+1.414\times10^{-6}$ & $+4.4\times10^{-5}$ & $-1.85\times10^{-4}$ \\
GAPW-GTH one-centre & $-2.726\times10^{-6}$ & $-1.977\times10^{-6}$ & $-3.2\times10^{-5}$ & $-1.7\times10^{-5}$ \\
\bottomrule
\end{tabular}
\end{table}

These checks test the self-consistent total-energy derivative, including the reconstruction and moving-grid terms, rather than only model differentiation at fixed fields. The smaller steps do not uniformly reduce the error because energy subtraction amplifies finite SCF and numerical noise. The test covers one Cartesian force component and one diagonal strain component in this periodic $\Gamma$-point system.

\FloatBarrier

\subsection{Full and symmetry-reduced meshes}
\label{esi-sec:symmetry-meshes}

Symmetry reduction changes the quadrature used to evaluate a given Brillouin-zone integral. CP2K's k-point implementation accounts for integration weights, rotations of orbital components, and Bloch-phase conventions.\cite{CP2KDynamics2026,Alizadeh2026PeriodicGFN2} For finite-difference forces, each displaced geometry must be evaluated either on the full k-point mesh or using symmetry operations valid for that geometry.

We test symmetry reduction for C and MgO by comparing total energies obtained with the full $4\times4\times4$ k-point mesh and its symmetry-reduced representation containing 10 irreducible points. Geometry and all other computational settings are identical within each comparison. The full point-group reduction uses a symmetry tolerance of $10^{-8}$. Table~\ref{esi-tab:symmetry} lists the absolute energy difference per reported cell, showing the identical AE carbon entries only once.

\begin{table}[htbp]
\centering
\caption{Selected full/reduced-grid energy checks. The two GAPW-AE/GTH variants reuse the GAPW-AE carbon calculation, which is shown only once. $|\Delta E|$ is in kJ~mol$^{-1}$ of simulation cells.}
\label{esi-tab:symmetry}
\begin{tabular}{llr}
\toprule
System & Representation & $|\Delta E|$\\
\midrule
C & GAPW-XC/GTH & 0.0200\\
C & GAPW-AE & 0.0229\\
MgO & GAPW-XC/GTH & 0.0291\\
MgO & GAPW-AE/GTH (direct) & 0.0493\\
MgO & GAPW-AE/GTH (1C) & 0.0378\\
\bottomrule
\end{tabular}
\end{table}

 Mixed MgO uses GTH Mg and AE O. Only the Mg valence-field representation changes between its direct and one-centre variants, while the O reconstruction is retained. The molecular-crystal production uses full point-group reduction of $3^3$ meshes.

\FloatBarrier

\section{Numerical settings and convergence}
\label{esi-sec:molecular-settings}

\subsection{Molecular cell padding}
\label{esi-sec:padding}

Isolated-system calculations with otherwise identical settings compare added extents of 22, 24, and 25~\AA{}. The maximum species-energy changes are 0.000417, 0.000496, and 0.000914~kJ~mol$^{-1}$ for 22--24, 24--25, and 22--25~\AA{}, respectively. The corresponding maximum reaction-energy changes are 0.000577, 0.000682, and 0.001247~kJ~mol$^{-1}$. Maximum changes of the atom-grid electron integral are $7.94\times10^{-8}$, $1.43\times10^{-7}$, and $9.72\times10^{-8}$ electrons. These comparisons retain the nonperiodic analytic electrostatic treatment.
\FloatBarrier

\subsection{Regular-grid cutoff}
\label{esi-sec:cutoff}


Table~\ref{esi-tab:ae-cutoff} examines the GAPW-AE total energy of H$_2$O at fixed atom quadrature. The notation $E(E_{\rm cut})$ denotes the total energy at plane-wave cutoff $E_{\rm cut}$. The ACONF series below instead evaluates a conformational energy difference.

\begin{table}[htbp]
\centering\small

\caption{Plane-wave cutoff dependence of the H$_2$O GAPW-AE total energy with the QZVPP-ae basis, 50 radial points and the 50-point Lebedev rule for both elements. The relative cutoff is fixed at 60~Ry. The molecule is placed in a cubic 20~\AA{} cell with nonperiodic boundary conditions and the analytic isolated-system Poisson solver. Differences are $\Delta E=E(E_{\rm cut})-E(1200~\mathrm{Ry})$.}
\label{esi-tab:ae-cutoff}
\begin{tabular}{rr}
\toprule
Cutoff/Ry & $\Delta E$/kJ~mol$^{-1}$\\
\midrule
400 & $-0.141062$ \\
600 & $+0.000101$ \\
800 & $+0.001984$ \\
1000 & $+0.008308$ \\
1200 & $0.000000$ \\
\bottomrule
\end{tabular}
\end{table}

The deviations at 600--1200~Ry are small but non-monotonic. The atom grid is held fixed here. Radial and angular quadrature are considered separately in Sec.~\ref{esi-sec:quadrature}.
\FloatBarrier

The conformational-energy comparison uses the same two ACONF geometries and the same Gaussian basis and GTH Hamiltonian for GPW and GAPW-XC. Each series is referenced to its own 800~Ry conformational energy. The table also shows the difference between representations at 800~Ry.

\begin{table}[htbp]
\centering\small
\caption{Energy difference between the two ACONF conformations in kJ mol$^{-1}$. Both GPW and GAPW-XC use the TZV2P-GTH basis. Each error is relative to its own 800 Ry result. The relative cutoff is 60 Ry.}
\label{esi-tab:aconf-cutoff}
\begin{tabular}{rrrrr}
\toprule
Cutoff/Ry & GPW & GAPW-XC & $\Delta$GPW & $\Delta$GAPW-XC\\
\midrule
240 & 3.242002 & 3.040638 & 0.114646 & -0.078099 \\
280 & 2.945569 & 3.093691 & -0.181786 & -0.025041 \\
320 & 3.126670 & 3.118586 & -0.000690 & -0.000146 \\
360 & 3.124540 & 3.119632 & -0.002816 & 0.000900 \\
400 & 3.136360 & 3.117980 & 0.009004 & -0.000757 \\
480 & 3.140075 & 3.118364 & 0.012715 & -0.000372 \\
560 & 3.129791 & 3.118335 & 0.002431 & -0.000397 \\
640 & 3.127929 & 3.118323 & 0.000569 & -0.000414 \\
800 & 3.127360 & 3.118733 & 0.000000 & 0.000000 \\
\bottomrule
\end{tabular}
\end{table}

\FloatBarrier

\subsection{Radial and angular quadrature}
\label{esi-sec:quadrature}

The radial and Lebedev angular grids of CP2K's atom-centred quadrature\cite{CP2KMadeSimple2026} are varied independently at fixed regular-grid settings. The tabulated $\Delta N$ is the atom-grid density integral minus the prescribed eight-electron count, without density renormalization.

\begin{table}[htbp]
\centering\small
\caption{Self-consistent GPW water quadrature test at 800/60 Ry. Total-energy shifts are relative to 200/974 and are in kJ~mol$^{-1}$ of water molecules. Electron-integral errors refer to eight explicit electrons.}
\label{esi-tab:water-quadrature}
\begin{tabular}{rrrr}
\toprule
Radial & Angular & $\Delta E$/kJ~mol$^{-1}$ & $\Delta N/e$\\
\midrule
50 & 50 & -0.969337 & +4.486e-04 \\
50 & 100 & +0.624157 & -4.318e-05 \\
50 & 194 & +0.692507 & -8.268e-06 \\
50 & 302 & +0.807142 & -1.074e-06 \\
50 & 434 & +0.644164 & -1.215e-07 \\
100 & 50 & -1.555769 & +4.491e-04 \\
100 & 100 & -0.028195 & -4.333e-05 \\
100 & 302 & +0.189414 & -8.919e-07 \\
100 & 434 & +0.021999 & -3.031e-07 \\
100 & 590 & +0.015974 & -2.206e-07 \\
150 & 434 & +0.033152 & -1.352e-07 \\
150 & 590 & +0.026630 & -2.079e-07 \\
150 & 770 & +0.037894 & -1.869e-07 \\
200 & 590 & +0.029718 & -2.275e-07 \\
200 & 770 & +0.041541 & -1.715e-07 \\
200 & 974 & +0.000000 & -1.347e-07 \\
\bottomrule
\end{tabular}
\end{table}

\subsection{Crystal basis, quadrature, k-point and device controls}
\label{esi-sec:crystal-numerical-controls}

The underlying crystal geometries and electronic protocols are specified in Sec.~\ref{esi-sec:crystal-protocol}. The CPU/CUDA labels distinguish model evaluation on the processor from evaluation on the accelerator through CUDA.

\label{esi-sec:urea-basis}
\label{esi-sec:crystal-basis}
Each basis comparison changes the basis for both the crystal and its isolated-molecule reference. The geometries, AE Hamiltonian, one-centre reconstruction, dispersion parameters, 800/60~Ry cutoffs, 200/974 quadrature, k-point reduction, and SCF threshold are held fixed. A separate urea comparison tests quadrature refinement with the TZVPP-ae basis. No geometry relaxation or reference-energy adjustment is involved. The three reported AE lattice energies use the QZVPP-ae basis, while the TZVPP-ae calculations quantify basis sensitivity.

\begin{table}[htbp]
\centering
\small
\caption{AE basis-set and integration-grid convergence tests. All energies are in kJ~mol$^{-1}$. The molar entity is the complete simulation cell for $E_{\rm crystal}$ and one molecule for $E_{\rm molecule}$ and $E_{\rm latt}$. Both phases use the same basis and grid. The QZVPP-ae calculations supply all three AE entries in main-text Table~\ref{tab:crystals}.}
\label{esi-tab:urea-basis}
\begin{tabular}{lllrrr}
\toprule
System & Basis & Radial/Lebedev & $E_{\rm crystal}$ & $E_{\rm molecule}$ & $E_{\rm latt}$\\
\midrule
CO$_2$ & TZVPP-ae & 200/974 & -1980161.77517989 & -495004.06742925 & -36.376366\\
CO$_2$ & QZVPP-ae & 200/974 & -1980775.68709177 & -495161.85483327 & -32.066940\\
NH$_3$ & TZVPP-ae & 200/974 & -594122.53140653 & -148491.88940781 & -38.743444\\
NH$_3$ & QZVPP-ae & 200/974 & -594188.48303155 & -148508.93620722 & -38.184551\\
Urea & TZVPP-ae & 150/770 & -1183023.65071038 & -591387.18779811 & -124.637557\\
Urea & TZVPP-ae & 200/974 & -1183024.03007136 & -591387.17216083 & -124.842875\\
Urea & QZVPP-ae & 200/974 & -1183215.53006767 & -591497.33756697 & -110.427467\\
\bottomrule
\end{tabular}
\end{table}

For urea with the TZVPP-ae basis, refining the quadrature for both crystal and isolated molecule from 150/770 to 200/974 shifts the lattice energy from $-124.637557$ to $-124.842875$~kJ~mol$^{-1}$, a change of $-0.205318$~kJ~mol$^{-1}$. Enlarging the basis in both phases to QZVPP-ae at fixed 200/974 gives $-110.427467$~kJ~mol$^{-1}$, a change of $+14.415408$~kJ~mol$^{-1}$. The same basis enlargement changes CO$_2$ from $-36.376366$ to $-32.066940$~kJ~mol$^{-1}$ and NH$_3$ from $-38.743444$ to $-38.184551$~kJ~mol$^{-1}$, with shifts of $+4.309426$ and $+0.558893$~kJ~mol$^{-1}$. Table~\ref{esi-tab:crystal-basis-errors} gives the corresponding DMC errors and basis shifts.

\label{esi-sec:crystal-basis-errors}
\begin{table}[htbp]
\centering\small
\caption{Signed lattice-energy errors and basis shifts for the AE crystal and isolated-molecule calculations using the same 200/974 quadrature in Table~\ref{esi-tab:urea-basis}, in kJ~mol$^{-1}$. DMC references are from Ref.~\citenum{DellaPia2024X23}. Negative reference errors indicate stronger binding. The MAEs use the same three crystals and unrounded energies at both basis levels.}
\label{esi-tab:crystal-basis-errors}
\begin{tabular}{lrrr}
\toprule
Crystal & TZVPP-ae--DMC & QZVPP-ae--DMC & QZVPP-ae--TZVPP-ae\\
\midrule
CO$_2$ & -6.976 & -2.667 & +4.309\\
NH$_3$ & -0.543 & +0.015 & +0.559\\
Urea & -16.343 & -1.927 & +14.415\\
\midrule
MAE & 7.954 & 1.537 & --\\
\bottomrule
\end{tabular}
\end{table}

\FloatBarrier

\begin{table}[htbp]
\centering
\caption{Numerical sensitivities in kJ\,mol$^{-1}$ per molecule. All unlisted scientific settings are held fixed. The atom-grid refinement uses energies of both the crystal and its isolated-molecule reference.}
\label{esi-tab:x23_controls_si}
\begin{tabular}{@{}p{0.27\textwidth}p{0.29\textwidth}p{0.25\textwidth}r@{}}
\toprule
System and representation & Change & Observable & Shift \\
\midrule
CO$_2$, GAPW-XC/GTH & 150/770 to 200/974 & $\Delta E_{\rm latt}$, both phases & $+0.044554$ \\
CO$_2$, GAPW-XC/GTH & 800 to 1200 Ry & $\Delta(E_{\rm crystal}/Z)$ only & $-0.015499$ \\
CO$_2$, GAPW-XC/GTH & $3^3$ to $5^3$ & $\Delta E_{\rm latt}$ & $+0.000391$ \\
NH$_3$, GAPW-XC/GTH & Symmetry tolerance $10^{-6}$ to $10^{-4}$ & $\Delta E_{\rm latt}$ & $-0.000429$ \\
CO$_2$, GAPW-XC/GTH & CPU minus CUDA & $\Delta E_{\rm latt}$ & $-0.000127$ \\
Urea, \GAPW{}-AE & CPU minus CUDA & $\Delta E_{\rm latt}$ & $-0.000379$ \\
\bottomrule
\end{tabular}
\end{table}

Refining the quadrature for both the CO$_2$ crystal and isolated molecule from 150/770 to 200/974 changes $E_{\rm crystal}/Z$ by $-0.110687$ and $E_{\rm molecule}$ by $-0.155241$~kJ~mol$^{-1}$. Their difference, $+0.044554$~kJ~mol$^{-1}$, measures the lattice-energy sensitivity. The $5^3$ crystal check uses 11 irreducible points out of 125, compared with four out of 27 at $3^3$. In the k-point, symmetry-tolerance, and device checks, the molecular reference is unchanged, so the crystal-energy shift per molecule equals the lattice-energy shift.

The 1200~Ry cutoff control concerns only the CO$_2$ crystal energy. No lattice-energy shift is reported because the separate gas-phase control lacks a complete runtime-provenance record. CO$_2$ CPU/CUDA and urea-AE CPU/CUDA controls use the same executable and computational settings within each comparison, changing only the device used for model evaluation. The device shifts are below 0.001~kJ~mol$^{-1}$ for these fixed inputs.

\FloatBarrier

\section{Selected molecular reaction comparison}
\label{esi-sec:molecular-data}

The comparison uses an identical set of 62 reactions from dietGMTKN55 for six protocols, namely GauXC GPW/GTH, GauXC GAPW-AE, PySCF, native GAPW-XC/GTH, native GAPW-AE with the TZVPP-ae basis, and native GAPW-AE with the QZVPP-ae basis.\cite{Gould2018,Goerigk2017,Poeschel2026MolecularSkala,Battaglia2026DietData} PySCF supplies independent all-electron Skala results with the def2-TZVP/ma-def2-TZVP basis protocol of GauXC GAPW-AE. The benchmark reference energies are the official dietGMTKN55 values, not the PySCF Skala results. The native AE calculations cover 140 distinct species.

The native molecular production calculations use plane-wave/relative cutoffs of 640/60~Ry for GAPW-AE and 400/60~Ry for GAPW-XC/GTH. The AE calculations use the TZVPP-ae and QZVPP-ae bases, whereas GAPW-XC/GTH uses the TZV2P-GTH basis. Both AE and GTH calculations use 100 radial points and the 434-point Lebedev rule. The production OT and diagonalization residual thresholds are $5\times10^{-7}$ and $5\times10^{-6}$, respectively.

\subsection{Cross-protocol statistics}
\label{esi-sec:cross-code}

Reactions are matched across datasets by molecular geometry, charge, multiplicity, and signed stoichiometric coefficients. The AE comparison is restricted to reactions whose constituent species are treated entirely all-electron, without mixed AE/GTH or PP substitutions. All selected CP2K AE inputs use \texttt{POTENTIAL ALL}. The two pure-GTH protocols use the same 62 reactions. The native AE calculations use the TZVPP-ae and QZVPP-ae bases with the intrinsic GAPW one-centre reconstruction.

GauXC GPW/GTH and native GAPW-XC/GTH use TZVP-GTH and TZV2P-GTH basis sets, respectively, both with GTH pseudopotentials. PySCF uses Becke partitioning without an atomic-radius adjustment. Total reaction energies retain each source's recorded D3(BJ) contribution.

\begin{table}[htbp]
\centering\small
\caption{Unweighted reference-error statistics for the identical selected 62 reactions, in kJ~mol$^{-1}$. ME is calculation minus reference. Med. and Max. denote absolute errors.}\label{esi-tab:common62-summary}
\begin{tabular}{@{}lrrrrrr@{}}
\toprule
Protocol & $N$ & ME & MAE & RMSE & Med. & Max. \\
\midrule
GauXC GPW/GTH & 62 & 0.628 & 4.978 & 9.619 & 2.452 & 60.435 \\
GauXC GAPW-AE (def2) & 62 & 1.199 & 4.053 & 9.938 & 1.162 & 62.372 \\
PySCF (def2) & 62 & 1.228 & 4.022 & 9.919 & 1.232 & 62.595 \\
Native GAPW-XC/GTH & 62 & -0.297 & 5.073 & 10.117 & 2.066 & 60.567 \\
GAPW-AE TZVPP-ae & 62 & 2.186 & 7.136 & 14.718 & 2.176 & 85.564 \\
GAPW-AE QZVPP-ae & 62 & 0.294 & 4.291 & 11.189 & 1.225 & 74.937 \\
\bottomrule
\end{tabular}
\end{table}

The def2-based GauXC protocol gives a lower molecular MAE than native TZVPP-ae (4.053 versus 7.136~kJ~mol$^{-1}$), while QZVPP-ae reaches 4.291~kJ~mol$^{-1}$.  The cross-implementation comparison includes differences in both orbital bases and numerical protocols.

Table~\ref{esi-tab:common62-summary} reports unweighted errors against the official references, whereas Table~\ref{esi-tab:common62-pairs} compares calculated reaction energies directly. Both tables use the same 62 reactions, whose individual energies are listed in Tables~\ref{esi-tab:common62-native} and \ref{esi-tab:common62-external}.

\begin{table}[htbp]
\centering\small
\caption{Differences between calculated reaction energies for the same 62 reactions, in kJ~mol$^{-1}$. Signed differences are first protocol minus second. MAD is the mean absolute difference between predictions, not the difference of their reference MAEs.}\label{esi-tab:common62-pairs}
\begin{tabular}{@{}lrrrr@{}}
\toprule
Pair & MD & MAD & RMSD & Max. \\
\midrule
Native GAPW-XC/GTH / GauXC GPW/GTH & -0.9247 & 1.9256 & 3.2918 & 16.5027 \\
TZVPP-ae / GauXC AE & 0.9870 & 5.7365 & 9.2836 & 34.8850 \\
QZVPP-ae / GauXC AE & -0.9051 & 3.5369 & 5.6717 & 23.1561 \\
GauXC AE / PySCF & -0.0290 & 0.1561 & 0.2293 & 0.8604 \\
\bottomrule
\end{tabular}
\end{table}

\begin{figure}[p]
\centering
\includegraphics[width=\textwidth]{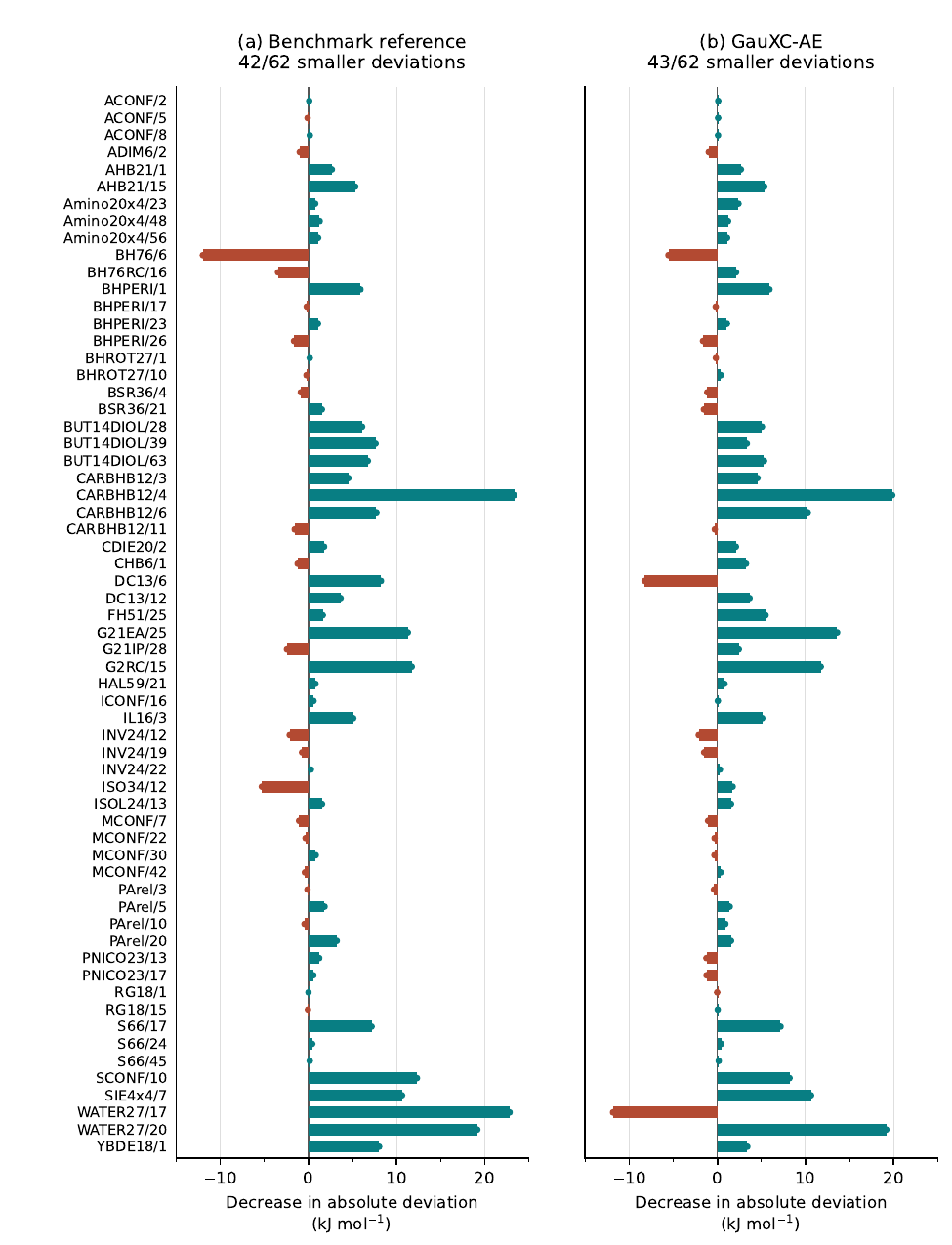}
\caption{Reaction-wise reduction in absolute energy deviations upon changing from the TZVPP-ae to the QZVPP-ae protocol, in kJ~mol$^{-1}$. Each bar is $|E_{\rm TZ}-E_c|-|E_{\rm QZ}-E_c|$, using (a) the benchmark reference or (b) GauXC-AE as $E_c$. All 62 reactions are retained in the same order in both panels. Positive values indicate smaller absolute deviations with the QZVPP-ae basis.}
\label{esi-fig:common62-basis-improvements}
\end{figure}

\FloatBarrier
\paragraph{Reaction-resolved basis-protocol comparison.}
Figure~\ref{esi-fig:common62-basis-improvements} shows how the basis change affects each reaction. Relative to TZVPP-ae, the selected QZVPP-ae protocol reduces the absolute benchmark error for 42 of the same 62 reactions and the absolute difference from GauXC-AE for 43. Positive values denote improvement and negative values deterioration.

\FloatBarrier
{\small
\begin{longtable}{@{}lrrrr@{}}
\caption{Native reaction energies on the selected 62-reaction population, in kJ~mol$^{-1}$. All three native columns use bases from the UZH protocol.}\label{esi-tab:common62-native}\\
\toprule
Reaction & Reference & GAPW-XC/GTH & GAPW-AE TZVPP-ae & GAPW-AE QZVPP-ae \\
\midrule
\endfirsthead
\toprule
Reaction & Reference & GAPW-XC/GTH & GAPW-AE TZVPP-ae & GAPW-AE QZVPP-ae \\
\midrule
\endhead
ACONF/2 & 2.56898 & 1.28293 & 2.85392 & 2.75568 \\
ACONF/5 & 2.48948 & 1.28689 & 2.45855 & 2.37104 \\
ACONF/8 & 4.92875 & 3.24638 & 5.11082 & 4.91425 \\
ADIM6/2 & 8.32616 & 11.41051 & 8.73566 & 9.70990 \\
AHB21/1 & -74.43336 & -92.83256 & -84.91928 & -82.25243 \\
AHB21/15 & -36.06608 & -48.50487 & -45.43956 & -40.11376 \\
Amino20x4/23 & 16.92428 & 15.43391 & 18.51236 & 16.11820 \\
Amino20x4/48 & 2.25936 & 1.15608 & 3.76777 & 2.47856 \\
Amino20x4/56 & 7.90358 & 8.09511 & 6.61567 & 7.72434 \\
BH76/6 & 74.47520 & 77.67614 & 74.74278 & 62.24167 \\
BH76RC/16 & 56.06560 & 45.57741 & 56.89353 & 60.33513 \\
BHPERI/1 & 147.69520 & 138.48789 & 164.22260 & 158.31086 \\
BHPERI/17 & 54.81040 & 45.25603 & 53.14221 & 52.96022 \\
BHPERI/23 & 116.31520 & 118.37341 & 110.77424 & 111.85745 \\
BHPERI/26 & 130.95920 & 131.38155 & 129.10353 & 127.47588 \\
BHROT27/1 & 11.42232 & 11.25676 & 11.86788 & 11.70391 \\
BHROT27/10 & 33.59752 & 35.31119 & 33.50158 & 33.90998 \\
BSR36/4 & 37.02840 & 35.31439 & 36.88456 & 38.02976 \\
BSR36/21 & 40.91952 & 34.89813 & 36.91690 & 38.44930 \\
BUT14DIOL/28 & 11.84072 & 11.29605 & 18.12852 & 11.65094 \\
BUT14DIOL/39 & 12.97040 & 13.02505 & 20.65342 & 13.02767 \\
BUT14DIOL/63 & 18.03304 & 17.14302 & 24.98665 & 17.83065 \\
CARBHB12/3 & 10.12946 & 11.28306 & 15.43051 & 10.87901 \\
CARBHB12/4 & 41.70193 & 41.73249 & 65.17513 & 41.58577 \\
CARBHB12/6 & 12.63986 & 12.34947 & 21.62676 & 11.37060 \\
CARBHB12/11 & 13.56034 & 16.42084 & 12.83593 & 15.83114 \\
CDIE20/2 & 18.40960 & 18.76933 & 20.91587 & 17.68792 \\
CHB6/1 & -144.05512 & -140.14019 & -143.03939 & -146.28856 \\
DC13/6 & -107.52880 & -134.02158 & -133.12497 & -124.88266 \\
DC13/12 & -245.60080 & -254.11771 & -280.98026 & -277.32449 \\
FH51/25 & -84.80968 & -90.48925 & -81.25774 & -86.72877 \\
G21EA/25 & 228.86480 & 213.17958 & 216.42552 & 230.02100 \\
G21IP/28 & 1019.67846 & 1003.55055 & 1020.42489 & 1022.85608 \\
G2RC/15 & -164.97512 & -168.08539 & -185.80314 & -174.07423 \\
HAL59/21 & 44.09936 & 50.65662 & 41.41951 & 42.23443 \\
ICONF/16 & 15.31344 & 17.05636 & 16.71153 & 16.12487 \\
IL16/3 & -489.15144 & -488.05716 & -501.85762 & -496.76800 \\
INV24/12 & 43.09520 & 46.39913 & 45.14792 & 47.26331 \\
INV24/19 & 25.94080 & 22.85371 & 25.54711 & 27.04270 \\
INV24/22 & 113.80480 & 114.44733 & 116.10448 & 115.81283 \\
ISO34/12 & 190.99960 & 201.39732 & 191.87815 & 184.85119 \\
ISOL24/13 & 138.15568 & 144.24655 & 144.97646 & 143.43009 \\
MCONF/7 & 12.21728 & 13.01917 & 10.94049 & 9.89263 \\
MCONF/22 & 22.21704 & 24.39856 & 19.82921 & 19.51291 \\
MCONF/30 & 20.33424 & 21.05847 & 21.62233 & 20.77903 \\
MCONF/42 & 30.62688 & 34.15212 & 30.69798 & 31.08516 \\
PArel/3 & 12.30096 & 12.14477 & 12.79273 & 11.70525 \\
PArel/5 & 30.08296 & 36.32561 & 32.42660 & 30.58488 \\
PArel/10 & 12.46832 & 11.04971 & 12.22813 & 13.14477 \\
PArel/20 & 8.99560 & 4.92488 & 4.69537 & 10.07264 \\
PNICO23/13 & 16.65232 & 17.44648 & 18.67000 & 17.43350 \\
PNICO23/17 & 29.70640 & 30.95232 & 30.60810 & 29.37639 \\
RG18/1 & 0.33472 & 0.73170 & 0.56026 & 0.54270 \\
RG18/15 & 1.00416 & 2.03319 & 1.18541 & 1.22589 \\
S66/17 & 71.88112 & 73.95584 & 80.23490 & 73.06240 \\
S66/24 & 11.79888 & 14.67413 & 14.49762 & 14.05858 \\
S66/45 & 7.32200 & 10.83949 & 8.51788 & 8.35847 \\
SCONF/10 & 23.63960 & 25.50964 & 36.29186 & 23.96906 \\
SIE4x4/7 & 130.95920 & 191.52574 & 216.52327 & 205.89603 \\
WATER27/17 & 320.07600 & 319.35748 & 346.31542 & 323.48213 \\
WATER27/20 & 111.29440 & 122.63740 & 141.26870 & 122.09978 \\
YBDE18/1 & 239.19928 & 234.27792 & 251.97581 & 243.94847 \\
\bottomrule
\end{longtable}
}

{\small
\begin{longtable}{@{}lrrrr@{}}
\caption{Reaction-matched earlier implementation results, in kJ~mol$^{-1}$. GauXC GAPW-AE and PySCF use def2-TZVP/ma-def2-TZVP. GauXC GPW/GTH uses a TZVP-GTH basis with GTH pseudopotentials.}\label{esi-tab:common62-external}\\
\toprule
Reaction & Reference & GauXC GPW/GTH & GauXC GAPW-AE & PySCF \\
\midrule
\endfirsthead
\toprule
Reaction & Reference & GauXC GPW/GTH & GauXC GAPW-AE & PySCF \\
\midrule
\endhead
ACONF/2 & 2.56898 & 1.51614 & 2.70488 & 2.88013 \\
ACONF/5 & 2.48948 & 1.33251 & 2.31403 & 2.20157 \\
ACONF/8 & 4.92875 & 3.25761 & 4.97115 & 4.81762 \\
ADIM6/2 & 8.32616 & 9.64238 & 7.74811 & 7.78158 \\
AHB21/1 & -74.43336 & -85.57961 & -72.79691 & -72.84224 \\
AHB21/15 & -36.06608 & -49.03304 & -36.79594 & -36.79087 \\
Amino20x4/23 & 16.92428 & 16.12294 & 15.52796 & 15.19663 \\
Amino20x4/48 & 2.25936 & 0.06588 & 2.51502 & 2.59257 \\
Amino20x4/56 & 7.90358 & 9.21476 & 8.53384 & 8.52032 \\
BH76/6 & 74.47520 & 79.98774 & 71.26673 & 71.76480 \\
BH76RC/16 & 56.06560 & 46.01286 & 59.67556 & 59.63146 \\
BHPERI/1 & 147.69520 & 143.26150 & 150.76083 & 150.95388 \\
BHPERI/17 & 54.81040 & 46.99134 & 54.38787 & 54.58550 \\
BHPERI/23 & 116.31520 & 118.89065 & 120.21677 & 120.49353 \\
BHPERI/26 & 130.95920 & 131.42996 & 132.04449 & 131.55239 \\
BHROT27/1 & 11.42232 & 11.84167 & 12.08683 & 12.10046 \\
BHROT27/10 & 33.59752 & 35.28863 & 34.84088 & 34.84112 \\
BSR36/4 & 37.02840 & 36.55882 & 34.10942 & 34.50168 \\
BSR36/21 & 40.91952 & 38.26966 & 33.46107 & 33.91601 \\
BUT14DIOL/28 & 11.84072 & 11.58231 & 12.36028 & 12.40093 \\
BUT14DIOL/39 & 12.97040 & 13.96245 & 15.16832 & 15.29769 \\
BUT14DIOL/63 & 18.03304 & 17.10798 & 18.76202 & 18.93717 \\
CARBHB12/3 & 10.12946 & 11.04613 & 10.68443 & 10.71231 \\
CARBHB12/4 & 41.70193 & 40.94177 & 43.46225 & 43.38295 \\
CARBHB12/6 & 12.63986 & 11.73896 & 9.91294 & 9.98562 \\
CARBHB12/11 & 13.56034 & 16.60086 & 14.19033 & 14.20384 \\
CDIE20/2 & 18.40960 & 18.93252 & 18.25009 & 18.28335 \\
CHB6/1 & -144.05512 & -134.14649 & -149.42828 & -149.48289 \\
DC13/6 & -107.52880 & -117.51884 & -133.99236 & -133.92558 \\
DC13/12 & -245.60080 & -259.56879 & -266.96065 & -266.10027 \\
FH51/25 & -84.80968 & -93.44890 & -88.45286 & -88.22484 \\
G21EA/25 & 228.86480 & 219.25079 & 237.07577 & 237.13020 \\
G21IP/28 & 1019.67846 & 1007.98305 & 1029.70514 & 1029.54667 \\
G2RC/15 & -164.97512 & -169.82124 & -150.91810 & -150.30841 \\
HAL59/21 & 44.09936 & 55.06964 & 42.59528 & 42.79198 \\
ICONF/16 & 15.31344 & 17.63662 & 16.38845 & 16.31562 \\
IL16/3 & -489.15144 & -485.93003 & -490.25729 & -490.64283 \\
INV24/12 & 43.09520 & 47.26901 & 43.63130 & 43.54607 \\
INV24/19 & 25.94080 & 25.76599 & 24.70678 & 24.73612 \\
INV24/22 & 113.80480 & 120.14363 & 113.10491 & 113.07369 \\
ISO34/12 & 190.99960 & 201.37424 & 187.49399 & 187.57034 \\
ISOL24/13 & 138.15568 & 142.88587 & 137.32918 & 137.05090 \\
MCONF/7 & 12.21728 & 14.25003 & 11.55123 & 11.53186 \\
MCONF/22 & 22.21704 & 25.05707 & 23.86352 & 23.67805 \\
MCONF/30 & 20.33424 & 21.03547 & 21.36303 & 21.23174 \\
MCONF/42 & 30.62688 & 33.82632 & 32.91529 & 32.83780 \\
PArel/3 & 12.30096 & 12.72470 & 12.44000 & 12.36893 \\
PArel/5 & 30.08296 & 37.95524 & 30.80324 & 30.77193 \\
PArel/10 & 12.46832 & 11.99059 & 14.03181 & 14.05480 \\
PArel/20 & 8.99560 & 6.52513 & 8.15940 & 8.46593 \\
PNICO23/13 & 16.65232 & 17.75198 & 19.13084 & 19.14686 \\
PNICO23/17 & 29.70640 & 31.38574 & 30.77153 & 30.92739 \\
RG18/1 & 0.33472 & 0.52885 & 0.56835 & 0.55532 \\
RG18/15 & 1.00416 & 1.43598 & 2.03724 & 2.05571 \\
S66/17 & 71.88112 & 69.41181 & 71.62485 & 71.58316 \\
S66/24 & 11.79888 & 12.19253 & 12.66511 & 12.61697 \\
S66/45 & 7.32200 & 10.41703 & 7.92550 & 7.98980 \\
SCONF/10 & 23.63960 & 26.07484 & 26.03616 & 25.93164 \\
SIE4x4/7 & 130.95920 & 191.39455 & 193.33166 & 193.55436 \\
WATER27/17 & 320.07600 & 319.44002 & 340.81766 & 340.55184 \\
WATER27/20 & 111.29440 & 127.42934 & 110.07609 & 109.82968 \\
YBDE18/1 & 239.19928 & 228.59282 & 246.26064 & 245.89214 \\
\bottomrule
\end{longtable}
}

\FloatBarrier

\section{Molecular-crystal lattice energies}
\label{esi-sec:x23_si}

\subsection{Molecular-crystal geometries and electronic protocol}
\label{esi-sec:crystal-protocol}

The crystal and gas-phase geometries are taken from the supporting information of Della Pia \emph{et al.}\cite{DellaPia2024X23} Crystal coordinates are wrapped into the supplied cells without relaxation or symmetrization. CO$_2$ and NH$_3$ have cubic cells with $a=5.624$ and 5.1305~\AA{} and four molecules each. Urea has a tetragonal cell with $a=b=5.565$, $c=4.684$~\AA{}, and two molecules. The separately supplied gas molecules are rigidly translated into centred cells whose dimensions are the molecular extents plus 25~\AA{}. They use nonperiodic boundary conditions for both the cell and electrostatics, together with the analytic isolated-system Poisson solver.

All four representations use Skala-1.1 Rev1 on a smoothly partitioned atom-composite quadrature and five multigrids with cutoffs of 800/60~Ry for both crystal and molecule. The reported AE energies for all three crystals and their isolated-molecule references use the QZVPP-ae basis with explicit nuclei, 200 radial points, and the 974-point Lebedev rule. GTH calculations use TZV2P-GTH bases, GTH potentials, 150 radial points, and the 770-point Lebedev rule. The TZVPP-ae calculations and the coarser 150/770 AE quadrature provide numerical controls. Matched TZVPP-ae/QZVPP-ae basis comparisons use 200/974 throughout (Sec.~\ref{esi-sec:crystal-basis}).

The augmented representations employ the small extended expansion and accurate augmented-XC integration.\cite{CP2K2020} GAPW-AE and GAPW-XC/GTH each retain their intrinsic reconstruction and are not subdivided into direct and one-centre variants. Only the two ordinary GAPW-GTH representations compare direct and one-centre reconstructed valence fields at fixed Hamiltonian and basis. All required hard-minus-soft fields are assembled before the joint model evaluation.

D3(BJ) is included, with $s_6=1$, $s_8=1.9889$, $a_1=0.3981$, $a_2=4.4211$, and the CP2K input parameter \texttt{R\_CUTOFF} set to 20~\AA{}.\cite{Grimme2011D3BJ} The factors $s_6$ and $s_8$ scale the sixth- and eighth-order pair-dispersion terms, while $a_1$ and $a_2$ set the Becke--Johnson damping. The B3LYP parameter alias selects these damping constants, not the electronic functional. All systems are neutral restricted singlets. The OT threshold is $5\times10^{-7}$. No density renormalization, energy alignment, or fitting to the reference lattice energies is applied.

Every crystal uses a $\Gamma$-centred $3^3$ mesh with full point-group reduction through spglib.\cite{Togo2024Spglib} With symmetry tolerance $10^{-6}$, the AE controls with the TZVPP-ae basis and 150/770 quadrature and GAPW-XC give 4, 10, and 10 irreducible points for CO$_2$, NH$_3$, and urea. The matched fine-grid CO$_2$ and NH$_3$ AE basis controls use $10^{-4}$ and four irreducible points in both bases, while urea retains ten. The GTH-direct and GTH-1C calculations use $10^{-4}$ and give 4, 4, and 6. These finite tolerances recognize near-symmetries in the decimal coordinates without moving the atoms. Their numerical sensitivity is distinct from a full-versus-reduced-grid comparison.

The electronic diffusion Monte Carlo (DMC) references have statistical uncertainties of 0.2, 0.1, and 0.3~kJ~mol$^{-1}$ for CO$_2$, NH$_3$, and urea. The estimated overall uncertainty is about 2~kJ~mol$^{-1}$ and exceeds these statistical error bars.\cite{DellaPia2024X23} Thermal and zero-point corrections to experimental sublimation enthalpies are not added to the DMC energies.

\subsection{Comparison with other electronic-structure methods}
\label{esi-sec:crystal-literature}

Table~\ref{esi-tab:x23-dft-context} compares the native results with the DMC references and published density-functional electronic lattice energies collected by Della Pia \emph{et al.}\cite{DellaPia2024X23,DellaPia2025MLIP} Their plane-wave calculations use hard projector augmented-wave (PAW) datasets, a 1000~eV cutoff, and dense system-specific k-point meshes. The PBE0+MBD values use FHI-aims. The MAE is the unweighted mean of $|E_{\rm latt}-E_{\rm latt}^{\rm DMC}|$ over CO$_2$, NH$_3$, and urea, using unrounded native energies and the reported literature values.

The periodic MP2 and SCS(MI)-MP2 values of Liang \emph{et al.}\cite{Liang2023CrystalMP2} use frozen-core correlation, cc-pVTZ/cc-pVQZ extrapolation to the complete-basis-set (CBS) limit, and extrapolated Brillouin-zone sampling. Their counterpoise-corrected cohesive energies include gas-phase monomer relaxation. We reverse the signs of their positive dissociation energies to express lattice energies as negative values. These calculations use PBE-TS-optimized X23 geometries, rather than the Della Pia geometries used here.

The multi-level calculation of Syty \emph{et al.}\cite{Syty2025CrystalCC} combines LNO-CCSD(T) for monomer relaxation and short-range dimers with RPA+ph for distant dimers and trimers and a periodic Hartree--Fock correction. Monomer and dimer energies use aug-cc-pVTZ/aug-cc-pVQZ extrapolation, while trimers use aug-cc-pVTZ. The geometries are taken from Della Pia \emph{et al.} without reoptimization. The tabulated total electronic lattice energies come from this composite method rather than a uniform periodic CCSD(T) treatment.

\begin{table}[htbp]
\centering
\small
\caption{Electronic lattice energies and three-crystal MAEs, all in kJ~mol$^{-1}$. DMC references are from Ref.~\citenum{DellaPia2024X23}, with statistical uncertainties in parentheses. Native Skala--D3(BJ) values are from main-text Table~\ref{tab:crystals}. Other density-functional values are transcribed from Tables S5, S8, and S22 of Ref.~\citenum{DellaPia2025MLIP}. Each dispersion model is part of the stated method. MP2 and SCS(MI)-MP2 values are from Table S2 of Ref.~\citenum{Liang2023CrystalMP2}, with reversed signs. Multi-level CCSD(T) denotes the LNO-CCSD(T)/RPA+ph/HF method from Table S5 of Ref.~\citenum{Syty2025CrystalCC}. MAEs summarize these three crystals.}
\label{esi-tab:x23-dft-context}
\begin{tabular}{lrrrr}
\toprule
Method & CO$_2$ & NH$_3$ & Urea & MAE\\
\midrule
DMC reference & -29.4(2) & -38.2(1) & -108.5(3) & --\\
\midrule
Skala--D3(BJ), GAPW-XC/GTH & -37.41 & -40.07 & -113.73 & 5.04\\
Skala--D3(BJ), GAPW-AE & -32.07 & -38.18 & -110.43 & 1.54\\
Skala--D3(BJ), GAPW-GTH direct & -37.38 & -40.09 & -113.80 & 5.06\\
Skala--D3(BJ), GAPW-GTH one-centre & -37.39 & -40.08 & -113.72 & 5.03\\
\midrule
PBE & -6.51 & -28.24 & -73.24 & 22.70\\
PBE+D3 & -25.07 & -43.76 & -109.90 & 3.76\\
PBE+MBD & -22.74 & -42.10 & -109.11 & 3.72\\
revPBE+D3 & -22.60 & -38.32 & -99.88 & 5.18\\
SCAN+rVV10 & -31.47 & -41.69 & -113.55 & 3.54\\
vdW-DF2 & -32.75 & -39.70 & -103.85 & 3.17\\
PBE0+MBD & -23.74 & -38.99 & -109.40 & 2.45\\
MP2/CBS & -27.81 & -38.34 & -108.66 & 0.63\\
SCS(MI)-MP2/CBS & -25.59 & -33.31 & -101.42 & 5.26\\
Multi-level CCSD(T) & -29.4 & -37.6 & -111.0 & 1.03\\
\bottomrule
\end{tabular}
\end{table}

\FloatBarrier

\section{Ice lattice energies and relative phase energies}
\label{esi-sec:ice-basis}

The AE comparison includes all thirteen DMC-ICE13 phases at both orbital-basis levels, namely Ih, II, III, IV, VI, VII, VIII, IX, XI, XIII, XIV, XV, and XVII. Their electronic lattice energies provide a second test of the basis sensitivity observed for the molecular crystals. The DMC references are from Della Pia \emph{et al.}\cite{DellaPia2022Ice}

\subsection{ICE13 electronic protocol and molecular reference}
\label{esi-sec:ice-protocol}

Both basis levels use the same fixed crystal geometries, Skala--D3(BJ), explicit-electron Hamiltonian, joint one-centre reconstruction, 800/60~Ry cutoffs, and 150/770 atom quadrature. At each basis level, the crystal and isolated water reference use matched basis, cutoff, and quadrature settings, with nonperiodic electrostatics for the molecule. The SCF threshold is $5\times10^{-7}$ with OT. The QZVPP-ae crystals use $\Gamma$-centred $3^3$ meshes with full point-group reduction through spglib.\cite{Togo2024Spglib} The inputs specify the symmetry tolerances and irreducible meshes for each phase.

The TZVPP-ae calculations use CP2K's K290 symmetry backend, including its inversion fallback for ice IX. Separate same-basis spglib checks for IX and VII change the energy by $-0.000035$ and $+0.000200$~kJ~mol$^{-1}$ per molecule, respectively. These controls distinguish the symmetry-backend change from the much larger basis shifts. For XIII, the irreducible meshes contain 14 points with the TZVPP-ae basis and 10 with the QZVPP-ae basis, both with full point-group reduction. Its TZVPP-ae and QZVPP-ae protocols differ in orbital basis, symmetry backend, and symmetry tolerances.

The gas-phase reference is a centred isolated water molecule at the reconstructed Partridge--Schwenke equilibrium geometry, with OH distance 0.95865~\AA{} and HOH angle $104.348^\circ$.\cite{Gillan2012WaterClusters} Its total energies are $-200621.54001019$ and $-200683.26812332$~kJ~mol$^{-1}$ with the TZVPP-ae and QZVPP-ae bases, respectively. This geometry follows the equilibrium molecular reference, although a direct comparison with the original DMC monomer coordinates was not possible because that file was unavailable. Any molecular-reference mismatch would shift all absolute ice lattice energies together. Relative phase energies cancel that molecular reference exactly.

\subsection{Absolute and relative lattice energies}
\label{esi-sec:ice-results}

\begin{table}[htbp]
\centering\small
\caption{Complete matched DMC-ICE13 comparison. Electronic lattice energies and signed residuals are in kJ~mol$^{-1}$ per water molecule, with $\delta=E_{\mathrm{latt}}-E_{\mathrm{DMC}}$. Parentheses give DMC statistical uncertainties. MAE, mean signed error (ME), and root-mean-square error (RMSE) use all thirteen phases at each basis level.}
\label{esi-tab:ice-basis}
\begin{tabular}{lrrrrr}
\toprule
Phase & TZVPP-ae & QZVPP-ae & DMC & $\delta_{\rm TZ}$ & $\delta_{\rm QZ}$\\
\midrule
Ih & -76.948 & -57.727 & -59.45(7) & -17.498 & 1.723\\
II & -80.829 & -57.918 & -59.14(7) & -21.689 & 1.222\\
III & -76.814 & -55.868 & -58.20(7) & -18.614 & 2.332\\
IV & -77.539 & -54.809 & -55.62(7) & -21.919 & 0.811\\
VI & -80.663 & -56.275 & -57.67(7) & -22.993 & 1.395\\
VII & -80.495 & -55.128 & -54.46(7) & -26.035 & -0.668\\
VIII & -79.642 & -54.803 & -55.22(8) & -24.422 & 0.417\\
IX & -78.841 & -57.209 & -58.85(7) & -19.991 & 1.641\\
XI & -77.043 & -57.760 & -59.29(8) & -17.753 & 1.530\\
XIII & -79.942 & -57.036 & -57.33(7) & -22.612 & 0.294\\
XIV & -81.015 & -57.142 & -57.75(7) & -23.265 & 0.608\\
XV & -80.474 & -56.286 & -57.71(7) & -22.764 & 1.424\\
XVII & -75.684 & -56.674 & -57.70(8) & -17.984 & 1.026\\
\midrule
MAE & -- & -- & -- & 21.349 & 1.161\\
ME & -- & -- & -- & -21.349 & 1.058\\
RMSE & -- & -- & -- & 21.511 & 1.292\\
\bottomrule
\end{tabular}
\end{table}

Table~\ref{esi-tab:ice-basis} reports the phase-resolved lattice energies, signed residuals, and aggregate errors. All thirteen TZVPP-ae energies overbind. With the QZVPP-ae basis, only VII overbinds and III has the largest absolute residual. Thus, the large systematic TZVPP-ae overbinding is not retained with the enlarged AE basis and fully reduced k-point protocol.  As for the three molecular crystals, enlarging the AE basis improves the balance between crystal and gas-phase energies. The remaining errors generally exceed the DMC statistical uncertainties quoted in the table.

\begin{table}[htbp]
\centering\small
\caption{Relative electronic energies with respect to ice Ih, in kJ~mol$^{-1}$ per water molecule. The MAE uses all twelve nontrivial differences, excluding the identically zero Ih row. DMC differences are calculated from the rounded absolute energies in Table I of Ref.~\citenum{DellaPia2022Ice}, so the XI value is 0.16 rather than its separately rounded 0.15. The gas-phase reference cancels exactly.}
\label{esi-tab:ice-relative}
\begin{tabular}{lrrrr}
\toprule
Phase & TZVPP-ae & QZVPP-ae & DMC & QZVPP-ae--DMC\\
\midrule
Ih & 0.000 & 0.000 & 0.00 & 0.000\\
II & -3.881 & -0.190 & 0.31 & -0.500\\
III & 0.133 & 1.859 & 1.25 & 0.609\\
IV & -0.592 & 2.918 & 3.83 & -0.912\\
VI & -3.716 & 1.452 & 1.78 & -0.328\\
VII & -3.547 & 2.600 & 4.99 & -2.390\\
VIII & -2.694 & 2.924 & 4.23 & -1.306\\
IX & -1.894 & 0.518 & 0.60 & -0.082\\
XI & -0.095 & -0.033 & 0.16 & -0.193\\
XIII & -2.994 & 0.692 & 2.12 & -1.428\\
XIV & -4.068 & 0.585 & 1.70 & -1.115\\
XV & -3.527 & 1.442 & 1.74 & -0.298\\
XVII & 1.263 & 1.054 & 1.75 & -0.696\\
\midrule
MAE & 4.173 & 0.821 & -- & --\\
\bottomrule
\end{tabular}
\end{table}

The relative energies in Table~\ref{esi-tab:ice-relative} test the balance among condensed phases without a gas-phase reference. For a phase $p$ containing $N_p$ water molecules per cell, we evaluate
\begin{equation}
 \Delta E_p = E_p/N_p-E_{\mathrm{Ih}}/N_{\mathrm{Ih}}.
\end{equation}
Here, $E_p$ is the total cell energy of phase $p$, and $E_{\mathrm{Ih}}$ and $N_{\mathrm{Ih}}$ are the corresponding energy and molecule count for ice Ih, so $\Delta E_p$ is a relative energy per water molecule.
The smaller relative-energy errors show that the basis improvement is not merely a common shift in absolute binding. The relative-energy RMSE decreases from 4.857 to 1.036~kJ~mol$^{-1}$.  A systematic bias remains in the relative energies, as QZVPP-ae underestimates eleven of the twelve differences, with a mean signed error of $-0.720$~kJ~mol$^{-1}$. Ice XIII illustrates how close absolute agreement can coexist with a larger relative-energy error. II and XI remain slightly below Ih with the QZVPP-ae basis, unlike the DMC central values, and VII has the largest relative residual.  The comparison concerns electronic energies at fixed geometries, not finite-temperature phase stability, which also requires vibrational and pressure contributions.

\FloatBarrier

\section{Ten-solid equations of state}
\label{esi-sec:solids}

\subsection{Structures and equation of state}
\label{esi-sec:solid-structures}

The selected solids are AlN, AlP, BN, BP, C, LiCl, MgO, MgS, Si, and SiC from the Goldzak set.\cite{Goldzak2022Solids} All use conventional cubic eight-atom cells. Energies are evaluated at volumes around the static-lattice experimental volume, with the same volume set in all four representations for each solid. Third-order Birch--Murnaghan fits give the equilibrium lattice constant and elastic response under isotropic compression. The two GAPW-AE/GTH variants reuse the GAPW-AE curves for BN and C.

The energy is fitted to
\begin{equation}
 E(V)=E_0+\frac{9V_0B_0}{16}
 \left[B'_0(x-1)^3+(6-4x)(x-1)^2\right],\qquad
 x=(V_0/V)^{2/3},
\end{equation}
where $E(V)$ is the total energy of a cell of volume $V$, $E_0$ and $V_0$ are the equilibrium energy and volume, $B_0$ is the bulk modulus, and $B'_0=(\partial B/\partial P)_{P=0}$ is its pressure derivative at zero pressure $P$. The dimensionless variable $x$ expresses the volume ratio, and the fitted cubic lattice constant is $a_0=V_0^{1/3}$.

\subsection{Ten-solid electronic protocol}
\label{esi-sec:solid-settings}

All calculations use 800/60~Ry cutoffs, 150 radial points, the 770-point Lebedev rule, and the small extended one-centre basis. All are neutral, spin-restricted calculations with UZH 2026.2 basis and potential definitions.\cite{Mirhosseini2026UZH} GAPW-XC/GTH uses TZV2P-GTH bases and the corresponding element-specific GTH potentials for every atom. GAPW-AE uses TZVPP-ae orbital bases and explicit core electrons for every element. Both series cover all ten solids.

The two GAPW-AE/GTH variants instead use a fixed element-wise selection. Li, B, C, N, and O retain the TZVPP-ae basis, while Mg, Al, Si, P, S, and Cl use the TZV2P-GTH basis. This selection was not required by missing AE bases, which are available for every LC10 element. GAPW-AE/GTH (direct) uses direct GTH valence fields. GAPW-AE/GTH (1C) uses smooth GTH valence fields plus one-centre hard-minus-soft contributions. The pseudopotentials and orbital bases are identical between these two variants. Their direct/1C distinction applies only to GTH atoms, and intrinsic AE reconstruction is retained in both.

The fixed selection gives genuinely mixed AE/GTH calculations for AlN, BP, LiCl, MgO, and SiC, all-GTH calculations for AlP, MgS, and Si, and the same all-AE calculation for BN and C. The last two therefore provide no independent direct/one-centre test. The eight independent comparisons between direct and one-centre GTH treatments comprise five mixed and three all-GTH systems. Every reported energy includes D3(BJ). Diagonalization and OT use residual thresholds of $5\times10^{-6}$ and $5\times10^{-7}$, respectively. Exact inputs specify the remaining dispersion and integration conventions.

The actual $\Gamma$-centred meshes are listed in Table~\ref{esi-tab:actual-k-meshes}. The mesh is identical across representations at each volume but can change with volume within an EOS. The selected full/reduced-mesh checks in Sec.~\ref{esi-sec:symmetry-meshes} constrain symmetry reduction for the tested geometries and meshes.

\subsection{Basis sensitivity of structural observables}
\label{esi-sec:solid-basis}

TZVPP-ae and QZVPP-ae are compared at identical volumes for Si, diamond, and MgO. We define the basis-induced energy shift as
\begin{equation}
 \delta E(V)=E_{\mathrm{QZ}}(V)-E_{\mathrm{TZ}}(V).
\end{equation}
Here, $E_{\mathrm{TZ}}$ and $E_{\mathrm{QZ}}$ are evaluated with the TZVPP-ae and QZVPP-ae bases, respectively, at the same cell volume $V$.
A constant $\delta E$ changes only the energy origin. Its volume dependence can instead alter the equilibrium volume, curvature, and higher derivatives that determine the elastic response.

Geometry, full k-point mesh, core treatment, cutoffs, and quadrature are unchanged at each common volume. The QZVPP-ae calculations use full point-group reduction, through spglib for Si and diamond and the native K290 backend for MgO, and diagonalization with an SCF threshold of $5\times10^{-6}$. Same-basis checks reproduce the TZVPP-ae energies to $4.59\times10^{-4}$~kJ~mol$^{-1}$ of cells for Si and $1.59\times10^{-5}$~kJ~mol$^{-1}$ of cells for diamond. The corresponding MgO check differs by $6.72\times10^{-3}$~kJ~mol$^{-1}$ of cells. These small differences separate the basis effect from changes in the executable and symmetry path.

\begin{table}[htbp]
\centering\small
\caption{AE basis-set comparison. Lattice constants are in \AA{} and bulk moduli in GPa. Both bases use the same volume set. Experimental static-lattice references are from Ref.~\citenum{Goldzak2022Solids}. These three QZVPP-ae controls do not replace the uniform ten-solid TZVPP-ae comparison.}
\label{esi-tab:solid-basis}
\begin{tabular}{llrr}
\toprule
Solid & Basis/reference & $a_0$ & $B_0$\\
\midrule
Si & TZVPP-ae & 5.3747 & 100.36\\
Si & QZVPP-ae & 5.3551 & 106.06\\
Si & Experiment & 5.4210 & 100.30\\
C & TZVPP-ae & 3.5292 & 491.16\\
C & QZVPP-ae & 3.5365 & 475.81\\
C & Experiment & 3.5530 & 453.30\\
MgO & TZVPP-ae & 4.0935 & 198.84\\
MgO & QZVPP-ae & 4.0904 & 197.45\\
MgO & Experiment & 4.1890 & 173.00\\
\bottomrule
\end{tabular}
\end{table}

The response is material dependent (Table~\ref{esi-tab:solid-basis}). For diamond, QZVPP-ae increases $a_0$ by 0.00735~\AA{} and lowers $B_0$ by 15.35~GPa, improving both relative to experiment. For Si, $a_0$ decreases by 0.01950~\AA{} and $B_0$ increases by 5.70~GPa, worsening both errors. MgO changes much less, with shifts of $-0.00307$~\AA{} and $-1.39$~GPa. Its lattice constant remains substantially shorter than experiment, while its bulk modulus improves slightly but remains too large. Across the sampled volumes, the span of $\delta E(V)$ is 1.798, 2.316, and 0.211~kJ~mol$^{-1}$ of atoms for diamond, Si, and MgO, respectively. These nonconstant shifts explain why the basis changes affect the fitted structure as well as the absolute energy.

The basis-set tests for molecular crystals and ice in Secs.~\ref{esi-sec:crystal-basis} and \ref{esi-sec:ice-basis} identify substantial basis contributions to crystal--molecule energy differences. The structural observables instead depend on how the crystal energy varies with volume. The uniform ten-solid comparison below uses the TZVPP-ae protocol.

\subsection{Matched literature comparisons}
\label{esi-sec:solid-literature}

All methods are compared against the same primary static-lattice experimental values from Goldzak \emph{et al.}\cite{Goldzak2022Solids} Their Hartree--Fock (HF), second-order M{\o}ller--Plesset perturbation theory (MP2), spin-component-scaled MP2 (SCS-MP2), and scaled opposite-spin MP2 (SOS-MP2) results cover all ten selected solids (Table~\ref{esi-tab:literature-Goldzak2022}). Zhang \emph{et al.}\cite{Zhang2018Solids} supply local-density approximation (LDA), PBE, PBEsol, M06-L, SCAN, and HSE06 values for nine of the selected solids (Table~\ref{esi-tab:literature-Zhang2018}). AlN is omitted from this comparison because it is not reported in that work. We use their uncorrected static electronic values, not predictions with zero-point corrections. Their SCAN calculations were non-self-consistent on PBE densities.

\begin{figure}[htbp]
\centering
\includegraphics[width=\textwidth]{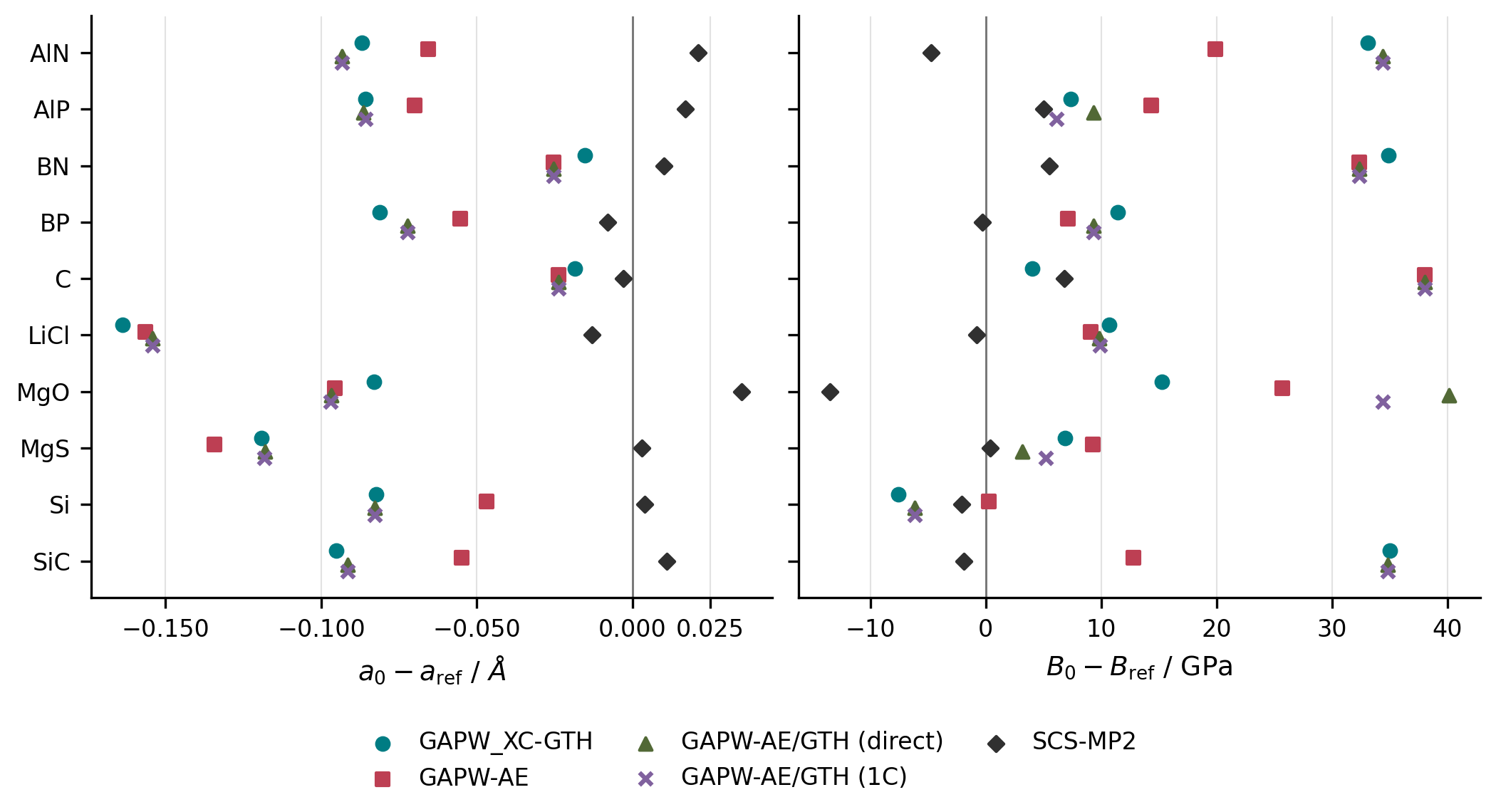}
\caption{Signed lattice-constant and bulk-modulus errors against the common Goldzak static-lattice reference. SCS-MP2 provides a correlated-method comparison on the same solids. Systematic native lattice contraction persists across representations.}
\label{esi-fig:periodic-residuals}
\end{figure}

SCS-MP2 and SOS-MP2 give matched ten-solid lattice MAEs of 0.0125 and 0.0149~\AA{}, and bulk-modulus MAEs of 4.10 and 3.88~GPa. The native ranges are 0.0728--0.0844~\AA{} and 16.61--21.77~GPa (main-text Table~\ref{tab:lc10-primary} and Table~\ref{esi-tab:lc10-reconstruction}). The SCS-MP2 comparison in Fig.~\ref{esi-fig:periodic-residuals} highlights the systematic native lattice contraction across this set. Established semilocal and screened-hybrid methods also have smaller lattice errors on the material sets shared with those studies. These results identify a limitation of the present structural predictions. The comparisons do not uniquely separate functional and numerical contributions.

Alternative experimental moduli listed by Goldzak \emph{et al.} for BN, BP, and MgO are 410.2, 168.0, and 169.8~GPa, instead of the primary 388.5, 176.5, and 173.0~GPa. Simultaneously replacing all three lowers each native ten-solid modulus MAE by 1.00~GPa and leaves the qualitative conclusion intact.

\begin{table*}[t]
\centering
\small
\caption{GAPW-AE/GTH equilibrium lattice constants $a_0$ and bulk moduli $B_0$ using direct and one-centre GTH valence fields. Experimental static-lattice references follow Ref.~\citenum{Goldzak2022Solids}. Each MAE uses all ten solids. Both variants use the GAPW-AE results for BN and C. Across the eight systems containing GTH atoms in the GAPW-AE/GTH variants, the mean absolute direct/one-centre difference is 1.39~GPa, with a maximum of 5.75~GPa for MgO.}
\label{esi-tab:lc10-reconstruction}
\begin{tabular}{lrrrrrr}
\toprule
& \multicolumn{3}{c}{$a_0$ (\AA{})} & \multicolumn{3}{c}{$B_0$ (GPa)} \\
\cmidrule(lr){2-4}\cmidrule(lr){5-7}
Solid & Experiment & Direct & One-centre & Experiment & Direct & One-centre \\
\midrule
AlN & 4.3680 & 4.2747 & 4.2746 & 206.00 & 240.42 & 240.40 \\
AlP & 5.4480 & 5.3616 & 5.3621 & 87.00 & 96.36 & 93.12 \\
BN & 3.5930 & 3.5675 & 3.5675 & 388.50 & 420.87 & 420.87 \\
BP & 4.5250 & 4.4527 & 4.4527 & 176.50 & 185.86 & 185.86 \\
C & 3.5530 & 3.5292 & 3.5292 & 453.30 & 491.35 & 491.35 \\
LiCl & 5.0720 & 4.9178 & 4.9179 & 37.30 & 47.13 & 47.16 \\
MgO & 4.1890 & 4.0923 & 4.0919 & 173.00 & 213.13 & 207.37 \\
MgS & 5.1880 & 5.0700 & 5.0697 & 81.00 & 84.17 & 86.22 \\
Si & 5.4210 & 5.3383 & 5.3383 & 100.30 & 94.13 & 94.13 \\
SiC & 4.3470 & 4.2556 & 4.2556 & 228.90 & 263.73 & 263.71 \\
\midrule
MAE & -- & 0.0844 & 0.0844 & -- & 21.77 & 21.07 \\
\bottomrule
\end{tabular}
\end{table*}

\begin{table*}[t]
\centering
\small
\caption{Matched comparison with Ref.~\citenum{Goldzak2022Solids} on the ten-solid set. Every row uses the identical intersection and the Goldzak static-lattice experimental reference. MAE($a_0$) is in \AA{} and MAE($B_0$) in GPa. A dash denotes unavailable data, not zero error. Native rows retain the numerical qualifications discussed in the text.}
\label{esi-tab:literature-Goldzak2022}
\begin{tabular}{lrrr}
\toprule
Method & Solids & MAE($a_0$) & MAE($B_0$) \\
\midrule
GAPW-XC/GTH & 10 & 0.0831 & 16.61 \\
GAPW-AE & 10 & 0.0728 & 16.89 \\
GAPW-AE/GTH (direct) & 10 & 0.0844 & 21.77 \\
GAPW-AE/GTH (1C) & 10 & 0.0844 & 21.07 \\
HF & 10 & 0.0563 & 15.23 \\
MP2 & 10 & 0.0201 & 5.71 \\
SCS-MP2 & 10 & 0.0125 & 4.10 \\
SOS-MP2 & 10 & 0.0149 & 3.88 \\
\bottomrule
\end{tabular}
\end{table*}

\begin{table*}[t]
\centering
\small
\caption{Matched comparison with Ref.~\citenum{Zhang2018Solids} on nine solids. AlN is omitted because no values are reported for it in that work. Every row uses the identical intersection and the Goldzak static-lattice experimental reference. MAE($a_0$) is in \AA{} and MAE($B_0$) in GPa. A dash denotes unavailable data, not zero error. Native rows retain the numerical qualifications discussed in the text.}
\label{esi-tab:literature-Zhang2018}
\begin{tabular}{lrrr}
\toprule
Method & Solids & MAE($a_0$) & MAE($B_0$) \\
\midrule
GAPW-XC/GTH & 9 & 0.0827 & 14.78 \\
GAPW-AE & 9 & 0.0737 & 16.56 \\
GAPW-AE/GTH (direct) & 9 & 0.0834 & 20.36 \\
GAPW-AE/GTH (1C) & 9 & 0.0835 & 19.59 \\
LDA & 9 & 0.0324 & 4.66 \\
PBE & 9 & 0.0451 & 12.96 \\
PBEsol & 9 & 0.0113 & 4.72 \\
M06-L & 9 & 0.0128 & 3.46 \\
SCAN & 9 & 0.0123 & 3.08 \\
HSE06 & 9 & 0.0170 & 5.86 \\
\bottomrule
\end{tabular}
\end{table*}

\begin{table*}[t]
\centering
\small
\caption{$\Gamma$-centred k-point meshes used along each EOS. The mesh is identical across representations at a given volume but can change with volume. The accompanying inputs give the volume-dependent assignments.}
\label{esi-tab:actual-k-meshes}
\begin{tabular}{lr}
\toprule
Solid & Meshes \\
\midrule
AlN & $5^3$ \\
AlP & $4^3$ \\
BN & $6^3$, $7^3$ \\
BP & $5^3$ \\
C & $6^3$, $7^3$ \\
LiCl & $5^3$ \\
MgO & $5^3$, $6^3$ \\
MgS & $4^3$, $5^3$ \\
Si & $4^3$, $5^3$ \\
SiC & $5^3$ \\
\bottomrule
\end{tabular}
\end{table*}

\FloatBarrier

\section{Electronic band gaps}
\label{esi-sec:bandgaps}

\subsection{Material selection}
\label{esi-sec:bandgap-selection}

The dataset comprises 50 band-gap results for 28 materials with experimental references taken from the literature, including 20 AE, 23 GAPW-XC/GTH, and seven mixed AE/GTH results (Table~\ref{esi-tab:bandgap-outcomes}).\cite{Lee2021Gaps} The primary representations are GAPW-XC/GTH and GAPW-AE wherever the available AE bases cover all atoms. Mixed AE/GTH calculations use \texttt{PAW\_ONE\_CENTER}.

\subsection{Band-gap geometries and electronic protocol}
\label{esi-sec:bandgap-protocol}

AE atoms use TZVPP-ae and \texttt{POTENTIAL ALL}. GTH atoms use TZV2P-GTH and GTH potentials. The seven mixed AE/GTH calculations treat P/As in InP/InAs, S/Se in CdS/CdSe, S/Se in BaS/BaSe, and Mg in MgTe as AE. Their complementary elements use GTH potentials with 13 electrons for In, 12 for Cd, 10 for Ba, and 6 for Te. The Li GTH inputs use three explicit electrons. AE treatment is nonrelativistic and no explicit spin--orbit coupling is included.

The available phases use two-atom primitive FCC cells and half-shifted $6\times6\times6$ Monkhorst--Pack meshes\cite{Monkhorst1976} with 28 irreducible points. The cubic geometries follow the experimental lattices and phases tabulated by Heyd \emph{et al.}\cite{Heyd2005Gaps} LiH uses rocksalt $a=4.084$~\AA{} from the room-temperature geometry reported by Nolan \emph{et al.}\cite{Nolan2009LiH} LiF uses rocksalt $a=4.010$~\AA{} from Matsushita \emph{et al.}\cite{Matsushita2011Gaps}   Zincblende MgS here is distinct from the rocksalt MgS structural test.

The calculations use fixed insulating occupations and k-point diagonalization, with SCF thresholds of $10^{-6}$ or tighter.

Fixed-density band evaluation combines the irreducible $\Gamma$-centred $12^3$ mesh with symmetry-path sampling, giving 198 points. We determine the generalized Kohn--Sham gap from the conduction-band minimum and valence-band maximum across all calculated $k$ points and also report the smallest direct gap. The band extrema need not occur at high-symmetry points.

\subsection{Calculated band gaps}

\begingroup\small\setlength{\tabcolsep}{3pt}
\begin{longtable}{llrrrrrrrr}
\caption{Native and representative published band gaps in eV for all 28 materials. AE, GTH, and Mixed 1C denote GAPW-AE, GAPW-XC/GTH, and mixed AE/GTH one-centre calculations, respectively. D, Z, and R denote diamond, zincblende, and rocksalt. Experimental, PBE, and SCAN values follow Ref.~\citenum{Lee2021Gaps}. HSE and PBE0 follow Ref.~\citenum{Lee2022Hybrids}. A dash denotes an unavailable value.}\label{esi-tab:bandgap-outcomes}\\
\toprule
Material & Phase & Exp. & AE & GTH & Mixed 1C & PBE & SCAN & HSE & PBE0 \\
\midrule
\endfirsthead
\toprule
Material & Phase & Exp. & AE & GTH & Mixed 1C & PBE & SCAN & HSE & PBE0 \\
\midrule
\endhead
\bottomrule
\endfoot
C & D & 5.48 & 4.895 & 4.735 & -- & 4.33 & 4.64 & 5.32 & 6.07 \\
Si & D & 1.17 & 1.284 & 0.951 & -- & 0.66 & 0.93 & 1.15 & 1.79 \\
Ge & D & 0.74 & 0.827 & 0.229 & -- & 0.00 & 0.20 & 0.69 & 1.28 \\
SiC & Z & 2.42 & 2.036 & -- & -- & 1.50 & 1.81 & 2.26 & 2.97 \\
BN & Z & 6.22 & 5.828 & 5.358 & -- & 4.64 & 5.03 & 5.80 & 6.55 \\
BP & Z & 2.40 & 1.949 & 1.712 & -- & 1.38 & 1.67 & 2.00 & 2.69 \\
BAs & Z & 1.46 & 1.763 & 1.476 & -- & 1.34 & 1.57 & 1.82 & 2.49 \\
AlP & Z & 2.51 & 2.399 & -- & -- & 1.67 & 1.99 & 2.29 & 2.96 \\
AlAs & Z & 2.23 & 2.269 & -- & -- & 1.58 & 1.86 & 2.03 & 2.68 \\
$\beta$-GaN & Z & 3.30 & 2.260 & 1.854 & -- & 1.79 & 1.86 & 2.72 & 3.42 \\
GaP & Z & 2.35 & 2.122 & 1.923 & -- & 1.66 & 1.85 & 2.25 & 2.92 \\
GaAs & Z & 1.52 & 1.525 & 0.852 & -- & 0.51 & 0.62 & 1.18 & 1.79 \\
InP & Z & 1.42 & -- & 1.226 & 1.380 & 0.61 & 0.57 & 1.33 & 1.96 \\
InAs & Z & 0.41 & -- & 0.239 & 0.388 & 0.00 & 0.00 & -- & -- \\
ZnS & Z & 3.66 & 3.514 & 2.536 & -- & 2.12 & 2.38 & 3.08 & 3.77 \\
ZnSe & Z & 2.70 & 2.570 & 1.650 & -- & 1.33 & 1.60 & 2.09 & 2.74 \\
CdS & Z & 2.55 & -- & 1.652 & 1.813 & 1.15 & 1.22 & 1.96 & 2.63 \\
CdSe & Z & 1.90 & -- & 1.147 & 1.243 & 0.64 & 0.72 & 1.29 & 1.92 \\
MgS & Z & 5.40 & 4.580 & 4.234 & -- & 3.57 & 4.14 & -- & -- \\
MgTe & Z & 3.60 & -- & 2.944 & 2.979 & 2.59 & 3.16 & -- & -- \\
MgO & R & 7.22 & 5.900 & 5.660 & -- & 4.93 & 5.59 & -- & -- \\
MgSe & R & 2.47 & 2.675 & 2.431 & -- & 2.01 & 2.58 & -- & -- \\
BaS & R & 3.88 & -- & 2.544 & 2.639 & 2.26 & 2.58 & -- & -- \\
BaSe & R & 3.58 & -- & 2.226 & 2.350 & 2.07 & 2.39 & -- & -- \\
BaTe & R & 3.08 & -- & 1.782 & -- & 1.74 & 2.03 & -- & -- \\
LiH & R & 4.90 & 4.103 & -- & -- & 3.01 & 3.61 & 4.01 & 4.69 \\
LiF & R & 14.20 & 10.761 & -- & -- & 9.33 & 10.08 & 11.44 & 12.22 \\
LiCl & R & 9.40 & 7.828 & 7.333 & -- & 6.40 & 7.21 & 7.78 & 8.52 \\
\end{longtable}
\endgroup
\begin{table}[htbp]
\centering
\caption{Native errors in eV on the available population for each method and literature comparison. The semilocal and hybrid XC benchmark populations follow Refs.~\citenum{Lee2021Gaps} and \citenum{Lee2022Hybrids}, respectively. The material lists below define the population for each row.}\label{esi-tab:bandgap-native-statistics}
\begin{tabular}{llrrrrl}
\toprule
Comparison & Method & $n$ & ME & MAE & RMSE & Maximum \\
\midrule
Semilocal & GAPW-AE & 20 & -0.533 & 0.608 & 0.988 & 3.439 (LiF) \\
Semilocal & GAPW-XC/GTH & 23 & -0.835 & 0.837 & 0.987 & 2.067 (LiCl) \\
Semilocal & Mixed AE/GTH (1C) & 7 & -0.650 & 0.650 & 0.794 & 1.241 (BaS) \\
Hybrid XC & GAPW-AE & 17 & -0.513 & 0.578 & 1.002 & 3.439 (LiF) \\
Hybrid XC & GAPW-XC/GTH & 15 & -0.776 & 0.778 & 0.926 & 2.067 (LiCl) \\
Hybrid XC & Mixed AE/GTH (1C) & 3 & -0.478 & 0.478 & 0.570 & 0.737 (CdS) \\
\bottomrule
\end{tabular}
\end{table}
\FloatBarrier
\paragraph{Available populations.}
Semilocal comparison, GAPW-AE ($n=20$) includes AlAs, AlP, BAs, BN, BP, C, GaAs, GaP, Ge, LiCl, LiF, LiH, MgO, MgS, MgSe, Si, SiC, ZnS, ZnSe, $\beta$-GaN.

Semilocal comparison, GAPW-XC/GTH ($n=23$) includes BAs, BN, BP, BaS, BaSe, BaTe, C, CdS, CdSe, GaAs, GaP, Ge, InAs, InP, LiCl, MgO, MgS, MgSe, MgTe, Si, ZnS, ZnSe, $\beta$-GaN.

Semilocal comparison, Mixed AE/GTH (1C) ($n=7$) includes BaS, BaSe, CdS, CdSe, InAs, InP, MgTe.

Hybrid XC comparison, GAPW-AE ($n=17$) includes AlAs, AlP, BAs, BN, BP, C, GaAs, GaP, Ge, LiCl, LiF, LiH, Si, SiC, ZnS, ZnSe, $\beta$-GaN.

Hybrid XC comparison, GAPW-XC/GTH ($n=15$) includes BAs, BN, BP, C, CdS, CdSe, GaAs, GaP, Ge, InP, LiCl, Si, ZnS, ZnSe, $\beta$-GaN.

Hybrid XC comparison, Mixed AE/GTH (1C) ($n=3$) includes CdS, CdSe, InP.
\begin{table}[htbp]
\centering
\caption{Native errors in eV on matched core-treatment subsets. AE-full contains the fifteen materials with both pure-AE and GTH results, AE-XC overlap its twelve-material overlap with the hybrid XC benchmark, and Mixed the seven compounds with mixed AE/GTH and GTH results. The material lists below define these sets.}\label{esi-tab:bandgap-core-subsets}
\begin{tabular}{lrlrrrl}
\toprule
Set & $n$ & Method & ME & MAE & RMSE & Maximum \\
\midrule
AE-full & 15 & Skala AE & -0.398 & 0.493 & 0.679 & 1.572 (LiCl) \\
AE-full & 15 & Skala GTH & -0.837 & 0.839 & 1.008 & 2.067 (LiCl) \\
AE-XC overlap & 12 & Skala AE & -0.336 & 0.421 & 0.609 & 1.572 (LiCl) \\
AE-XC overlap & 12 & Skala GTH & -0.816 & 0.819 & 0.976 & 2.067 (LiCl) \\
Mixed & 7 & Skala mixed 1C & -0.650 & 0.650 & 0.794 & 1.241 (BaS) \\
Mixed & 7 & Skala GTH & -0.766 & 0.766 & 0.885 & 1.354 (BaSe) \\
\bottomrule
\end{tabular}
\end{table}
\FloatBarrier
\paragraph{Matched core-treatment populations.}
AE-full comprises BAs, BN, BP, C, GaAs, GaP, Ge, LiCl, MgO, MgS, MgSe, Si, ZnS, ZnSe, $\beta$-GaN.
AE-XC overlap comprises BAs, BN, BP, C, GaAs, GaP, Ge, LiCl, Si, ZnS, ZnSe, $\beta$-GaN.
Mixed comprises BaS, BaSe, CdS, CdSe, InAs, InP, MgTe.

\subsection{Reference populations and statistical interpretation}

The experimental values and ten local or semilocal comparators are taken from Table~1 of the author manuscript in Ref.~\citenum{Lee2021Gaps}. The twelve hybrid comparators use Table~AII of the author manuscript in Ref.~\citenum{Lee2022Hybrids}, specifically the truncated-Coulomb band results, not its Madelung-only table. The local and semilocal study uses GTH-LDA and unc-def2-QZVP-GTH. The hybrid study uses GTH with the uncontracted basis, a Madelung correction during SCF, and truncated Coulomb exchange for band evaluation.

The SC/40 reference compilation underlying most of these experimental entries includes alloy-extrapolated values and weighted averages of measured spin--orbit splittings intended to approximate scalar-relativistic gaps.\cite{Heyd2005Gaps} The reference column is therefore a historical literature benchmark, not a uniform collection of raw optical onsets at one temperature.

The local and semilocal comparison contains LDA, five GGAs (PBE, PBEsol, revPBE, BLYP, and B97-D), and four meta-GGAs (SCAN, M06-L, MN15-L, and B97M-rV). The classification describes the underlying XC form and does not remove the dispersion or nonlocal correlation terms in the named approximations. All twelve hybrid comparators include exact exchange. They comprise global hybrid GGAs, global hybrid meta-GGAs, the screened hybrid HSE, and range-separated hybrids.

For each population, gap errors are $E_g^{\rm calc}-E_g^{\rm ref}$. ME, MAE, and RMSE follow the definitions in Sec.~\ref{esi-sec:scope}, and the maximum is the largest absolute error. Table~\ref{esi-tab:bandgap-native-statistics} summarizes the native errors on the largest available sets, which differ between methods and publications. The machine-readable dataset contains all four measures and individual signed errors for every maximum-available and matched population, including the full literature comparisons and the local or semilocal functionals evaluated on the hybrid-benchmark populations. The matched tables below use identical material sets to separate differences in accuracy from differences in coverage.

Of the 28 materials in Table~\ref{esi-tab:bandgap-outcomes}, 22 have both GAPW-XC/GTH and the preselected AE-or-mixed one-centre result. The combined series comprises fifteen pure-AE and seven genuinely mixed results. Table~\ref{esi-tab:bandgap-core-subsets} reports the pure-AE/GTH and mixed/GTH comparisons on separate matched subsets, including the pure-AE overlap with the hybrid benchmark.

Fifteen of the 22 materials also have published hybrid-functional results. This overlap contains twelve pure-AE and three mixed results (InP, CdS, and CdSe). BaS, BaSe, InAs, MgO, MgS, MgSe, and MgTe have only local or semilocal comparators here. Table~\ref{esi-tab:bandgap-paired-combined-Lee2021} reports all error measures on the full 22-material population. Main-text Table~\ref{tab:bandgap-method-errors} uses exactly the fifteen-material overlap for the local, semilocal, and hybrid functionals.

\begingroup\small\setlength{\tabcolsep}{4pt}
\begin{table}[htbp]
\centering
\caption{Matched errors in eV on the 22 materials with both AE-or-mixed and GTH results in Table~\ref{esi-tab:bandgap-outcomes}. Skala AE / mixed 1C comprises fifteen AE and seven mixed results. All ten local or semilocal comparators use exactly this population.\cite{Lee2021Gaps} The set comprises BAs, BN, BP, BaS, BaSe, C, CdS, CdSe, GaAs, GaP, Ge, InAs, InP, LiCl, MgO, MgS, MgSe, MgTe, Si, ZnS, ZnSe, $\beta$-GaN.}\label{esi-tab:bandgap-paired-combined-Lee2021}
\begin{tabular}{llrrrl}
\toprule
Method & XC class & ME & MAE & RMSE & Maximum \\
\midrule
Skala AE / mixed 1C & Native & -0.478 & 0.543 & 0.718 & 1.572 (LiCl) \\
Skala GTH & Native & -0.814 & 0.816 & 0.971 & 2.067 (LiCl) \\
LDA & LDA & -1.452 & 1.452 & 1.612 & 3.390 (LiCl) \\
PBE & GGA & -1.220 & 1.220 & 1.379 & 3.000 (LiCl) \\
PBEsol & GGA & -1.361 & 1.361 & 1.518 & 3.220 (LiCl) \\
revPBE & GGA & -1.125 & 1.125 & 1.280 & 2.790 (LiCl) \\
BLYP & GGA & -1.105 & 1.117 & 1.275 & 2.840 (LiCl) \\
B97-D & GGA & -1.024 & 1.035 & 1.181 & 2.570 (LiCl) \\
SCAN & meta-GGA & -0.924 & 0.944 & 1.075 & 2.190 (LiCl) \\
M06-L & meta-GGA & -0.875 & 0.928 & 1.115 & 2.320 (LiCl) \\
MN15-L & meta-GGA & -1.013 & 1.087 & 1.219 & 1.920 (LiCl) \\
B97M-rV & meta-GGA & -0.850 & 0.889 & 1.034 & 2.180 (LiCl) \\
\bottomrule
\end{tabular}
\end{table}
\FloatBarrier
\endgroup

On the 22-material population, the AE-or-mixed and GAPW-XC/GTH MAEs are 0.543 and 0.816~eV, compared with 1.220 for PBE, 0.944 for SCAN, and 0.889~eV for B97M-rV. Their signed MEs of $-0.478$ and $-0.814$~eV indicate predominantly underestimated gaps. On the fifteen-material overlap, the native MAEs become 0.432 and 0.778~eV. HSE, PBE0, and B3LYP give 0.435, 0.402, and 0.405~eV. The AE-or-mixed RMSE of 0.602~eV exceeds their respective 0.578, 0.502, and 0.510~eV values.

The 22-material comparison uses the published MgO reference of 7.22~eV.\cite{Lee2021Gaps} Rafferty and Brown report a direct EELS threshold of $7.22\pm0.03$~eV in their Table~I, alongside a literature value of 7.8~eV.\cite{Rafferty1998Gaps}  Substituting the alternative literature reference of 7.83~eV raises the AE-or-mixed and GAPW-XC/GTH MAEs from 0.543/0.816 to 0.571/0.844~eV. PBE and SCAN change from 1.220/0.944 to 1.248/0.971~eV on the same population. The two reference values originate from EELS and optical measurements, respectively.

The fifteen-material population shared by AE and GTH retains Lee's MgO reference of 7.22~eV. Using 7.83~eV instead raises the respective MAEs from 0.493/0.839 to 0.534/0.880~eV. The PBE, SCAN, and B97M-rV MAEs on that same population change from 1.255/0.937/0.885 to 1.295/0.978/0.925~eV. The baseline statistics use the original reference values.

For cubic BAs, Yue \emph{et al.} determine an indirect absorption edge of 1.82~eV at room temperature, whereas both Lee tables use 1.46~eV.\cite{Yue2020BAs} Table~\ref{esi-tab:bandgap-BAs-sensitivity} changes only this reference for every method on the same populations. The 22-material AE-or-mixed/GTH MAEs become 0.532/0.831~eV. On the fifteen-material hybrid overlap, the AE-or-mixed MAE becomes 0.416~eV, compared with 0.411 for HSE, 0.378 for PBE0, and 0.381~eV for B3LYP. Which of Skala and HSE has the lower MAE therefore depends on the reference convention.

\begin{table}[htbp]
\centering
\caption{Separate BAs reference sensitivity in eV for representative comparators. Only the BAs reference changes from Lee's 1.46 to the 1.82~eV room-temperature absorption estimate of Yue \emph{et al.}\cite{Yue2020BAs} The populations remain the same 22 and fifteen materials as in Table~\ref{esi-tab:bandgap-paired-combined-Lee2021} and main-text Table~\ref{tab:bandgap-method-errors}, respectively. All other references and calculated gaps are unchanged. All comparator statistics and separate pure-AE subsets are retained in the machine-readable dataset. The Lee-reference tables use the unmodified published references.}\label{esi-tab:bandgap-BAs-sensitivity}
\begin{tabular}{rlrrrr}
\toprule
$n$ & Method & ME & MAE & RMSE & Maximum \\
\midrule
22 & Skala AE / mixed 1C & -0.495 & 0.532 & 0.715 & 1.572 \\
22 & Skala GTH & -0.831 & 0.831 & 0.973 & 2.067 \\
22 & PBE & -1.236 & 1.236 & 1.382 & 3.000 \\
22 & SCAN & -0.940 & 0.950 & 1.076 & 2.190 \\
22 & B97M-rV & -0.866 & 0.887 & 1.034 & 2.180 \\
15 & Skala AE / mixed 1C & -0.389 & 0.416 & 0.597 & 1.572 \\
15 & Skala GTH & -0.800 & 0.800 & 0.930 & 2.067 \\
15 & PBE & -1.205 & 1.205 & 1.345 & 3.000 \\
15 & SCAN & -0.971 & 0.971 & 1.087 & 2.190 \\
15 & B97M-rV & -0.869 & 0.869 & 1.016 & 2.180 \\
15 & HSE & -0.411 & 0.411 & 0.571 & 1.620 \\
15 & PBE0 & +0.261 & 0.378 & 0.459 & 0.880 \\
15 & B3LYP & +0.073 & 0.381 & 0.468 & 1.170 \\
\bottomrule
\end{tabular}
\end{table}
\FloatBarrier

The experimental reference set includes estimated band gaps and excitonic absorption energies.\cite{Wolverson2001MgS,Zollweg1958Absorption}

For both Ge and MgS, the sampled band gap is indirect in the AE calculations but direct with GAPW-XC/GTH, indicating that the AE/GTH protocol can influence not only the gap magnitude but also the relative positions of the band extrema (Table~\ref{esi-tab:bandgap-edges}).

\begin{longtable}{llrrl}
\caption{Sampled minimum and minimum direct gaps in eV. D denotes agreement within $10^{-6}$~eV, I a direct-minus-minimum splitting of at least $10^{-3}$~eV, and N an intermediate numerical near-degeneracy. The labels describe the sampled band edges.}\label{esi-tab:bandgap-edges}\\
\toprule
Material & Method & Minimum gap & Direct gap & Type \\
\midrule
\endfirsthead
\toprule
Material & Method & Minimum gap & Direct gap & Type \\
\midrule
\endhead
\bottomrule
\endfoot
AlAs & GAPW-AE & 2.269 & 3.420 & I \\
AlP & GAPW-AE & 2.399 & 4.460 & I \\
BAs & GAPW-AE & 1.763 & 3.875 & I \\
BAs & GAPW-XC/GTH & 1.476 & 3.761 & I \\
BN & GAPW-AE & 5.828 & 10.058 & I \\
BN & GAPW-XC/GTH & 5.358 & 10.057 & I \\
BP & GAPW-AE & 1.949 & 4.233 & I \\
BP & GAPW-XC/GTH & 1.712 & 4.228 & I \\
BaS & GAPW-XC/GTH & 2.544 & 2.665 & I \\
BaS & Mixed AE/GTH (1C) & 2.639 & 2.755 & I \\
BaSe & GAPW-XC/GTH & 2.226 & 2.451 & I \\
BaSe & Mixed AE/GTH (1C) & 2.350 & 2.561 & I \\
BaTe & GAPW-XC/GTH & 1.782 & 2.176 & I \\
C & GAPW-AE & 4.895 & 6.716 & I \\
C & GAPW-XC/GTH & 4.735 & 6.202 & I \\
CdS & GAPW-XC/GTH & 1.652 & 1.652 & D \\
CdS & Mixed AE/GTH (1C) & 1.813 & 1.813 & D \\
CdSe & GAPW-XC/GTH & 1.147 & 1.147 & D \\
CdSe & Mixed AE/GTH (1C) & 1.243 & 1.243 & D \\
GaAs & GAPW-AE & 1.525 & 1.525 & D \\
GaAs & GAPW-XC/GTH & 0.852 & 0.852 & D \\
GaP & GAPW-AE & 2.122 & 2.761 & I \\
GaP & GAPW-XC/GTH & 1.923 & 2.247 & I \\
Ge & GAPW-AE & 0.827 & 1.039 & I \\
Ge & GAPW-XC/GTH & 0.229 & 0.229 & D \\
InAs & GAPW-XC/GTH & 0.239 & 0.239 & D \\
InAs & Mixed AE/GTH (1C) & 0.388 & 0.388 & D \\
InP & GAPW-XC/GTH & 1.226 & 1.226 & D \\
InP & Mixed AE/GTH (1C) & 1.380 & 1.380 & D \\
LiCl & GAPW-AE & 7.828 & 7.828 & D \\
LiCl & GAPW-XC/GTH & 7.333 & 7.333 & D \\
LiF & GAPW-AE & 10.761 & 10.761 & D \\
LiH & GAPW-AE & 4.103 & 4.103 & N \\
MgO & GAPW-AE & 5.900 & 5.900 & D \\
MgO & GAPW-XC/GTH & 5.660 & 5.660 & D \\
MgS & GAPW-AE & 4.580 & 4.664 & I \\
MgS & GAPW-XC/GTH & 4.234 & 4.234 & D \\
MgSe & GAPW-AE & 2.675 & 4.136 & I \\
MgSe & GAPW-XC/GTH & 2.431 & 3.601 & I \\
MgTe & GAPW-XC/GTH & 2.944 & 2.944 & D \\
MgTe & Mixed AE/GTH (1C) & 2.979 & 2.979 & D \\
Si & GAPW-AE & 1.284 & 3.212 & I \\
Si & GAPW-XC/GTH & 0.951 & 2.865 & I \\
SiC & GAPW-AE & 2.036 & 5.271 & I \\
ZnS & GAPW-AE & 3.514 & 3.514 & D \\
ZnS & GAPW-XC/GTH & 2.536 & 2.536 & D \\
ZnSe & GAPW-AE & 2.570 & 2.570 & D \\
ZnSe & GAPW-XC/GTH & 1.650 & 1.650 & D \\
$\beta$-GaN & GAPW-AE & 2.260 & 2.260 & D \\
$\beta$-GaN & GAPW-XC/GTH & 1.854 & 1.854 & D \\
\end{longtable}

\subsection{Direct and one-centre valence representations}

The controls in Table~\ref{esi-tab:bandgap-controls} change the GTH valence representation within ordinary GAPW, not the AE reconstruction or the primary protocol. The Si entries use GTH pseudopotentials for all atoms, while BP, MgO, and LiCl use the mixed AE/GTH treatments listed below.  For InP, CdS, and BaS, the direct and one-centre calculations use identical scientific settings apart from \texttt{PAW\_ONE\_CENTER} versus \texttt{DIRECT\_VALENCE}. The executable environments and thread counts are recorded with the input data.

\begin{table}[htbp]
\centering
\caption{Ordinary-GAPW direct and one-centre gaps in eV for seven materials. The final column gives the one-centre-minus-direct difference.}\label{esi-tab:bandgap-controls}
\begin{tabular}{llrrr}
\toprule
Material & Actual core treatment & Direct & One-centre & One-centre minus direct \\
\midrule
Si & All GTH & 0.950857 & 0.950856 & -0.00000038 \\
BP & B AE / P GTH & 1.794049 & 1.794045 & -0.00000405 \\
MgO & O AE / Mg GTH & 5.913617 & 5.922130 & +0.00851297 \\
LiCl & Li AE / Cl GTH & 7.338573 & 7.338544 & -0.00002818 \\
InP & Mixed AE/GTH & 1.378229 & 1.379532 & +0.001303 \\

CdS & Mixed AE/GTH & 1.812347 & 1.812683 & +0.000337 \\

BaS & Mixed AE/GTH & 2.638700 & 2.638570 & -0.000131 \\

\bottomrule
\end{tabular}
\end{table}
\FloatBarrier

Direct and one-centre gaps are compared for seven materials, comprising six genuinely mixed AE/GTH systems (BP, MgO, LiCl, InP, CdS, and BaS) and one all-GTH system (Si). The absolute direct-minus-one-centre gap differences are below 0.009~eV throughout this subset. The largest change is 0.008513~eV for MgO, while the direct and one-centre results for InP, CdS, and BaS differ by at most 0.001303~eV. These results show a small sensitivity to the valence representation for the tested systems and numerical settings.

\FloatBarrier

\end{document}